\documentclass[useAMS,usenatbib]{mn2e}

\usepackage{epsfig}
\usepackage{multirow}
\usepackage{multicol}
\usepackage{longtable}
\usepackage{threeparttablex}
\usepackage{lscape}
\usepackage{amssymb,amsmath}
\usepackage{url}

\title[Subaru Telescope adaptive optics observations of gravitationally lensed quasars in the SDSS]{Subaru Telescope adaptive optics observations of gravitationally lensed quasars in the Sloan Digital Sky Survey\thanks{Based on data collected at Subaru Telescope, which is
operated by the National Astronomical Observatory of Japan.}}
\author[C.E.~Rusu et al.]
{Cristian E. Rusu,$^{1,2}$\thanks{E-mail: cerusu@ucdavis.edu}
Masamune Oguri,$^{3,4,5}$\thanks{E-mail: masamune.oguri@ipmu.jp} 
Yosuke Minowa,$^{6,8}$ 
Masanori Iye,$^{2,7,8}$
 \newauthor
Naohisa Inada,$^{9}$ 
Shin Oya,$^{6}$
Issha Kayo,$^{10}$
Yutaka Hayano,$^{6}$
Masayuki Hattori,$^{6}$
 \newauthor
Yoshihiko Saito,$^{6}$
Meguru Ito,$^{6}$
Tae-Soo Pyo,$^{6}$
Hiroshi Terada,$^{6}$
Hideki Takami$^{6}$
 \newauthor
and
Makoto Watanabe$^{11}$ \\
$^1$ Department of Physics, University of California, Davis, CA 95616, USA\\ 
$^2$Optical and Infrared Astronomy Division, National Astronomical
Observatory of Japan, 2-21-1 Osawa, Mitaka, Tokyo 181-8588, \\ 
Japan\\ 
$^3$Research Center for the Early Universe, University of Tokyo,
7-3-1 Hongo, Bunkyo-ku, Tokyo 113-0033, Japan\\
$^4$Department of Physics, University of Tokyo, 7-3-1 Hongo,
Bunkyo-ku, Tokyo 113-0033, Japan\\
$^5$Kavli Institute for the Physics and Mathematics of the Universe
(Kavli IPMU, WPI), University of Tokyo, Chiba 277-8583, Japan\\
$^6$Subaru Telescope, National Astronomical Observatory of Japan,
	       650 North A'ohoku Place, Hilo, Hawaii 96720, USA\\
$^7$Department of Astronomy, University of Tokyo, 
7-3-1 Hongo, Bunkyo-ku, Tokyo 113-0033, Japan\\
$^8$Department of Astronomical Science, The Graduate University
                for Advanced Studies (SOKENDAI), National Astronomical\\ 
                Observatory of Japan, 2-21-1,Osawa, Mitaka, Tokyo 181-8588, Japan\\                 
$^9$Department of Physics, Nara 
	      National College of Technology, Yamatokohriyama,
	      Nara 639-1080, Japan\\    
$^{10}$Department of Liberal Arts, Tokyo University of Technology, 
           5-23-22 Nishikamata, Ota-ku, Tokyo 114-8650, Japan\\    
$^{11}$Department of Cosmosciences, Hokkaido University,
	       Kita 10, Nishi 8, Kita-ku, Sapporo, Hokkaido 060-0810, Japan\\    	      
}

\begin{document}


\pagerange{\pageref{firstpage}--\pageref{lastpage}} \pubyear{2014}

\maketitle

\label{firstpage}

\begin{abstract}
We present the results of an imaging observation campaign conducted with the Subaru Telescope adaptive optics system (IRCS+AO188) on 28 gravitationally lensed quasars and candidates (23  doubles, 1 quad, 1 possible triple and 3 candidates) from the SDSS Quasar Lens Search. We develop a novel modelling technique that fits analytical and hybrid point spread functions (PSFs), while simultaneously measuring the relative astrometry, photometry, as well as the lens galaxy morphology. We account for systematics by simulating the observed systems using separately observed PSF stars. The measured relative astrometry is comparable with that typically achieved with the Hubble Space Telescope, even after marginalizing over the PSF uncertainty. We model for the first time the quasar host galaxies in 5 systems, without a-priory knowledge of the PSF, and show that their luminosities follow the known correlation with the mass of the supermassive black hole. For each system, we obtain mass models far more accurate than those previously published from low-resolution data, and we show that in our sample of lensing galaxies the observed light profile is more elliptical than the mass, for ellipticity $\gtrsim0.25$. We also identify eight doubles for which the sources of external and internal shear are more reliably separated, and should therefore be prioritized in monitoring campaigns aimed at measuring time-delays in order to infer the Hubble constant.
\end{abstract}



\section{Introduction}\label{section:intro}

The first gravitationally lensed quasar \citep{walsh79} has been discovered more than 30 years ago, turning gravitational lensing from an  obscure theoretical field into a mainstream observational one. More than 100 strongly lensed quasars have been discovered to date
, and it has convincingly been demonstrated that these objects provide insights into various topics in astrophysics and cosmology, as well as being a unique tool for studying the dark universe. Applications include the study of the quasar host galaxies at high redshift \citep[e.g.,][]{peng06-2}, dark matter substructures and luminous satellites \citep[e.g.,][]{chiba05,mckean09,minezaki09,vegetti12}, the structure and evolution of massive galaxies \citep[e.g.,][]{kochanek00,rusin05,koopmans09,oguri14}, and microlensing applied in the study of the structure of quasar accretion disk \citep[e.g.,][]{agol99,dai10}, broad line regions\citep[e.g.,][]{richards04,sluse012}, as well as to measure the stellar mass fractions in the lens \citep[e.g.,][]{schechter02,mediavilla09,schechter14}. Following early work by e.g. \citet[][]{turner84,fukugita90,kochanek96}, analyses of statistically well-defined samples of lensed quasars (i.e. samples in which lens candidates are selected by a homogeneous method whose completeness is known) can now constrain the cosmological constant/dark energy by comparing the number and distribution of image separation of lensed quasars with theoretical models \citep[e.g.,][]{oguri12}. Time delay measurements between quasar images constrain the Hubble constant free of the calibration in the cosmic distance ladder \citep[e.g.,][]{oguri07}. Finally, the distribution of lensed image separations, from galaxy to cluster mass scales, reflects the hierarchical structure formation and the effects of baryon cooling \citep[e.g.,][]{kochanek01}.


 
 \medskip
 
 The Sloan Digital Sky Survey Quasar Lens Search \citep[SQLS;][]{oguri06-1,inada12} is a systematic survey for lensed quasars, aiming to construct a large sample of gravitationally lensed quasars at optical wavelengths. It relies on the large homogeneous sample of
spectroscopically-confirmed quasars from the Sloan Digital Sky Survey \citep[SDSS;][]{york00}. The techniques employed by the SQLS to identify lensed quasar candidates are described in the references above. We followed up all candidates with deeper imaging (usually with the University of Hawaii 2.2m Telescope; UH88) to detect the lensing galaxy. We then performed follow-up spectroscopy of the most promising candidates, to confirm their lensing nature. SQLS is at present the prominent search for lensed quasars in the optical, comprising of 62 lensed quasars to date (December 2014), $\sim$2/3 of which are new discoveries (\citealt{kayo10,inada12})\footnotemark \footnotetext{A publicly available list of all lensed quasars in the SQLS is available at \url{http://www-utap.phys.s.u-tokyo.ac.jp/~sdss/sqls/}}. It has also produced the largest statistically complete sample of lensed quasars (26 objects; \citealt{inada12}).
 
 A disadvantage of SQLS, like other ground-based optical strong lens surveys, is its poor detail in imaging lensed quasars. Even when performing follow-up observations with the UH88 telescope, the pixel scale $\sim0.22''$ is large, and the seeing $\sim0.8''$ is similar to the image separation of a galaxy-scale strong lens ($\sim 1''-2''$). Therefore, high-resolution imaging of these quasar lenses is the key to turning each lens into a highly useful astrophysical and cosmological probe. This is necessary for obtaining accurate relative astrometry and point/surface photometry for the quasar images, lensing galaxy, and quasar host galaxy (in case the latter is detected), which are used to constrain both the light and the mass distribution in these systems.
 
In the following, we highlight three of the applications enabled by the high-resolution images of a large sample of objects, such as the sample provided by our work.
 
{\it Estimating the Hubble constant from high resolution and time delay measurements:} Although monitoring observations, for which only relative fluxes are relevant, can be performed with small telescopes aided by image deconvolution, to determine time delays measurements between multiple images, \citep[e.g.,][]{eulaers13,rathna13}, high resolution single epoch observations are still required to construct an accurate lens model \citep[e.g.,][]{suyu12}. As time delays are currently being measured by the COSmological MOnitoring of GRAvItational Lenses \citep[COSMOGRAIL;][]{eigenbrod05} for many of the SQLS lenses, high resolution data is in demand. For example, \citet{courbin97} obtained high resolution images of a four-image lensed quasar with an early adaptive optics systems, resulting in a relative lens galaxy position three times more precise than before, which allowed to measure the Hubble constant two times more precisely than in previous studies.


{\it Quasar galaxy hosts and the correlation with $M_\mathrm{BH}$:}
The tight correlations found between the masses of supermassive black holes $M_\mathrm{BH}$ and overall properties of the host galaxy bulges, such as velocity dispersion, luminosity and stellar mass \citep[e.g.,][]{kormendy95,ferrarese00,haring04,gultekin11} suggest that the black hole growth is coupled to the galaxy evolution. These correlations have been established for local galaxies, based on spatially resolved kinematics. Since spatially resolved kinematics are very difficult to obtain at $z\gtrsim1$, the most straightforward correlation that can be studied at large redshifts is the one between $M_\mathrm{BH}$ and the bulge galaxy luminosity. However, in this case AGNs or quasars must be employed, as the nuclear activity allows the estimation of $M_\mathrm{BH}$ using the virial technique \citep[e.g.,][]{ho99,vestergaard02}. 

The difficulty then lies in accurately decomposing the faint quasar host, which is subject to the cosmological surface brightness dimming, from the bright nuclear source. As was demonstrated by \citet{peng06-2}, this task is facilitated for gravitationally lensed quasars by two effects. First, strong gravitational lensing produces natural magnification typically by several factors, by increasing the total flux and the apparent size of the host galaxy, while conserving surface brightness. Second, lensing distorts the host into an arc and gives it a dramatically different morphology from that of the Point Spread Function (PSF), making it significantly less susceptible to systematic problems in the PSF model (an inaccurately known PSF). High-resolution observations are paramount in measuring the surface brightness of the lensed host galaxy.


{\it The ellipticity and orientation of mass and light in the lensing galaxies:} An ongoing question is to what extent there are correlations between the azimuthal luminous profile of the lens galaxies and their overall (dark + luminous) mass profile. Strong lensing provides a unique tool for testing these correlations. There exists a known correlation between the mass and light orientation, within $\sim10$ deg \citep[e.g.,][]{ferreras08,treu09}. It is not presently known however if the scatter in the relation is intrinsic, or due to measurement errors. On the other hand, there is less consensus on a correlation between the mass and light ellipticity. For example, such a tight correlation was found by \citet{koopmans06} for the Sloan Lens ACS Survey \citep[SLACS;][]{bolton05} sample of galaxy-galaxy lenses (i.e. foreground galaxies acting as strong lenses for background galaxies), although \citet{ferreras08} did not find a correlation for a selected subsample of the SLACS lenses. A correlation was also found for the SL2S galaxy-galaxy lenses \citep{gavazzi12}, but has significantly more scatter. There are also recent results from different samples \citep[e.g.,][]{dye14,kostrzewa14}. The difference between the SLACS and SL2S lenses is that the former lie in an environment with significantly less average external shear $\langle \gamma \rangle \lesssim 0.035$, and also that they have smaller Einstein radius $R_\mathrm{Ein}$ relative to the characteristic scale (effective radius) of the galaxy $R_\mathrm{eff}$, resulting in the more relevant role played by stellar mass in defining the potential within the critical curve ($R_\mathrm{Ein}/R_\mathrm{eff}\sim1.1$, 0.5 for SL2S and SLACS, respectively). For much larger values of $R_\mathrm{Ein}/R_\mathrm{eff}$, studies based on weak lensing \citep[e.g.][]{hoekstra04} found that the dark matter haloes, which dominate the mass profile at these radii, are flatter than the luminous profile, in agreement with theoretical expectations.

For the lenses of strongly lensed quasars, initial Hubble Space Telescope ($HST$) observations have failed to find a correlation between the mass and light ellipticity \citep{keeton98}. More recent results employing a uniform analysis technique and deconvolution of $HST$ data are presented in \citet{sluse12} and show that there is a correlation for their sample of four-image lensed quasars (quads). 
The SLACS lenses and the lenses of strongly lensed quasars are known to probe different populations of lenses, with lensed quasars typically lying in richer environments \citep[e.g.,][]{oguri05}. In addition, for most lensed quasars $R_\mathrm{Ein}$ is on the order of a few $R_\mathrm{eff}$, therefore probing an intermediate range. The new correlation found by \citet{sluse12}, after eliminating the outliers, holds only for ellipticities $\lesssim0.25$. For these small ellipticities, as the authors note, one possible bias resulting in the correlation is that the effect of a rounder dark matter halo on the total mass distribution may be harder to detect. 
  
 
 
  \medskip
 
 The goal of this work is to enhance the value of the SQLS gravitationally lensed quasar sample through the use of high-resolution observations. Among the 62 lensed quasars in the SQLS sample, only $\sim22$ have available high-resolution observations, mainly supplied by $HST$. However, recent progress of adaptive optics (AO), especially the use of laser guide stars
(LGS), makes it possible to obtain high-resolution images using ground-based telescopes. Due to the smaller diffraction limit, near-infrared imaging with Subaru Telescope AO can potentially achieve {\it three times} the spatial resolution of $HST$, while also employing a finer pixel scale. 

AO has previously been used in the literature in the study of gravitationally lensed quasars. \citet{ledoux98}, \citet{auger08} and \citet{sluse08} used AO to study galaxy scale two-image lensed quasars (doubles). \citet{rusin00,courbin02,meylan05} discovered additional lensing galaxies in known lensed quasars, using AO. \citet{faure11} observed a quad in order to look for substructure, which \citet{mckean09} was successful in detecting for another quad. In addition, \citet{marshall07} and \citet{suyu11} have used AO to study the arcs of strongly lensed galaxies, reconstruct the source, or study in high detail the lensing galaxy. However, these have been isolated studies focused on single systems. The observational campaign described in this paper is the largest dedicated study of a large sample of lensed quasars with AO.

In Section \ref{section:obsall} we present our observing strategy and describe acquired data including data reduction. In Section \ref{section:technique} we present the morphological modelling technique we employed, and in Section \ref{section:lensmodels} the mass modelling technique. We continue with a comparison of our results with those found for the same systems in the literature (Section \ref{sect:discuss}), and a concluding discussion of the technique we used, in Section \ref{sect:discutionoverall}. Section \ref{section:hostfacts} shows results obtained on the quasar host galaxies. Finally, Section \ref{section:lightmass} studies the lens environment and the relation between mass and light for the systems in our sample, and Section \ref{sect:concl} concludes this work. In the main appendix, we describe in detail the analysis of each individual system. 
\mbox{}

Throughout this work, the concordance cosmology with $H_0=100 h\ \mathrm{km}^{-1}\  \mathrm{s}^{-1}\ \mathrm{Mpc}^{-1}$, $h=0.70$, $\Omega_m=0.27$ and $\Omega_\Lambda=0.73$ is assumed. For SDSS~J1001+5027 and SDSS~J1206+4332, where time delay measurements are available, the \citet{planck13} result, $h=0.673\pm0.012$, is assumed. Magnitudes are given in the Vega system. All object coordinates assume J2000. All observed images and residual plots are in logarithmic scale.


\section{Observational strategy and acquired data}\label{section:obsall}

\subsection{Observational strategy}\label{section:obs}

Sky coverage is a limiting factor when using AO, as all targets must be located close to a tip-tilt (TT) star of suitable brightness in order to perform AO correction, even in the case that LGS is used. Subaru Telescope however presently has the most relaxed constraints on TT star, among comparable systems on 8-10 m telescopes (Section \ref{section:ircsao}). This makes it an ideal tool to use in the observational campaign of the SQLS quasars. Indeed, 54 of the 62 SQLS quasars are accessible to the Subaru Telescope AO system.

\begin{figure}
\includegraphics[width=85mm]{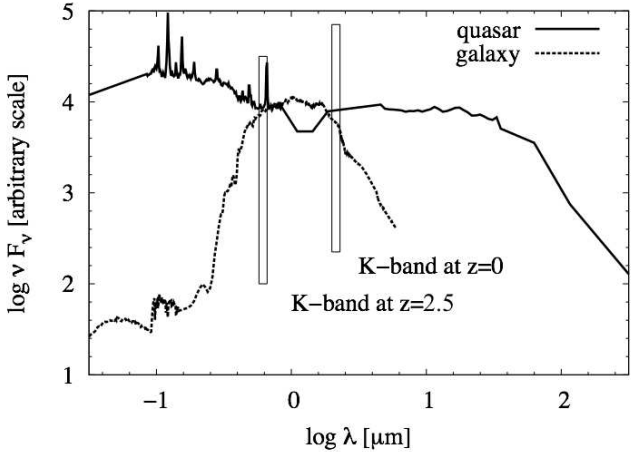}
\caption{ Typical galaxy (E-type) and radio quiet quasar spectrum, with the location of the $K$ band at redshifts 0 and 2.5 marked. The composite quasar spectrum is from \citet{shang11}. For the galaxy template, the mean spectrum of local galaxies from \citet{coleman80} is extrapolated using the evolutionary models of \citet{bruzual93}.
\label{fig:qsogal}}
\end{figure}

The AO correction functions better at longer wavelengths \citep[e.g.][]{davies08}, making $K$-band the natural choice among the widely used $JHK$ near-infrared bands. An additional reason to use $K$-band in the study of quasars and their host galaxies is that for redshifts $\lesssim 2.5$, $K$-band falls in a wavelength region where the quasar host galaxy is brightest compared to the nuclear source (Figure \ref{fig:qsogal}). Moreover, for the purpose of modelling gravitationally lensed quasars, extinction, intrinsic variability and microlensing are all weaker at progressively longer wavelengths, contaminating the true image flux ratio less \citep[e.g.][]{yonehara08}. However, exposure times in the $K$-band are limited by the strong sky background emission. On the other hand, $K'$-band provides a good compromise between background level, sensitivity, and AO performance, and we therefore used this band throughout the observation campaign.

As one of the goals of the campaign is to measure accurate lens galaxy morphologies and potentially resolve the quasar host galaxies from the point-like nuclear source, a good knowledge of the PSF is essential. For the majority of the systems, we used the SDSS CasJobs\footnote{\protect\url{http://skyserver.sdss3.org/casjobs/}}  search query to identify bright PSF and guide star pairs. These were selected such that the guide stars of the PSF and the target have similar brightness, and the separations between the target and its guide star, as well as the PSF and its respective guide star, are similar. The PSFs are close on the sky to the target (within $\sim30'$), and they were observed immediately before or after the target, in the same natural guide star (NGS) or LGS mode. This was done to ensure as much as possible that the atmospherical turbulence characteristics and therefore the AO correction is relatively constant between the PSF and the target. However, as we will discuss below, these separately observed PSF stars turn out not to be suitable for the accurate characterisation of the target image PSFs. SDSS~J1515+1511 and SDSS~J1002+4449 are the only observed systems that have a bright star in the detector field of view (FOV), which can potentially be used as PSF. Here we centred the FOV, and therefore the LGS, between the star and the target. This was done, again, so that the AO correction at the location of the target and the PSF star is similar.



\subsection{Acquired data and data reduction}\label{section:data}

\subsubsection{IRCS+AO188}\label{section:ircsao}

The AO imaging observations were performed with The Infrared Camera and Spectrograph \citep[IRCS;][]{kobayashi00} at the 
Subaru telescope, along with the Laser Guide Star Adaptive Optics system \citep[LGS$+$AO188;][]{hayano08,hayano10,minowa10}. IRCS uses two $1024\times 1024$ ALLADIN III arrays for imaging and spectroscopy, and in imaging mode provides two plate scales, of 52 and 20 mas per pixel. After the distortion correction described in Section \ref{section:distort}, the pixel scales are  $0.05284''\pm0.000017''$ and $0.02053''\pm0.000005''$, respectively. The available FOVs are $54''$ and $21''$, respectively. 

 AO188 uses a curvature sensor with 188 control elements, operating at 2000 Hz, and a 188 element bimorph deformable mirror (BIM188). It operates at the Nasmyth focus of the Subaru Telescope. Both NGS and LGS modes are available. The recommended guide star constraints in NGS mode are brightness larger than $R=16.5$ mag, within $30''$ from the target. The LGS mode uses an artificial sodium laser guide star \citep{hayano06} for high-order wavefront sensing. The LGS is typically less bright than $\sim10.5$ mag in $R$-band. The recommended tip-tilt star constraints in LGS mode are brightness larger than $R=18$ mag, within $60''$ (acceptable up to $90''$) from the target. More information about AO188 is provided on the instrument web page\footnotemark. \footnotetext{\protect\url{http://www.naoj.org/Observing/Instruments/AO/system.html}}
 

\subsubsection{Acquired data}\label{section:obsdata}

The observations were performed between 2011-2014. In total, $\sim7$ nights were assigned for the present campaign, about half of which were lost due to telescope/instrument trouble or cloud coverage. Observation priority was given based on the scientific interest of each target and the degree to which the targets are suitable for AO observations in terms of TT star brightness and separation, and distance to zenith/airmass. The 28 objects that were successfully observed, the observation modes, exposure times, filters, pixel scales, typical airmass, photometric stars and observation dates are given in Table \ref{tab:followup-data}. Two of the objects, SDSS~J0820+0812 and SDSS~J1206+4332, were observed as back-up, without AO.

All observations were performed with 5 or 9-point dithering, in order to remove bad pixels and cosmic rays, and to allow for flat frame and sky frame creation from the data. Although dome flats were obtained for some observations, these have proven to be of inferior quality. Although the targets were chosen to be accessible to AO, many of these border the recommended limit in terms of the TT brightness and separation. As such, low Strehl ratios $\lesssim10\%$ (Table \ref{tab:analpsf}) were typically obtained. The 52 mas pixel scale mode was typically used, in order to increase the signal-to-noise (S/N) and not to avoid significantly oversampled PSFs. The list of observed TT and PSF stars is given in Table \ref{tab:AO}. For 12 objects, PSF star observations were skipped because of rapidly changing atmospheric conditions. Typical seeing during the observations can be inferred from Table \ref{tab:analpsf}, and was generally $\sim0.5''-1''$.


\subsubsection{Data reduction and calibration}\label{section:reduction}

Data reduction was performed with IRAF\footnotemark  \footnotetext{IRAF is distributed by the National Optical Astronomy Observatory, which is operated by the Association of Universities for Research in Astronomy (AURA) under cooperative agreement with the National Science Foundation.}, using the IRCS IMGRED\footnotemark \footnotetext{The package is available at \protect\url{http://www.naoj.org/Observing/DataReduction/index.html}} package designed to reduce data obtained with IRCS. The reduction consisted of the following steps:

\begin{table*}
\rotatebox{90}{
 \centering
 \begin{minipage}{155mm}
  \caption{Summary of observations}
  \begin{tabular}{@{}llccclll@{}}
  \hline 
\multirow{2}{*}{Object} &
\multirow{2}{*}{Guide star} &
Exposure & 
\multirow{2}{*}{Filter} &
Pixel scale &
\multirow{2}{*}{Airmass} &
Photometric &
Observation date \\ 
& & [s] & & [mas] & & calibration & (UTC) \\
 \hline
SDSS~J0743+2457 & NGS & $20\times60$   & $K'$ & 52 & 1.13 & yes$^a$ & 2011 November 29 \\
SDSS~J0819+5356 & NGS & $14\times60$   & $K'$ & 52 & 1.21 & yes$^c$ & 2012 February 19 \\
SDSS~J0820+0812 & w/o AO & $14\times60$ & $K'$ & 52 & 1.38-1.75 & yes$^b$ & 2012 February 19 \\
SDSS~J0832+0404 & LGS & $16\times60$ & $K'$ & 52 & 1.07 & FS127 & 2014 January 14 \\
SDSS~J0904+1512 & LGS & $17\times60$ & $K'$ & 52 & 1.05 & FS126 & 2013 April 28 \\
SDSS~J0926+3059$^f$ & LGS & $11\times60$ & $K'$ & 52 & 1.06 & no & 2014 January 15 \\
SDSS~J0946+1835 & NGS & $6\times60$   & $K'$ & 52 & 1.08 & yes$^a$ & 2011 May 17 \\
SDSS~J0946+1835 & NGS & $33\times240$   & $K'$ & 20 & 1.03-1.27 & P550C & 2012 February 2 \\
SDSS~J1001+5027 & LGS & $27\times40$   & $K'$ & 52 & 1.25-1.38 & FS127 & 2012 February 20 \\
SDSS~J1002+4449 & LGS & $16\times50$ & $K'$ & 52 & 1.10 & yes$^a$ & 2011 November 29 \\
SDSS~J1054+2733 & LGS & $16\times60$ & $K'$ & 52 & 1.02 & FS127
 & 2014 January 14 \\
SDSS~J1055+4628 & LGS & $8\times60$ & $K'$ & 52 & 1.12 & FS126 & 2013 April 27 \\
SDSS~J1128+2402 & LGS & $15\times60$ & $K'$ & 52 & 1.02 & FS127
 & 2014 January 14 \\
SDSS~J1131+1915 & LGS & $7\times60$ & $K'$ & 52 & 1.00 & FS127 & 2012 February 20 \\
SDSS~J1206+4332 & w/o AO & $14\times60$ & $K'$ & 52 & 1.09 & FS126 & 2013 April 28 \\
SDSS~J1216+3529 & LGS & $9\times60$ & $K'$ & 52 & 1.05 & FS126 & 2013 April 28 \\
SDSS~J1254+2235 & LGS & $30\times60$ & $K'$ & 52 & 1.04-1.12 & FS23 & 2012 February 21 \\
SDSS~J1313+5151 & LGS & $48\times25$ & $K'$ & 52 & 1.19 & FS23 & 2012 February 21 \\
SDSS~J1320+1644 & LGS & $31\times60$ & $K'$ & 52 & 1.00 & no & 2013 April 28 \\
SDSS~J1322+1052 & LGS & $24\times60$ & $K'$ & 52 & 1.26-1.40 & FS126 & 2013 April 27 \\
SDSS~J1330+1810 & LGS & $12\times60$ & $K'$ & 52 & 1.09-1.13 & FS127 & 2012 February 20 \\
SDSS~J1330+1810 & LGS & $13\times240$ & $K'$ & 20 & 1.05 & FS126 & 2013 February 27 \\
SDSS~J1334+3315 & LGS & $10\times60$ & $J$ & 52 & 1.07 & P272-D$^d$ & 2011 February 18 \\
SDSS~J1334+3315 & LGS & $5\times60$ & $H$ & 52 & 1.09 & P272-D$^d$ & 2011 February 18 \\
SDSS~J1334+3315 & LGS & $10\times60$ & $K'$ & 52 & 1.05 & P272-D$^d$ & 2011 February 18 \\
SDSS~J1353+1138 & LGS & $23\times60$ & $K'$ & 52 & 1.31-1.47 & FS126 & 2013 April 27 \\
SDSS~J1400+3134 & LGS & $15\times60$ & $K'$ & 52 & 1.48-1.67 & FS126 & 2013 April 27 \\
SDSS~J1405+0959 & LGS & $19\times60$ & $J$ & 52 & 1.02 & FS126 & 2013 April 27 \\
SDSS~J1405+0959 & LGS & $26\times60$ & $H$ & 52 & 1.02-1.07 & FS126 & 2013 April 27 \\
SDSS~J1405+0959 & LGS & $14\times60$ & $K'$ & 52 & 1.10-1.14 & FS23 & 2012 February 21 \\
SDSS~J1406+6126 & LGS & $21\times60$ & $K'$ & 52 & 1.46 & no & 2014 January 15 \\
SDSS~J1455+1447 & LGS & $23\times60$ & $K'$ & 52 & 1.08-1.18 & no & 2014 January 15 \\
SDSS~J1515+1511 & LGS & $22\times40$ & $K'$ & 52 & 1.01 & FS23 & 2012 February 21 \\
SDSS~J1620+1203 & LGS & $2\times60$ & $K'$ & 52 & 1.13 & no & 2011 August 8 \\
SDSS~J1620+1203 & LGS & $16\times60$ & $K'$ & 52 & 1.16-1.22 & 2MASS$^e$ & 2013 April 27 \\
\hline
\end{tabular}
\\ 
{\footnotesize Exposure times show the number of frames that were combined in the final science frame. $^a$ Based on 1 star in the FOV; $^b$ based on 1 star and 1 galaxy in the FOV; $^c$ based on 3 stars in the FOV; $^d$ dark sky pattern may affect model photometry; $^e$ compared the total magnitude with the measurement in 2MASS; $^f$ refers to SDSS J092634.56+305945.9; 
}
\label{tab:followup-data}
\end{minipage}}
\end{table*}

 \begin{enumerate}
  \item Each frame was checked for dark patterns by dividing it to the next observed frame. Also, the number of counts at the location of the target was checked for linearity. All targets are well inside linearity limit, except for the core of the star inside the FOV for  SDSS~J0743+2457 (within 5\%), and cores of the bright quasar images in SDSS~J0904+1512 (within 3\%), SDSS~J1322+1052 (within 2\%) and SDSS~J1353+1138 (within 4\%).
  \item As persistent fringe artifacts were found in the dome flats, sky flats were created by masking bad pixels and bright objects, then median-combining the dithered frames which were divided by the mean count value. Each raw frame was corrected for different pixel sensitivities by division to the sky flat.
  \item The geometric distortion correction map described in Section \ref{section:distort} was applied.                      
  \item Sky-background frames were generated by median-combining the flat-fielded frames. As the near-infrared sky is highly time-variable, the sky-background frames were generated for each dither sequence individually. Pixel values for the masked pixels were interpolated from the combined ones. Before subtracting the sky frame, each frame was normalized so that their median values match each other.
   \item Position offsets between the images were obtained via cross-correlations between regions that contain bright objects, typically the targets. Cosmic rays falling in the regions used to perform the cross-correlations were manually masked. The full-width at half-maximum (FWHM) of star-like objects in each frame was measured by fitting a Moffat function to the radial profile with the IRAF IMEXAM task, and the frames with significantly different values were discarded. The frames were average-combined using rejection algorithms (average sigma clipping or clipping based on the CCD parameters). Pixel binning was performed for a few frames with significantly oversampled PSF, as mentioned in Appendix \ref{section:objectsdescript}. 
   \item Photometric zero-point calibration was performed, typically using standard stars. All objects were corrected for Galactic extinction \citep{schlegel98}, and atmospheric extinction relative to the standard star\footnotemark . \footnotetext{According to the Mauna Kea summit extinction values (\protect\url{http://www.jach.hawaii.edu/UKIRT/astronomy/utils/exts.html}) the atmospheric extinction is of $\sim$ 0.1 mag/airmass for $K$-band.} It must be noted that the standard star catalogue magnitudes at $K-$band were used to calibrate the photometry at $K'-$band. However, the expected differences based on interpolation\footnotemark \footnotetext{\protect\url{http://www2.keck.hawaii.edu/inst/nirc/UKIRTstds.html}} are small ($0.01-0.03$ mag). 
\end{enumerate}
 

\section{Morphological modelling technique}\label{section:technique}

\begin{figure}
\includegraphics[width=80mm]{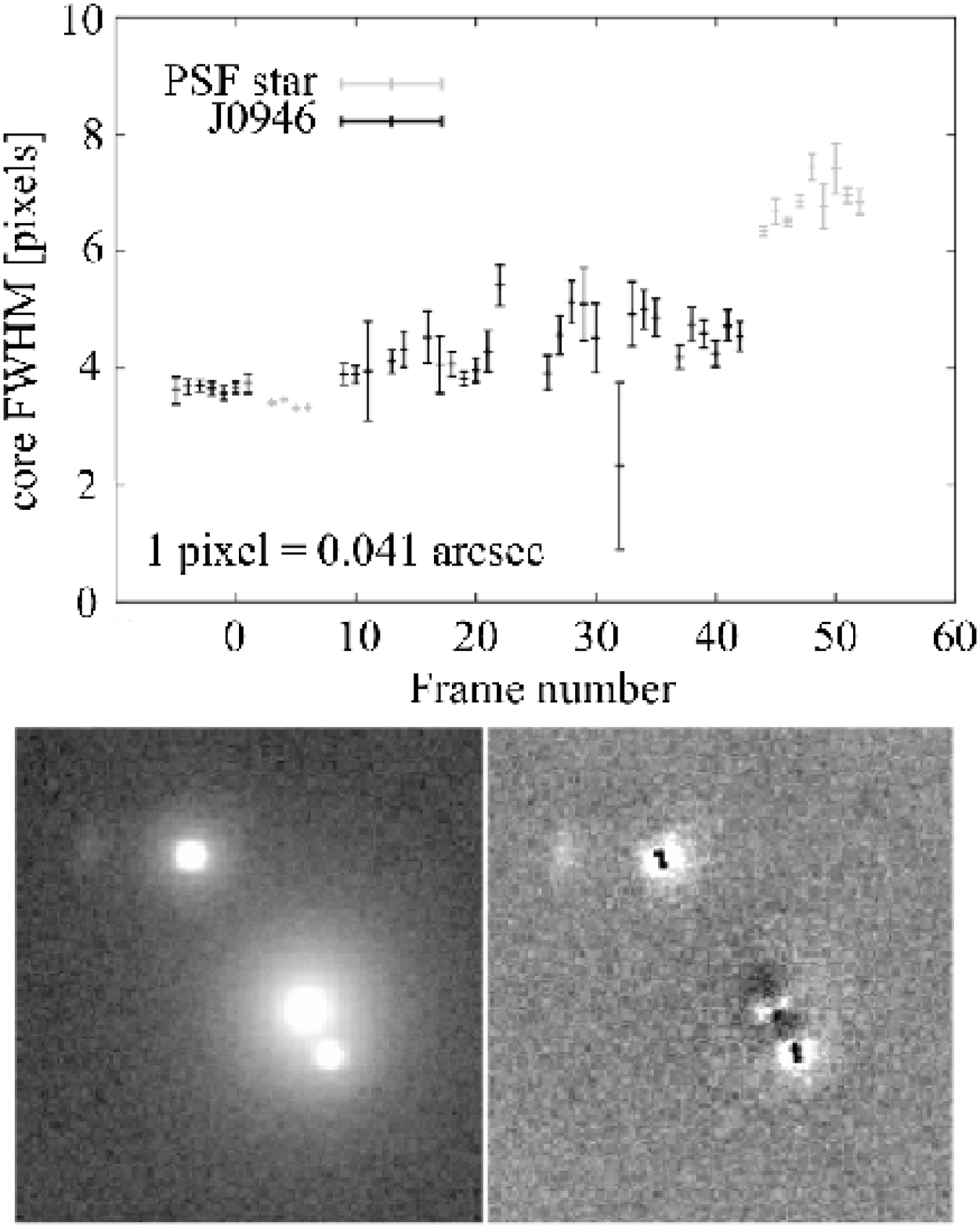}
\caption{\emph{Top:} PSF FWHM for each individual frame in the SDSS~J0946+1835, as well as for the separate PSF star. Frames are numbered chronologically. \emph{Bottom left}: original SDSS~J0946+1835 observed frame; \emph{bottom right}: residuals after modelling the system with the separate PSF star, coadded from its first four frames.
\label{fig:0946psf}}
\end{figure}

After reducing the data, it became clear that none of the separately-observed PSF stars represent suitable PSFs to model the corresponding systems. This was also concluded by other studies employing separately observed PSF calibrators \citep[e.g.,][]{kuhlbrodt05}. The reason is PSF variability, both in the science targets and the separately-observed stars. This is caused by the rapid change of the atmospheric turbulence characteristics, which induce different responses from the AO system. Figure \ref{fig:0946psf} shows the time variability of the core component FWHM in the PSF for both the target and the separately-observed star in SDSS~J0946+1835, as well as the outstanding residuals obtained if the star is used as a PSF to model the system. 

It is imperative that we construct a well-characterised PSF for each system, in order to reach the science goal, i.e. obtain an accurate astrometrical, photometrical and morphological characterisation of the multiple components of each observed system. In order to guide the reader, here we provide a short summary of the following subsections, in which we illustrate in detail our PSF reconstruction and modelling technique. In section \ref{aopsf} we give general considerations on the AO PSF. In Section \ref{section:newmethod} we briefly mention the AO PSF modelling techniques encountered in the literature. In Section \ref{Hostlens} we introduce our technique of modelling analytically all components of a gravitationally lensed quasar simultaneously, as their light profiles overlap, and derive the most suitable analytical profile. We describe the implementation of this technique, where we take special care to properly explore the parameter space. In Section \ref{subsection:hybrid} we show that for selected systems we can go beyond an analytical PSF, and we derive a hybrid PSF that specifically accounts for the non-analytical components. In the next Sections (\ref{section:hostfit} and \ref{section:simulate}, respectively), we describe how we need to modify out technique to model systems where we detect a quasar host galaxy, as its profile fitting requires a mass model. We then insure that the error estimates for each derived parameter accounts for systematics introduced by our modelling technique. Finally, in Section \ref{section:distort} we address the issue of how the instrument field distortion impacts our derived astrometry.

\subsection{The adaptive optics PSF}\label{aopsf}

For a seeing-limited image, spatial resolution is usually characterised by the FWHM of a point-like stellar profile. This characterisation becomes unreliable near the diffraction limit, as the FWHM measurement is complicated due to the diffraction rings. In the diffraction limit, the shape of the PSF is described by a two-dimensional Airy function (ignoring the telescope spider), with the first diffraction ring at an angular distance of $1.22 \lambda/D$ from the centre; here $\lambda$ denotes wavelength, and $D$ is the diameter of the telescope mirror. In this case the Strehl ratio \citep{strehl02} is used, and is defined as the ratio of the intensity at the peak of the observed seeing disc divided by the intensity at the peak of the theoretical Airy disc. 

In general the AO PSF is described as two components: a nearly diffraction-limited core of FWHM $\sim\lambda/D$, and a seeing limited halo/wing of FWHM $\sim\lambda/r_0$, where $r_0$ is Fried's parameter \citep{fried65}. The core is typically approximated by a Gaussian \citep[e.g.,][]{law06}, whereas the halo is approximated by a Moffat profile \citep[][Section \ref{Hostlens}]{moffat69}, which has a more extended wing than a Gaussian.

As a product of the atmosphere as well as the AO system, the AO PSF exhibits both temporal and spatial variability. The spatial variability  (anisoplanatism) causes the PSF to vary across the FOV. In LGS AO, there are three sources of anisoplanatism \citep[e.g.,][]{vandam06}. Angular anisokinetism (or tip-tilt angular anisoplanatism) results from the difference between the tip-tilt component of the wavefronts of the TT star and the science object, and is radially symmetric around the TT star. Focal anisoplanatism (the cone effect)  occurs because the LGS (located at a finite altitude) samples the cone of turbulence between the LGS and the telescope, while the turbulence experienced by the science object is distributed in a cylinder. Finally, angular anisoplanatism results from the difference in the higher order wavefront terms between LGS and the science object. 

 As the distance of the objects from the TT star is increased, the  FWHM of the core PSF component broadens, and its major axis points towards the TT star. 

In the course of this work, the PSF is assumed not to vary spatially for objects in close proximity of each other, such as the multiple components of a gravitationally lensed system. Indeed, these objects are separated by $\sim 2''$, much less than the isoplanatic angle in the LGS mode. 


\subsection{Techniques used in the literature}\label{section:newmethod}

There are several examples in the literature of modelling AO lensed quasars without an a priori known PSF. These made use of the typical structure of a lensed quasar: two or four point-like images with a lensing galaxy in between. While the quasar point-source is safely  unresolved, since it is typically located at high redshift, the quasar host galaxy may or may not be resolved. In all cases, only doubles where the host galaxy is unresolved were modelled in the literature, and in the case of quads, the host was actively removed during modelling, without estimating any of its physical parameters.

\citet{sluse08} modelled the double SDSS J0806+2006 with the PSF estimated from the more isolated, brighter image, and employing image deconvolution. They do not mention however if any systematics are introduced by this technique. \citet{davies04} however argued for the use of convolution against deconvolution, since the PSF is poorly defined, the S/N is limited, and smooth extended emission (in the case of hosts) makes the results uncertain. When there is a physical model to test, they advise that the best approach is to convolve that model with the best estimate of the PSF and then compare the result to the observations. \citet{faure11} took advantage of the cross configuration of the quad SDSS J0924+0219, where the host galaxy has a different morphology at the location of each image, and estimated the PSF by combining the superposed and normalised images, in an iterative procedure. \citet{lagattuta10} analysed the quad B0128+437 by iteratively fitting a PSF modelled as three Gaussian components, together with the lensing galaxy modelled with a Sersic profile. \citet{koptelova13} modelled the PSF for a lensed double by using a hybrid composed of the observed brightness distribution of the bright image at the centre, and analytical wings.

\subsection{Modelling with an analytical PSF}\label{Hostlens}

There is enough information \citep[e.g.,][]{marshall07} in the multiple images of a lensed system to reconstruct the PSF. In this paper, we combined the approach of \citet{lagattuta10} and \citet{koptelova13} into a new technique, by adopting an analytical PSF model and simultaneously optimising for the global PSF parameters as well as the individual objects comprising the system: multiple point sources, and the lensing galaxy modelled as a Sersic profile convolved with the PSF. Where possible, we take an additional step where we refine this fit by using a hybrid PSF (Section \ref{subsection:hybrid}). The optimisation is performed using a downhill simplex method \citep{press92} to find the minimum goodness-of-fit $\chi^2$ in parameter space, and is implemented in a new software, Hostlens, which was developed by one of us (M. Oguri). Hostlens combines the functionality of Galfit with that of Glafic \citep{oguri10}. Like Galfit, it allows the simultaneous fitting of both point and extended sources, but in addition can simultaneously optimise the parameters of an analytical PSF. Like Glafic, it can fit an observed lensing configuration with a mass model, while simultaneously fitting analytically the morphology of an extended lensed source, such as a quasar host galaxy (Section \ref{section:hostfit}). We note that most of the functionalities of Hostlens are currently implemented in Glafic. 

\begin{figure*}
\includegraphics[width=175mm]{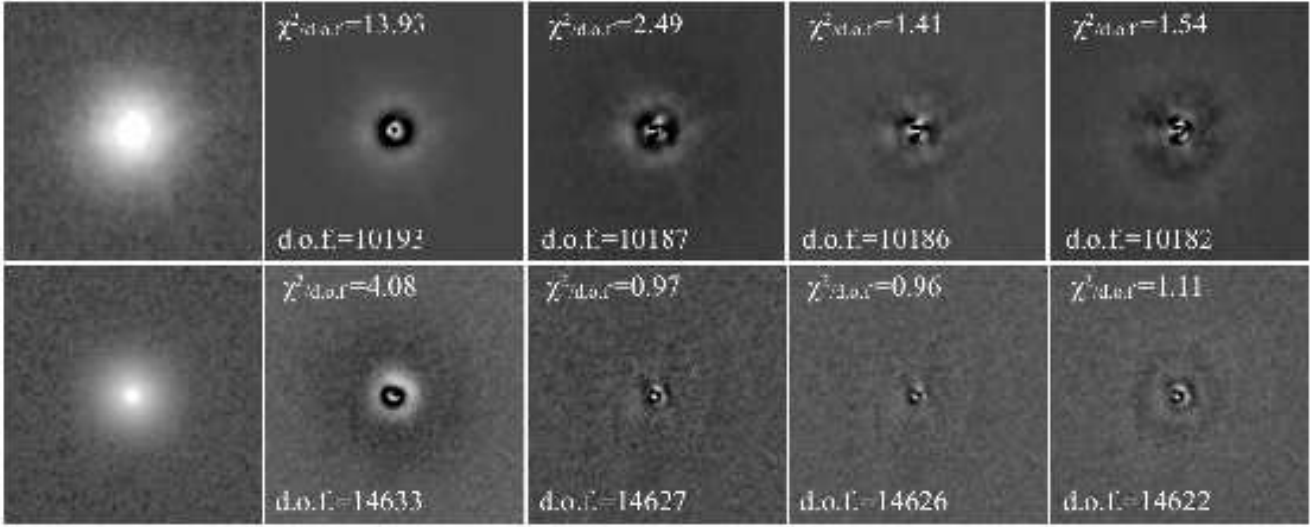}
\caption{Residuals after analytically modelling with Galfit the PSF stars for SDSS~J0946+1835 (January 2012) $\mathit{(top)}$ and SDSS~J1515+1511 $\mathit{(down)}$. From left to right: original star, residuals after subtracting 1 Moffat, Gaussian + Moffat, 2 Moffat, and 3 Gaussians models, respectively. The d.o.f. and $\chi^2/\mathrm{d.o.f.}$ corresponding to each fit are shown.
\label{fig:starresid2D}}
\end{figure*}
\begin{figure*}
\includegraphics[width=175mm]{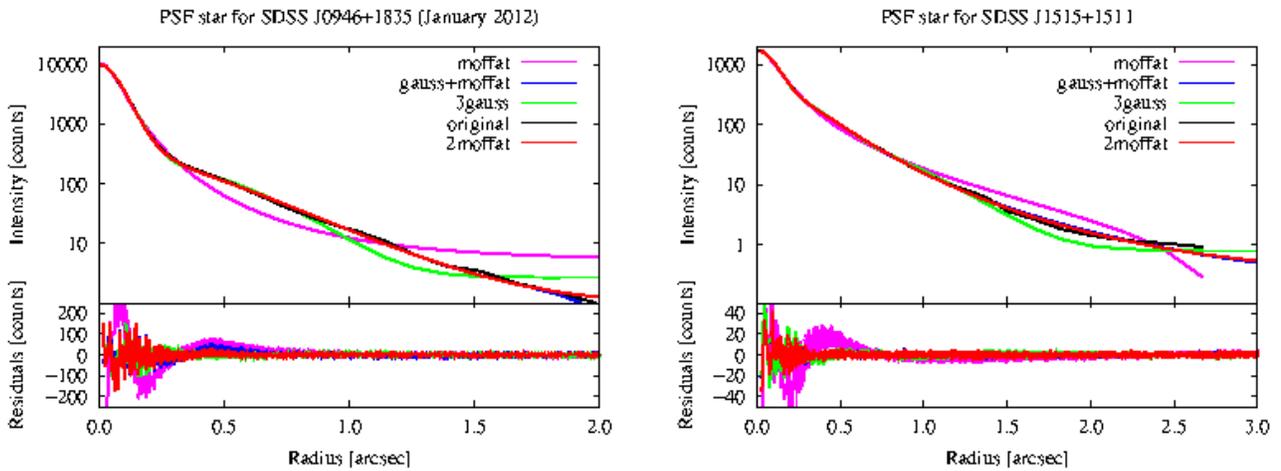}
\caption{Radial profiles and residuals after analytically modelling the SDSS~J0946+1835 (January 2012) $\mathit{(left)}$ and SDSS~J1515+1511 $\mathit{(right)}$ PSF stars. The radial profiles were generated with the IRAF task ELLIPSE, and the residuals with the task PRADPROF. \it{A colour version of this figure is available in the online version of the paper.}
\label{fig:starresid1D}}
\end{figure*}

We considered several analytical PSF models to use with Hostlens. According to Section \ref{aopsf}, the AO PSF consists of two components, one describing the compact PSF core, and the other one the extended seeing wings. \citet{akiyama08} used a single Moffat profile, whose wings are more extended than that of a Gaussian, to approximate the full AO PSF. \citet{davies10} stated that the PSF is generally matched quite well by the sum of a narrow Gaussian for the core, and either a Moffat or another Gaussian for the halo. \citet{falomo08} used a Gaussian for the core and an exponential function for the wings. \citet{lagattuta10}, mentioned above, used three Gaussians: one component represents the diffraction-limited core, another represents the seeing-limited diffuse PSF, and the third encodes structure in the PSF. In order to find the most appropriate model, we fitted the majority of bright stars observed during the observation runs (i.e. the PSF stars) with analytical profiles. Two of these fits are shown in Figures \ref{fig:starresid2D} and \ref{fig:starresid1D}. The first figure shows a two-dimensional fit and the subsequent residuals after subtracting elliptical but concentric 1 Moffat, 1 Gaussian + 1 Moffat, 2 Moffat, and 3 Gaussian profiles, respectively. The second shows the fitted radial plot corresponding to the same models. As can be seen, both in terms of the formal $\chi^2/$d.o.f., as well as in terms of the appearance of residuals, and radial profile matches, the 2 Moffat profile provides the best fit. This is found in the overwhelming majority of fitted stars, although in some cases, such as the second example shown, the 1 Gaussian + 1 Moffat profile provides a similarly good fit. The residuals in the other models all show a concentric halo, and do not match the overall radial profiles. On the other hand, none of the models (including 2 Moffat) is able to model the non-analytical components close to the centre. This entirely empirical analysis resulted in the decision to use the 2 Moffat profile as the analytical PSF for Hostlens. The better fit of a Moffat core, which is more extended than a Gaussian, may be due to the very small Strehl ratios $\lesssim10$\% typically obtained in the observations. Indeed, the typical core FWHM is $\sim0.2''$, about three times that of the diffraction-limited core.

 The Moffat profile \citep{moffat69} has surface brightness of the form
 
 \begin{equation}
I(r) = \left[1+\left( 2^{1/\beta} - 1    \right)   (2r/\mathrm{FWHM})^2     \right]^{-\beta}\ \ , \ \ r\equiv\sqrt{x^2+\left( \frac{y}{q}\right)^2}\ ,
\end{equation}

\noindent  where $r$ is the elliptical radial distance, $x\ (y)$ is the distance to the centre along the semimajor (semiminor) axis and $q$ is the ratio of the semiminor to the semimajor axis; $\beta$ is a parameter specifying the shape of the profile, related to the core-to-halo flux ratio. The Moffat profile results into a Gaussian profile as the $\beta=\infty$ case. In Hostlens, the PSF is modelled with 9 analytical parameters: two for each of the four components of an individual Moffat profile: FWHM, ellipticity, position angle and $\beta$, as well as one parameter specifying the relative distribution of flux, flux1/(flux1 + flux2). The two Moffat profiles are concentric. Each point source is modelled with three parameters: the positions along X and Y, and the total flux. In addition, every Sersic profile has ellipticity, orientation, effective radius and Sersic index parameters. Finally, there is a parameter for the sky value. Therefore, a typical two-image lensed system requires 24 parameters to model analytically. We modelled each system with Hostlens, at least at an initial stage. The derived analytical parameters of all PSFs are shown in Table \ref{tab:analpsf}. We selected suitably large cut regions (typically $\sim7\arcsec\times7\arcsec$) around the targets, and masked all objects that were not simultaneously fitted by Hostlens, in order to allow reliable sky value estimates and model fitting. 

Due to the large number of parameters and the possibility that the parameters might get trapped in a local $\chi^2$ minimum, we chose 500 sets of different initial parameter values from a range of plausible values, e.g. from a flat distribution in the case of position angles, and a gaussian distribution around the IRAF IMEXAM-derived value for the core FWHM component. We chose the model with the lowest value of $\chi^2$ (we also checked the second-best model to be similar), and in addition we performed a parameter search around it using Markov-chain Monte Carlo (MCMC), for the purpose of both refining the parameter values, and estimating confidence intervals (error bars). We ran ten MCMC chains, and removed a fraction of $\sim 10\%-30\%$ of the points from the head of each chain (the ``burn-in stage" points; i.e sensitive to the start values). To test the convergence of these chains, we compared the variance of the distribution of points in each chain to the variance of the combined distribution, following \citet{gelman95}. After applying a smoothing technique, the combined chains provided 68\% confidence intervals on the analytical parameters ($1 \sigma$ in the Gaussian approximation).

Throughout this work, unless otherwise stated, for all the systems where we did not detect a quasar host galaxy, and which we fitted with a Hostlens-produced analytical PSF, we also checked the modelling results for consistency, with Galfit. In order to produce with Galfit error bars that account for the uncertainties of the analytical PSF, we produced 500 realisations of the Hostlens-produced PSF, by drawing at random from the probability distribution of the analytical parameters of the PSF, obtained during the MCMC runs above; we then ran Galfit with each of the 500 PSFs, and took the standard deviations of the resulting parameters as an error estimate. We did not use Galfit for the systems with host galaxy detections, as it lacks the functionality to fit this component (see Section \ref{section:hostfit}).

We note that Hostlens and Glafic measure the effective radius of a Sersic profile using a circularised radius, while Galfit measures the radius along the semi-major axis. The effective radii measured in this paper are reported following the Galfit definition. For the lensing galaxies, we considered both an unconstrained Sersic index $n$ and $n=4$ (Table \ref{tab:lensmorphology}), in order to ease comparison with other samples of lenses where the light profile is modelled as a de Vaucouleurs profile \citep[see][and references therein]{oguri14}.

\subsection{Hybrid PSF}\label{subsection:hybrid}

\begin{figure*}
\includegraphics[width=100mm]{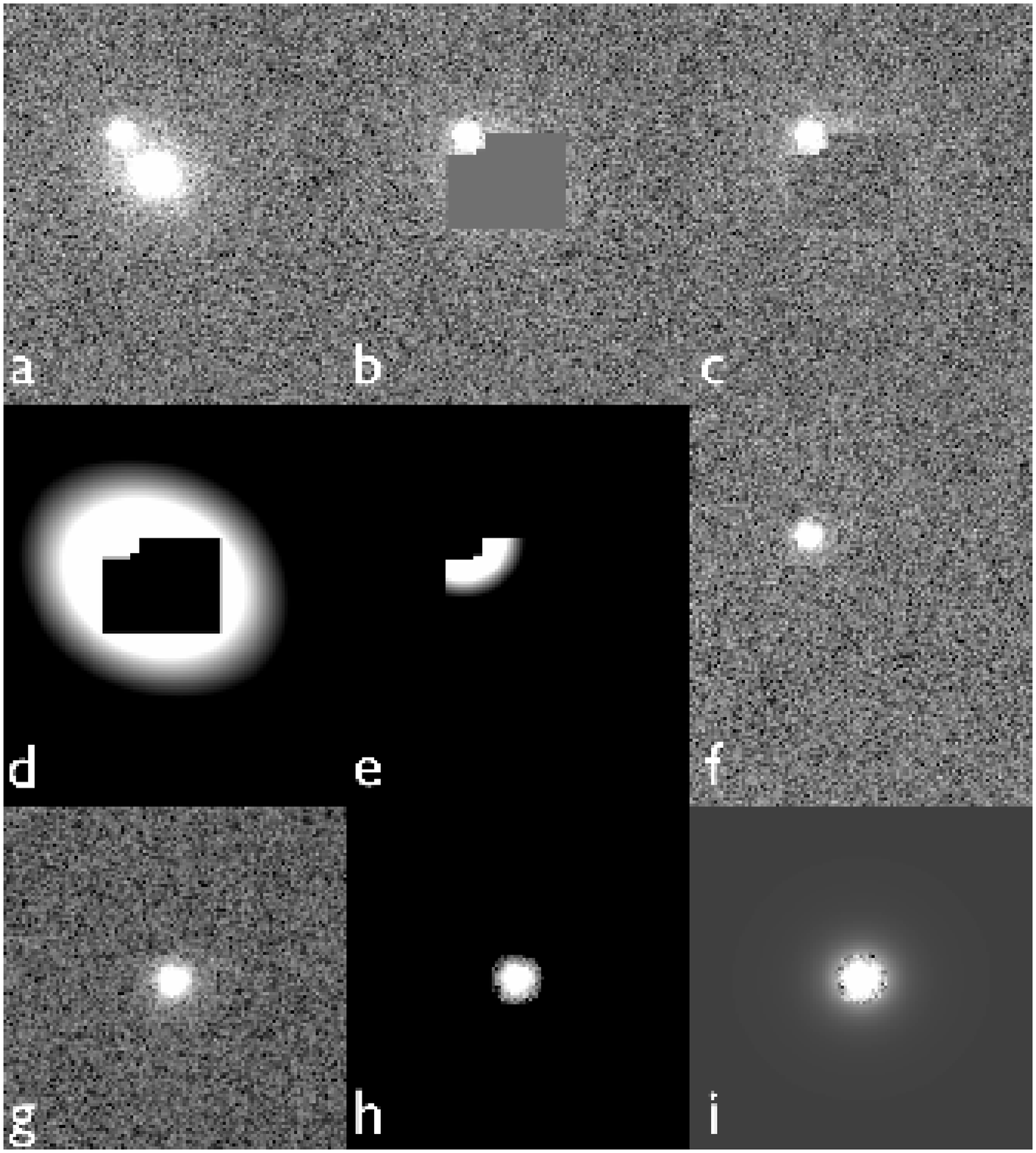}
\caption{ Example of a hybrid PSF creation. $\mathit{(a)}$ Original frame. $\mathit{(b)}$ The section of the frame containing most of the light with the exception of the bright quasar image is cut. $\mathit{(c)}$ The cut section is replaced with nearby blank sky. $\mathit{(d)}$ The best-fit analytical models of all objects except the bright quasar image, which were created with an analytical PSF, are removed from the rest of the frame. $\mathit{(e)}$ The analytical wings of the bright quasar image are added to the section where the sky was replaced. $\mathit{(f)}$ The resulting frame is now clear from all objects except the bright quasar image. $\mathit{(g)}$ The resulting PSF frame is centred, and the sky pedestal is removed. The resulting image is fitted analytically. $\mathit{(h)}$ Only a circular region around the PSF centre is kept, large enough to contain all visible non-analytical residuals. $\mathit{(i)}$ The portion of the frame that was removed is replaced with the best-fit analytical wings.
\label{fig:hybrid}}
\end{figure*}

The observed systems can be classified into two categories, according to the degree in which the bright image A is separated from the other objects, and the visual appearance of the modelling residuals. For systems with small separation where the non-analytical residuals at the centre clearly overlap with a large fraction of the galaxy light (e.g. SDSS~J1254+2235 and SDSS~J1334+3315), as well as for objects where, due to the low S/N, no conspicuous core residuals are seen, data modelling stops as described above. For the other systems, it is possible to model away the faint quasar image and the lensing galaxy using the analytical Hostlens profiles, and obtain a clean image A to be used as the ``actual PSF'' (panels a-g in Figure \ref{fig:hybrid}). The advantage in this approach is that the estimated PSF is the ``true'' PSF (accounting for all non-analytical features), albeit a noisy one. Next, we tested whether this PSF produces better statistical results. A simple comparison with the analytical fit in terms of $\chi^2$ cannot be used, as image A is fitted perfectly by design. Therefore, we compared the standard deviation of the residual pixel values in a region containing the rest of the objects (typically B and G). We found that in almost all cases, the analytical PSF produced better results. This is due to large noise in the wing of the ``actual PSF''. Therefore, we used another approach, previously employed by \citet{koptelova13}. This consists in using the observed brightness distribution of the bright image only in its central regions, and replacing the wings after a certain cut radius with analytical wings from the 2 Moffat fit (panels h-i in Figure \ref{fig:hybrid}). This ``hybrid PSF'' approach succeeded in producing fits at least as good as the analytical PSF in terms of the standard deviation of the residuals. 

We considered two sources of systematics for the hybrid PSF approach. First, because the contributions from B and G are initially modelled using the analytical PSF, they may be improperly subtracted at the location of A, and therefore affect the hybrid PSF estimation. This can be checked by building the hybrid PSF in an iterative approach, where in each subsequent step the hybrid PSF is used to model away B and G, resulting in a better hybrid PSF estimate. We did this for SDSS~J1405+0959 and obtained that subsequent PSFs are virtually identical. This is expected, as we applied the hybrid PSF technique only to systems with isolated bright images. Secondly, the final models may depend on the size of the hybrid cut region. We chose the original cut radius as the radius at which all non-analytical components disappear, of $\sim10$ pixels. For all hybrid PSF systems (excluding the ones with a quasar host galaxy detection, as described below, or otherwise specified), we included into the error budget the scatter between the results obtained from this cut size, and another one $50\%-100\%$ larger. 

We used the analytical and/or hybrid PSF approaches described above for all systems where a quasar host galaxy was not detected. Table \ref{tab:morphologytechnique} shows whether the analytical or the hybrid PSF was used to model each system.

\subsection{Fitting the quasar host galaxy}\label{section:hostfit}

\begin{figure}
\includegraphics[width=85mm]{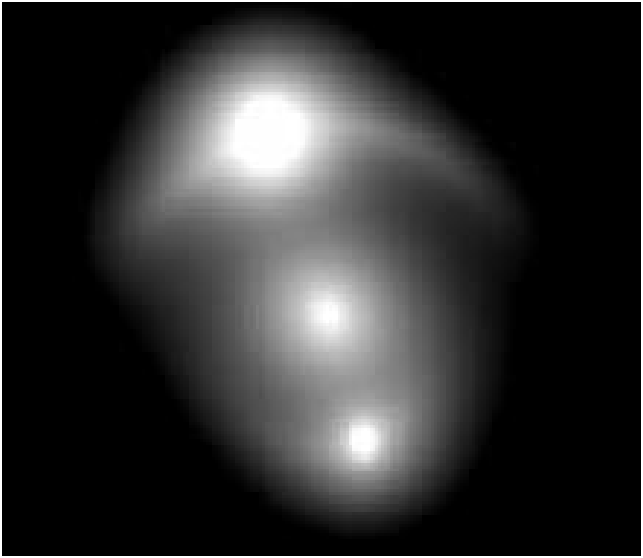}
\caption{ A simulated typical two-image gravitationally lensed quasar with a detected quasar host galaxy (color-inverted image). 
\label{fig:simulatedsystem}}
\end{figure}

A novel undertaking in this paper is the fitting of quasar host galaxies, for several systems, without an a-priori known PSF. This is only possible for gravitationally lensed quasars, because the host galaxies are tangentially stretched around the critical curves by the lensing effect, resulting in arcs that are distinct from the PSF shape, and can be modelled analytically (Figure \ref{fig:simulatedsystem}).

In order to model the spatially extended flux distribution of the lensed images, it is necessary to model both the lens mass distribution and the extended source distribution. This is done using the approach described in \citet{peng06-2}, and makes use of the fact that  gravitational lensing conserves surface brightness. 

Hostlens was designed with the functionality to fit host galaxies modelled with a Sersic profile. 
When fitting a host galaxy with Hostlens, the parameters typically used to model the individual images are replaced by a single source profile. This profile consists of the parameters of a Sersic model, its flux as well as the flux of the central point source, and the lensing parameters: the Einstein radius, shear/ellipticity and their orientation angles, where a singular isothermal mass model is assumed (Section \ref{section:lensmodels}). The luminous lens galaxy is fitted simultaneously, as before, with a Sersic profile.  

There are five cases where the quasar host galaxy was clearly detected and fitted. We define a clear detection as a system for which there are clear visible arcs similar in orientation to the critical curves of the lensing model, and which are convincingly removed by fitting a host galaxy. 

In the case of two systems with host detections, SDSS~J0904+1512 and SDSS~J1322+1052, there were outstanding residuals when using an analytical PSF, and therefore we employed a hybrid PSF. This was possible through an iterative process, where an analytical PSF is initially used to obtain a first rough approximation of the host profile. The host galaxy is then subtracted from the original observed frame, and subsequently a hybrid PSF is created as described in Section \ref{subsection:hybrid}. This hybrid PSF is used in the next step of the iteration to refit the original system, and refine the host model. Subsequent steps lead to new hybrid PSFs. The iteration stops once the $\chi^2$ stops decreasing. 

\begin{figure*}
\includegraphics[width=120mm]{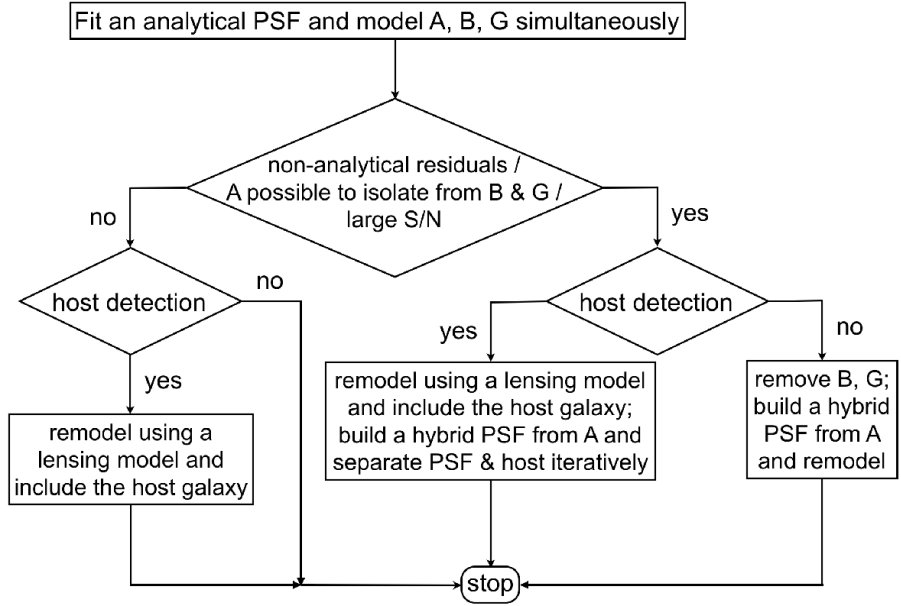}
\caption{ Flowchart of the overall approach in the morphological modelling of the imaging data.
\label{fig:flowchart}}
\end{figure*}

To summarise, the general approach to the morphological modelling of the imaging data in this paper, whether the quasar host galaxy is detected or not, is described in the flowchart of Figure \ref{fig:flowchart}. 

\subsection{Simulations for testing the reliability of the technique}\label{section:simulate}

To test the reliability of the morphological fitting with Hostlens using the analytical or hybrid PSF methods, we performed a series of simulations for each system. 
In these simulations, we used a separately observed PSF star, or a star in the FOV that is the most similar in terms of analytical parameters to the PSF of the respective system (prioritising for the core FWHM parameter). Table \ref{tab:morphologytechnique} shows which star was used as PSF template for each system. With Hostlens and/or Galfit, we simulated each system using this corresponding PSF template, and the best-fit astrometry/morphology/photometry parameters we derived when modelling that particular observed system. Such a simulated system is shown in Figure \ref{fig:simulatedsystem}. From the final science frame of each observed (i.e. real) system, we identified 100 blank sky cuts, and added them as noise realisations to the simulated system. We then remodelled each of the 100 simulated noisy systems in the same way as we did the observed system. The scatter in the resulting 100 values for each fitted parameter, with respect to the known input value, represents an error estimate. The only exception is for the case when we simulated the system using the best-fit parameters, but remodelled it using Sersic index $n\equiv4$. In that case, we considered the scatter around the mean fitted value (Table \ref{tab:lensmorphology}).

We used two time-saving approximations in remodelling the simulated systems. First, we did not perform a search starting from random values in the parameter space for each individual simulated system, as we did for the original system, but instead we started the fit from the known (correct) values. Second, in cases where we used a hybrid PSF, we did not create it in each of the 100 simulations from that very same image, but created it from only one of the simulations, and then used it as the hybrid PSF to model all simulated frames. Also, for the few cases when a hybrid PSF is used to fit a detected host galaxy, we only performed a small number of simulations, as these are time consuming.

These simulations can account for the systematics introduced by the analytical modelling of a non-analytical PSF, or the use of hybrid PSFs. The orientation of the detector was typically the same for the science objects and the PSF stars, therefore diffraction patterns due to the telescope spider are similar. The noise cuts from the actual science frames, previously employed by \citet[e.g.][]{gabor09}, account for how the model parameters would depend on the existence of background objects, too faint to be visible, and are more realistic than artificially-generated noise. 

As final errors on astrometry/morphology/photometry, we used the largest error bars among those resulting from these simulations and all other methods described in the previous subsections. This is because the methods are not necessarily independent, and one typically dominates the error budget. 

\begin{table*}
 \centering
 \begin{minipage}{175mm}
\caption{Technique of morphology fitting}
  \begin{tabular}{@{}ll@{}}
  \hline
Object &
Technique\\ 
\hline
SDSS~J0743+2457 & analytical fitting / simulation with star in FOV \\
SDSS~J0819+5356 &  analytical fitting with host galaxy / simulation with star in FOV \\
SDSS~J0820+0812 & non-AO; fitting with star in FOV \\
SDSS~J0832+0404 & analytical fitting with host galaxy /  simulation with star in the SDSS~J0819+5356 FOV \\
SDSS~J0904+1512 & hybrid PSF created iteratively with host galaxy / simulation with self PSF \\
SDSS~J0926+3100 & analytical fitting \\
SDSS~J0946+1835 (2012) & hybrid PSF / simulation with the separately observed PSF star \\
SDSS~J0946+1835 (2011) & analytical fitting \\
SDSS~J1001+5027 & analytical fitting / simulation with self PSF \\
SDSS~J1002+4449 & analytical fitting / simulation with star in FOV \\
SDSS~J1054+2733 & analytical fitting \\
SDSS~J1055+4628 & analytical fitting / simulation with the separate PSF star from SDSS~J1405+0959 \\
SDSS~J1131+1915 & hybrid PSF / simulation produced with a cut around image A as PSF, refitted with the hybrid PSF \\
SDSS~J1206+4332 & analytical fitting / simulation with self PSF \\
SDSS~J1216+3529 & analytical fitting / simulation with the SDSS~J1001+5027 separate PSF star$^a$ \\
SDSS~J1254+2235 & analytical fitting / simulation with the separately observed PSF star \\
SDSS~J1313+5151 & hybrid PSF / simulation with self PSF  \\
SDSS~J1320+1644 & non-AO image fitting with star in FOV / AO image using image A cut as PSF \\
SDSS~J1322+1052 & hybrid PSF created iteratively with host galaxy / simulation with self PSF \\
SDSS~J1330+1810 (2012) & analytical fitting / simulation with the separate PSF star from SDSS~J1405+0959 \\
SDSS~J1330+1810 (2013) & analytical fitting / simulation with the separate PSF star from SDSS~J0946+1835 \\
SDSS~J1334+3315 ($J/H/K'$) & analytical fitting / simulation with the SDSS~J1131+1915 separate PSF star \\
SDSS~J1353+1138 & hybrid PSF / simulation with self PSF  \\
SDSS~J1400+3134 & analytical fitting / simulation with the SDSS~J0743+2457 bright star \\
SDSS~J1405+0959 ($J/H/K'$) & hybrid PSF / simulation with the separately observed PSF star \\
SDSS~J1406+6126 & non-AO image fitting with analytical PSF / simulation with the analytical PSF \\
SDSS~J1455+1447 & hybrid PSF / simulation with the SDSS~J0832+0404 separate PSF   \\
SDSS~J1515+1511 & fitting with star in FOV \\
SDSS~J1620+1203 (2011) & analytical fitting \\
SDSS~J1620+1203 (2012) & hybrid PSF / simulation with the separate PSF star from SDSS~J1405+0959 \\
\hline
\end{tabular}
{\footnotesize $^a$ Gaussian convolution was performed on the PSF star to more closely match the PSF of the system.}
\label{tab:morphologytechnique}
\end{minipage}
\end{table*}

\subsection{Accounting for the IRCS+AO188 field distortion}\label{section:distort}

The most important observational constraint for gravitational lens mass models of lensed quasars is provided by relative astrometry. In order to obtain accurate astrometry, it is necessary to account for the IRCS+AO188 field distortion, which results mainly from the pair of off-axis parabola mirrors.

\subsubsection{Creation of a distortion map}\label{subsection:distort}

The existing estimate of the geometrical distortion in the IRCS+AO188 was produced by matching stars in the October 2010 NGS AO observations of M15 with the high precision (residual distortion error $\sim0.5$ mas) $HST$/ACS globular cluster survey observation on May 2006 \citep{anderson08}. A catalogue of the stars (S/N$>$10) in the NGS AO, with positions measured using the IRAF IMEXAM task, was matched to the $HST$ catalogue, by using a third order polynomial transformation in X and Y (the IRAF GEOMAP task), along with pixel rescaling and rotation; a distortion map was obtained, with residual distortion estimated to be 7 mas/10 mas (4 mas/6 mas for the 20 mas detector mode) across the X and Y axes, respectively. \citet{mcnamara03} measured proper motions of stars in M15 that would amount to offsets of $\sim1.3$ mas during the time between observations, and can therefore be neglected; in addition, these are expected to have random vectors, without introducing systematics. A larger effect that has not been accounted for is the atmospherical refraction, which is not present in the $HST$ data and would stretch the image along the direction to the zenith. The effect would amount to a total of $\sim20$ mas or more across the FOV. 

\subsubsection{Estimating a more realistic residual distortion}\label{subsection:distortnew}

A more accurate distortion correction was not available at the time the bulk of the analysis presented in this paper was performed. While the measured residual distortion is an overall statistic measured over all pixels across a column of the detector, what matters for the purpose of this work is a measurement of the distortion that affects relative separations of objects located only a few arcseconds apart. 
We used two methods to estimate this.

First, we used a distinct observation of M15, also taken in October 2010, with the same detector configuration, in the 52 mas mode. We measured the relative separation of 13 bright star pairs as they are observed at different positions on the detector, with a 5-point dither of size  $3''$. The selected stars do not have nearby companions that may affect the measurement, and are not located close to cosmic rays or bad detector pixels. The field was distortion-corrected beforehand, using the distortion map described above. The measurements were performed with Galfit, by fitting two-dimensional Moffat profiles to the observed stars. 
The standard deviations of their measured separations across the X and Y axes are typically less than $\sim1$ mas, indicating that the residual distortion estimated above vastly overestimates the errors on the relative separation of close pairs. 


Second, we measured the separations between quasar images A and B of the actual AO data reported in this paper, in different dithered frames. This was more difficult, as the objects are intrinsically fainter, and there is also contamination from the lensing galaxy, usually located closer to the fainter image. To increase S/N, all dither frames where the objects fall on the same location on the detector were combined, by first calculating, via cross-correlation, necessary sub-pixel offsets due to telescope pointing errors. The observed systems were fitted with Hostlens, with the morphological parameters of the galaxies fixed at the values determined from the final, higher S/N science frame. The only free parameters were the positions and magnitudes of the objects, as well as the analytical 2 Moffat PSF parameters (since the PSF changes between frames). This method gave usable results (i.e. not affected by components poorly accounted for, or exceedingly large error bars) for 13 objects observed in the 52 mas mode, and one in the 20 mas mode. 
The resulting standard deviations are reported in Table \ref{tab:distort-data}. 

\begin{figure}
\includegraphics[width=75mm]{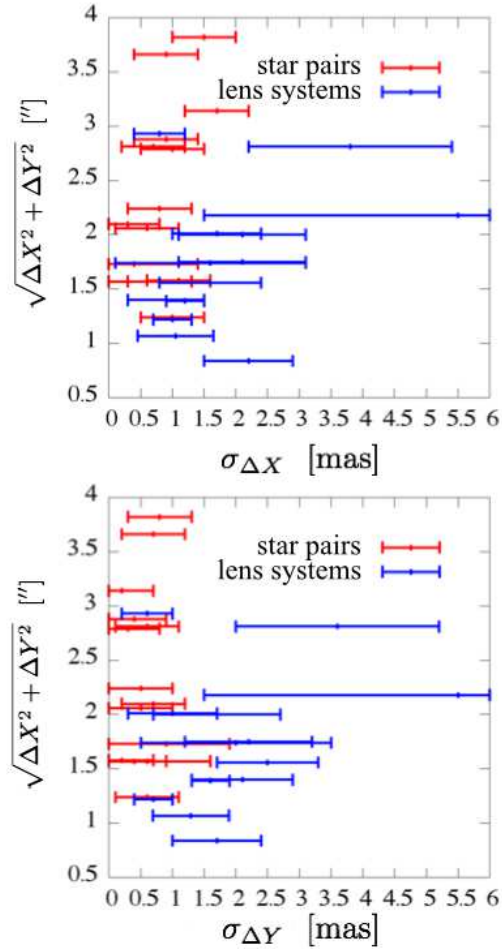}
\caption{ Star or quasar image separation on the X and Y axis, as a function of its measured standard deviation. The error bars are the typical measurement errors in one frame (in the case of the observed objects, the average of frames in the same dither position). \it{A colour version of this figure is available in the online version of the paper.}
\label{fig:histogrampairs}}
\end{figure}

The results of these two methods are compared in Figure \ref{fig:histogrampairs}. We found that the first method (the M15 star pairs) produces overall smaller standard deviations. This could be due either to the smaller measurement error bars (high S/N), or to the fact that the dither size was smaller, and therefore the locations of the pairs on the detector are closer to each other. In the latter case, it would be expected that objects with star pairs/quasar images with larger separations would also show comparatively larger standard deviations, but this is not obvious from the data. In view of the large error bars, the only statistical result that can be drawn is the average standard deviation, of $\sim1.5$ mas on both axes, for the quasar images. Here, two objects with much larger values were omitted. This value may be a reflection of the large measurement errors, or it may be due to actual residual distortion which is not corrected by the distortion map. As a result, this technique serves as an upper estimate of the mean residual distortion.

For the systems for which the second method gave reliable results, we used the measured standard deviations as estimates of the residual distortion (its upper estimates). The standard deviations are much smaller than the estimated residual distortion across the whole field, from Section \ref{subsection:distort}.  For the other objects, we used the average value of $\sim1.5$ mas. There are two additional factors that can bias a true measurement of the object separation (relative astrometry). First is the uncertainty in the pixel size. This affects the object separation as $\sqrt{N}\times\sigma_\mathrm{pixel}$, where N is the number of pixels separating the two objects. This effect is very small however and can be ignored (Table \ref{tab:distort-data}). 
The second factor is the atmospherical refraction. We estimated typical numerical values for its effect on each object based on \citet{helminiak09}, which studied the effects of differential atmospherical refraction on precise astrometry with AO in the infrared. We considered the resulting values, typically much less than 1 mas, as systematics, and reported them in Table \ref{tab:distort-data}. 

To conclude, we found that the residual field distortion affects the relative astrometry at a small level of $\sim1.5$ mas. We took three different systematics into account, added in quadrature and used them as additional errors on the relative astrometry of each system (including for the lensing galaxies located in between the quasar images): scatter in the positions at different dither positions, pixel size error, and atmospheric refraction. We present further considerations in Section \ref{sect:discutionoverall}.

\section{Lensing mass models}\label{section:lensmodels}

All objects (two-image lenses, in the great majority), unless otherwise required, are modelled with the singular isothermal sphere with external shear (SIS$+\gamma$), singular isothermal ellipsoid (SIE), and singular isothermal ellipsoid with shear SIE$+\gamma$) mass models for the lensing galaxy. For the latter model, the lens ellipticity and orientation were fixed at the observed values. The error on the flux ratio measured in $K'-$band was increased to 10\%, in order to account for microlensing, dust extinction, and time-delay \citep{yonehara08}. The models are fitted with the publicly available code Glafic \citep{oguri10}.

In cases where error bars are calculated on the best-fit model values, the error bars on the magnification factor and the time delay, which are not parameters but instead are results of lens modelling, are obtained from the histogram of magnification/time delay when fixing the model parameters to those obtained in the MCMC chains.

For several objects, the influence of nearby galaxies was considered by modelling them with SIS profiles, scaled to the mass of the main lens from their luminosity, through the Faber-Jackson law \citep{faber76} $L\propto\sigma^4$. This is in fact an upper limit to their influence, as it assumes that the galaxies and the main lens are at the same redshift, and they are early-type.

A description of the analysis and science results for each individual observed system is presented in Appendix \ref{section:objectsdescript} and \ref{section:appendobjects}. The astrometry/photometry, lens morphology and mass models for each system are provided in Tables \ref{tab:lensastrometry}, \ref{tab:lensmorphology}, and \ref{tab:massmodel}, respectively.

\section{Comparison with other available data}\label{sect:discuss}

Here we discuss how the derived astrometry and morphology results compare with previous data of the same objects, archival or available in the literature. The comparison is detailed for each object in Appendix \ref{section:objectsdescript}.

\subsection{Comparison with the previously available low-resolution data}

The astrometric results obtained from the current AO sample are generally not in agreement with the previously published low-resolution data available, within the formal error bars. The low-resolution data typically consists of observations in the optical and $I$ band, performed with the UH88/Tek2k (0.22 pixel$^{-1}$) camera. In addition to the IRCS+AO188 pixel size being $\sim4$ times finer (in the 52 mas mode), the seeing was also $\sim4$ times larger when compared to the AO observations, where the typical PSF core has FWHM $\sim0.2''$. The difference in astrometry is on the order of tens of milliarcseconds and larger for the lensing galaxies. 
There are several systems for which the position of the lensing galaxy is offset by 1-2 pixels (referring here to the UH88/Tek2k pixel size), significantly affecting the published mass models (for example, the lens lying closest to the bright quasar image, which is more unusual). In all of these cases (SDSS~J0832+0404, SDSS~J0904+1512, SDSS~J1001+5027, SDSS~J1131+1915, SDSS~J1216+3529 and SDSS~J1400+3134), the problem arises from the small separation between image B and the lensing galaxy G, of $\sim$1 pixel. As in all of these cases the peaks of the two objects are clearly separated in the AO data, the problem lies unequivocally with the inherent difficulties in the two-dimensional profile fitting multiple components in low-resolution data. 

In the cases where the lens galaxy morphology based on the low-resolution data is published, the results are typically in disagreement within the formal error bars. Again, this is most likely attributable to the poor resolution. The data do not however require a distinct interpretation, with a few exceptions. The main lenses in SDSS~J0743+2457 and SDSS~J1405+0959 are fitted in the discovery paper with a low and intermediate value Sersic index, respectively, whereas the new values in the AO data are larger, indicative of elliptical galaxies. Also, the lens in SDSS~J1330+1810 is originally fitted with Sersic index $\sim3.4$, but $\sim2.2$ in the new data, closer to the morphology of a disk-like galaxy. In addition, SDSS~J1254+2235, SDSS~J1353+1138 and SDSS~J1406+6126, originally modelled with a single Sersic component, require two concentric components in order to properly fit the AO data.

\subsection{Comparison with $HST$ data}\label{sect:hst}

There is a single object in the current AO sample, SDSS~J0832+0404, for which archival high-resolution data obtained with $HST$ is available (proposal ID 11289, PI. Jean-Paul Kneib). 
Unfortunately, the morphology of the lens galaxy is unreliable for a direct comparison with the AO data (ignoring also the fact that a galaxy may exhibit different morphology at visible and near-infrared wavelengths), due to the galaxy being very faint in the HST data, and in close proximity to the much brighter image B. Indeed, the morphological model diverges in terms of the Sersic index, effective radius and flux. Fixing the Sersic index at $n=4$, similar to the AO data, leaves residuals, and results in an effective radius that is $\sim10$ times larger than measured with AO. While a relative total flux comparison is problematic, the ratio of the peak pixel count at the core of B and G1 is $\sim15$, and it is B/G1 $\sim1$ in AO (figure \ref{fig:0832HSTresid}).

On the other hand, the relative astrometry is more straightforward to compare. The relative astrometry of A and B match at  $\lesssim1$ mas along the Y axis, and $\sim5$ mas along the X axis in the $HST$ and AO data. This is in good agreement considering the large error bars in Table \ref{tab:lensastrometry}. Even in case that the AO error bars are overestimated, the fact that the discrepancy along the X axis is larger can be expected due to the quasar host galaxy arc being more elongated along that axis, and affecting the measurement.



\onecolumn
\setlongtables
\begin{longtable}{ccccc}
  \caption{Relative Astrometry and Photometry of the Lens Systems} \\ 
  \hline 
  Object & 
  $\Delta X$ [$''$] & 
  $\Delta Y$ [$''$] & 
  $K'$ [mag] & 
  $K'$ ($n\equiv4)$ \\ 
  \hline
\multicolumn{5}{c}{SDSS J0743+2457 ($\theta=1.092''\pm0.020''$, $z_s=2.165$, $z_l=0.381$)} \vspace*{1.5mm} \\ \hline \vspace*{-2.0mm} \\ 
A & 0.000 $\pm$ 0.001 & 0.000 $\pm$ 0.001  & 16.88 $\pm$ 0.05 & \\
B & 0.807 $\pm$ 0.017 & 0.736 $\pm$ 0.011 & 18.93 $\pm$ 0.16 & \\
G & 0.676 $\pm$ 0.004 & 0.593 $\pm$ 0.004 & 17.32 $\pm$ 0.08 & 17.32 $\pm$ 0.06\\
\hline \multicolumn{5}{c}{SDSS J0819+5356 ($\theta=4.037''\pm0.003''$, $z_s=2.237$, $z_l=0.294$)} \vspace*{1.5mm} \\ \hline \vspace*{-2.0mm} \\ 
A & 0.000 $\pm$ 0.006 & 0.000 $\pm$ 0.003  & 16.39 $\pm$ 0.14$^a$  &  \\
B & $-$3.394 $\pm$ 0.005 & $-$2.185 $\pm$ 0.006 & 18.13 $\pm$ 0.14$^a$  &  \\
G1 & $-$1.991 $\pm$ 0.002 & $-$1.330 $\pm$ 0.005 & 13.65 $\pm$ 0.06 & 14.00 $\pm$ 0.01 \\
G2 & $\sim$$-23.18$ & $\sim$$-2.00$  & 15.07 $\pm$ 0.10 & \\
\hline \multicolumn{5}{c}{SDSS J0820+0812 ($\theta=2.090''\pm 0.030"$)$^{b}$, $z_s=2.024$, $z_l=0.803$} \vspace*{1.5mm} \\ \hline \vspace*{-2.0mm} \\ 
A & 0.000 $\pm$ 0.004 & 0.000 $\pm$ 0.004 & 17.21 $\pm$ 0.03 & \\
B & 0.918 $\pm$ 0.015 & 1.878 $\pm$ 0.025 & 17.98 $\pm$ 0.06 & \\
G & 0.690 $\pm$ 0.036 & 1.252 $\pm$ 0.063 & 16.81 $\pm$ 0.22 & \\
\hline \multicolumn{5}{c}{SDSS J0832+0404 ($\theta=2.182''\pm 0.006"$, $z_s=1.115$, $z_l=0.659$)} \vspace*{1.5mm} \\ \hline \vspace*{-2.0mm} \\ 
A & 0.000 $\pm$ 0.0015 & 0.000 $\pm$ 0.002 & 16.50 $\pm$ 0.06$^a$ & \\
B & -1.774 $\pm$ 0.004 & $-$1.270 $\pm$ 0.003 & 17.58 $\pm$ 0.06$^a$ & \\
G & $-$1.327 $\pm$ 0.004 & $-0.947$ $\pm$ 0.003 & 16.33 $\pm$ 0.04 & \\
\hline \multicolumn{5}{c}{SDSS J0904+1512 ($\theta=1.064''\pm 0.003"$, $z_s=1.826$, $z_l\sim0.2-0.5$)} \vspace*{1.5mm} \\ \hline \vspace*{-2.0mm} \\ 
A & 0.000 $\pm$ 0.000 & 0.000 $\pm$ 0.000  & 15.73 $\pm$ 0.03$^a$ & \\
B & $-$0.928 $\pm$ 0.001 & $-$0.521 $\pm$ 0.002 & 17.29 $\pm$ 0.03$^a$ & \\
G1 & $-$0.570 $\pm$ 0.001 & $-$0.423 $\pm$ 0.005 & 17.39 $\pm$ 0.19 & 17.45 $\pm$ 0.03 \\
G2 & 0.912 $\pm$ 0.017 & $-$4.967 $\pm$ 0.031 & 18.97 $\pm$ 0.09 & \\
\hline \multicolumn{5}{c}{SDSS J0946+1835 ($\theta=3.055\pm 0.002"$, $z_s=4.799$, $z_l=0.388$)$^c$} \vspace*{1.5mm} \\ \hline \vspace*{-2.0mm} \\ 
A & 0.000 $\pm$ 0.000 & 0.000 $\pm$ 0.000  & 17.10 $\pm$ 0.00 & \\
B & 1.754 $\pm$ 0.000 & $-$2.502 $\pm$ 0.000 & 17.57 $\pm$ 0.00 & \\
G1 & 1.459 $\pm$ 0.000 & $-$1.927 $\pm$ 0.001 & 15.00 $\pm$ 0.03 &15.29 $\pm$ 0.02 \\
G2$^d$ & 2.307 $\pm$ 0.008 & 2.751 $\pm$ 0.006 & 20.03 $\pm$ 0.05 & \\
G3 & $-$1.270 $\pm$ 0.007 & 0.080 $\pm$ 0.013 & 21.16 $\pm$ 0.20 & \\
\hline \multicolumn{5}{c}{SDSS J1001+5027 ($\theta=2.928''\pm 0.0015''$, $z_s=1.839$, $z_l=0.415$)} \vspace*{1.5mm} \\ \hline \vspace*{-2.0mm} \\ 
A & 0.000 $\pm$ 0.000 & 0.000 $\pm$ 0.001  & 15.84 $\pm$ 0.03 & \\
B & 2.489 $\pm$ 0.001 & 1.543 $\pm$ 0.001 & 16.10 $\pm$ 0.04 & \\
G1 & 2.054 $\pm$ 0.012 & 1.122 $\pm$ 0.009 & 15.87 $\pm$ 0.11 & 16.12 $\pm$ 0.03 \\
G2 & 1.799 $\pm$ 0.003 & $-$0.865 $\pm$ 0.005 & 17.55 $\pm$ 0.22 \\
\hline \multicolumn{5}{c}{SDSS J1055+4628 ($\theta=1.145''\pm 0.003''$, $z_s=1.249$, $z_l=0.388$)} \vspace*{1.5mm} \\ \hline \vspace*{-2.0mm} \\ 
A & 0.000 $\pm$ 0.001 & 0.000 $\pm$ 0.001  & 17.10 $\pm$ 0.03 & \\
B & $-$0.043 $\pm$ 0.002 & 1.145 $\pm$ 0.002 & 18.31 $\pm$ 0.06 &\\
G1 & $-$0.048 $\pm$ 0.002 & 0.850 $\pm$ 0.002 & 16.68 $\pm$ 0.11 &16.60 $\pm$ 0.15\\
G2$^d$ & $\sim$12.03 & $\sim$$-$6.05 & 16.53 $\pm$ 0.02 & \\
\hline \multicolumn{5}{c}{SDSS J1131+1915 ($\theta=1.403''\pm 0.002''$, $z_s=2.915$, $z_l\sim0.3$)} \vspace*{1.5mm} \\ \hline \vspace*{-2.0mm} \\ 
A & 0.000 $\pm$ 0.000 & 0.000 $\pm$ 0.000  & 15.97 $\pm$ 0.00 & \\
B & $-$1.396 $\pm$ 0.000 & $-$0.134 $\pm$ 0.000 & 17.55 $\pm$ 0.00 & \\
G1 & $-$0.946 $\pm$ 0.003 & $-$0.142 $\pm$ 0.003 & 17.90 $\pm$ 0.09 & 18.10$\pm$ 0.03\\
G2$^d$ & $\sim$$-$8.72 & $\sim$$-$3.85 & 17.94 $\pm$ 0.03 \\
\hline \multicolumn{5}{c}{SDSS J1206+4332 ($\theta=3.026''\pm 0.005''$, $z_s=1.789$, $z_l=0.748$)$^e$} \vspace*{1.5mm} \\ \hline \vspace*{-2.0mm} \\ 
A & 0.000 $\pm$ 0.003 & 0.000 $\pm$ 0.002  & 16.56 $\pm$ 0.02 & \\
B & $-$0.035 $\pm$ 0.003 & 3.027 $\pm$ 0.003 & 16.60 $\pm$ 0.03 &  \\
G1 & $-$0.508 $\pm$ 0.014 & 1.813 $\pm$ 0.012 & 15.85 $\pm$ 0.45 &  \\
G2 & 1.120 $\pm$ 0.029 & 6.070 $\pm$ 0.034 & 16.80 $\pm$ 0.76 &  \\
G3 & $-$1.853 $\pm$ 0.014 & 2.470 $\pm$ 0.060 & 17.44 $\pm$ 0.06 &  \\
G4 & 1.70 $\pm$ 0.11 & 4.75 $\pm$ 0.12 & 19.48 $\pm$ 0.81 &  \\
G5 & $-$7.23 $\pm$ 0.10 & $-$0.27 $\pm$ 0.11 & 18.38 $\pm$ 0.05 &  \\
\hline \multicolumn{5}{c}{SDSS J1216+3529 ($\theta=1.487''\pm0.013''$, $z_s=2.012$, $z_l\sim0.55$)} \vspace*{1.5mm} \\ \hline \vspace*{-2.0mm} \\ 
A & 0.000 $\pm$ 0.002 & 0.000 $\pm$ 0.002  & 16.82 $\pm$ 0.03 & \\
B & 1.484 $\pm$ 0.011 & 0.097 $\pm$ 0.006 & 17.87 $\pm$ 0.09 &\\
G1 & 1.165 $\pm$ 0.012 & 0.013 $\pm$ 0.014 & 17.37 $\pm$ 0.13 & 17.43 $\pm$ 0.10\\
G2$^d$ & $\sim$7.51 & $\sim$$-$12.63 & 17.46 $\pm$ 0.04 & \\
G3 & 1.781 $\pm$ 0.051 & 2.803 $\pm$ 0.066 & 18.47 $\pm$ 0.81 & \\
\hline \multicolumn{5}{c}{SDSS J1254+2235 ($\theta=1.561''\pm 0.004''$, $z_s=3.626$, $z_l\sim0.3$)} \vspace*{1.5mm} \\ \hline \vspace*{-2.0mm} \\ 
A & 0.000 $\pm$ 0.001 & 0.000 $\pm$ 0.002  & 16.95 $\pm$ 0.06 & \\
B & $-$1.446 $\pm$ 0.003 & $-$0.588 $\pm$ 0.002 & 17.82 $\pm$ 0.05 & \\
G1a & $-$0.871 $\pm$ 0.004 & $-$0.317 $\pm$ 0.003 & 17.58 $\pm$ 0.12 & 17.40 $\pm$ 0.15 \\
G1b & $-$0.871 $\pm$ 0.004 & $-$0.317 $\pm$ 0.003 & 17.54 $\pm$ 0.11 & 17.71 $\pm$ 0.17$^f$ \\
G1a+G1b & $-$0.871 $\pm$ 0.004 & $-$0.317 $\pm$ 0.003 & 16.80 $\pm$ 0.03 & \\
G2 & $-$2.456 $\pm$ 0.022 & $-$0.458 $\pm$ 0.025 & 21.01 $\pm$ 0.24 &  \\
\hline \multicolumn{5}{c}{SDSS J1313+5151 ($\theta=1.223''\pm 0.001''$, $z_s=1.875$, $z_l=0.194$)} \vspace*{1.5mm} \\ \hline \vspace*{-2.0mm} \\ 
A & 0.000 $\pm$ 0.000 & 0.000 $\pm$ 0.000  & 14.62 $\pm$ 0.02 & \\
B & 0.309 $\pm$ 0.000 & $-$1.184 $\pm$ 0.000 & 16.07 $\pm$ 0.01 & \\
G & 0.255 $\pm$ 0.001 & $-$0.977 $\pm$ 0.002 & 15.39 $\pm$ 0.03 & 15.69$\pm$ 0.03\\
\hline \multicolumn{5}{c}{SDSS J1322+1052 ($\theta=1.999''\pm 0.002''$, $z_s=1.716$, $z_l\sim0.55$)} \vspace*{1.5mm} \\ \hline \vspace*{-2.0mm} \\ 
A & 0.000 $\pm$ 0.000 & 0.000 $\pm$ 0.000  & 16.22 $\pm$ 0.06$^a$ & \\
B & 1.070 $\pm$ 0.001 & 1.689 $\pm$ 0.001 & 17.80 $\pm$ 0.06$^a$ & \\
G1 & 0.738 $\pm$ 0.0025 & 1.187 $\pm$ 0.001 & 15.24 (?) & 16.70 $\pm$ 0.02  \\
G2 & $\sim-6.27$ & $\sim$9.92 & 18.81 $\pm$ 0.02 &  \\
G3 & $-2.672$ $\pm$ 0.051 & 1.443 $\pm$ 0.020 & 19.28 $\pm$ 0.14 &  \\
G4 & $\sim$5.94 & $\sim$4.31 & 19.43 $\pm$ 0.08 & \\
G5 & $\sim$1.09 & $\sim$11.94 & 19.51 $\pm$ 0.02 & \\
\hline \multicolumn{5}{c}{SDSS J1330+1810 (max. sep. $1.710''\pm 0.003''$, $z_s=1.393$, $z_l=0.373$)} \vspace*{1.5mm} \\ \hline \vspace*{-2.0mm} \\ 
A & 0.000 $\pm$ 0.001 & 0.000 $\pm$ 0.000  & 16.72 $\pm$ 0.02 & \\
B & 0.414 $\pm$ 0.0015 & $-$0.011 $\pm$ 0.001 & 17.15 $\pm$ 0.03 & \\
C & 1.248 $\pm$ 0.002 & 1.168 $\pm$ 0.0015  & 17.61 $\pm$ 0.03 &  \\
D & $-$0.244 $\pm$ 0.002 & 1.579 $\pm$ 0.002 & 18.88 $\pm$ 0.02 &  \\
G & 0.221 $\pm$ 0.008 & 0.966 $\pm$ 0.0015 & 15.31 $\pm$ 0.06 &  \\
\hline \multicolumn{5}{c}{SDSS J1334+3315$^g$ ($0.838''\pm 0.002''$, $z_s=2.426$, $z_l=0.557$)} \vspace*{1.5mm} \\ \hline \vspace*{-2.0mm} \\ 
A & 0.001 $\pm$ 0.001 & 0.000 $\pm$ 0.000  & 18.74 $\pm$ 0.03 $J$ & \\
 & & & 18.31 $\pm$ 0.02 $H$ & \\
 & & & 17.67 $\pm$ 0.06 $K'$ & \\
B & $-$0.491 $\pm$ 0.001 & 0.680 $\pm$ 0.001 & 19.07 $\pm$ 0.04 $J$ & \\
 & & & 18.63 $\pm$ 0.03 $H$ & \\
 & & & 17.97 $\pm$ 0.07 $K'$ & \\
G & $-$0.134 $\pm$ 0.009 & 0.220 $\pm$ 0.009  & 20.53 $\pm$ 0.36 $J$ &  \\
 & & & & 19.93 $\pm$ 0.35 $H$ \\
 & & & 18.88 $\pm$ 0.38 $K'$ & \\
G1 & $-$3.171 $\pm$ 0.001 & $-$2.609 $\pm$ 0.001 & 18.74 $\pm$ 0.01 $J$ &  \\
 & & & 18.09 $\pm$ 0.02 $H$ & \\
 & & & 17.56 $\pm$ 0.01 $K'$ & \\
G2 & 1.151 $\pm$ 0.009 & 2.049 $\pm$ 0.009 & 23.11 $\pm$ 0.26 $J$ &  \\
 & & & 20.81 $\pm$ 0.10 $H$ & \\
 & & & 20.31 $\pm$ 0.07 $K'$ & \\
\hline \multicolumn{5}{c}{SDSS J1353+1138 ($\theta=1.392''\pm 0.0015"$, $z_s=1.629$, $z_l\sim0.25$)} \vspace*{1.5mm} \\ \hline \vspace*{-2.0mm} \\ 
A & 0.000 $\pm$ 0.000 & 0.000 $\pm$ 0.000  & 15.02 $\pm$ 0.01 & \\
B & 0.245 $\pm$ 0.000 & $-$1.371 $\pm$ 0.000 & 15.98 $\pm$ 0.00 & \\
G1a & 0.248 $\pm$ 0.001 & $-$1.042 $\pm$ 0.000 & 15.46 $\pm$ 0.02 & 16.00 $\pm$ 0.20 \\
G1b & [0.248 $\pm$ 0.001] & [$-$1.042 $\pm$ 0.000] & 17.13 $\pm$ 0.05 & 16.20 $\pm$ 0.28$^f$ \\
G1a+G1b & [0.248 $\pm$ 0.001] & [$-$1.042 $\pm$ 0.000]   & 15.25 $\pm$ 0.01 & 15.36 $\pm$ 0.01 \\
G2 & 2.824 $\pm$ 0.009 & 5.392 $\pm$ 0.005 & 18.82 $\pm$ 0.09 & \\
\hline \multicolumn{5}{c}{SDSS J1400+3134 ($\theta=1.744''\pm 0.003''$, $z_s=3.317$, $z_l\sim0.8$)} \vspace*{1.5mm} \\ \hline \vspace*{-2.0mm} \\ 
A & 0.000 $\pm$ 0.000 & 0.000 $\pm$ 0.000  & 16.94 $\pm$ 0.03 & \\
B & 1.032 $\pm$ 0.001 & 1.406 $\pm$ 0.001 & 17.42 $\pm$ 0.03 & \\
G1 & 0.327 $\pm$ 0.005 & 1.025 $\pm$ 0.004 & 17.81 $\pm$ 0.14 & \\
G2$^d$ & $\sim-$6.37 & $\sim$5.78 & 18.70 $\pm$ 0.05 &  \\
\hline \multicolumn{5}{c}{SDSS J1406+6126 ($\theta=2.029''\pm 0.011''$, $z_s=2.134$, $z_l=0.271$)$^h$} \vspace*{1.5mm} \\ \hline \vspace*{-2.0mm} \\ 
A & 0.000 $\pm$ 0.006 & 0.000 $\pm$ 0.002  & 17.43 $\pm$ 0.02 & \\
B & 1.686 $\pm$ 0.006 & $-$1.129 $\pm$ 0.006 & 17.99 $\pm$ 0.03 & \\
Ga & 1.160 $\pm$ 0.002 & $-$0.750 $\pm$ 0.003 & 16.53 $\pm$ 0.13 &  \\
Gb & 1.160 $\pm$ 0.002 & $-$0.750 $\pm$ 0.003 & 16.07 $\pm$ 0.11 &  \\
Ga+Gb & 1.160 $\pm$ 0.002 & $-$0.750 $\pm$ 0.003 & 15.52 $\pm$ 0.12 & \\
\hline \multicolumn{5}{c}{SDSS J1455+1447 ($\theta=1.753''\pm 0.001''$, $z_s=1.424$, $z_l\sim0.42$)$^h$} \vspace*{1.5mm} \\ \hline \vspace*{-2.0mm} \\ 
A & 0.000 $\pm$ 0.000 & 0.000 $\pm$ 0.000  & 15.81 $\pm$ 0.00 & \\
B & 1.609 $\pm$ 0.000 & 0.696 $\pm$ 0.000 & 16.95 $\pm$ 0.01 & \\
G1 & 1.003 $\pm$ 0.005 & 0.538 $\pm$ 0.004 & 16.05 $\pm$ 0.10 & 16.37 $\pm$ 0.02 \\
G2 & $\sim$10.75 & $\sim$2.10 & 18.03 $\pm$ 0.03 &  \\
\hline \multicolumn{5}{c}{SDSS J1515+1511 ($\theta=2.010''\pm 0.001"$, $z_s=2.054$, $z_l=0.742$)} \vspace*{1.5mm} \\ \hline \vspace*{-2.0mm} \\ 
A & 0.000 $\pm$ 0.001 & 0.000 $\pm$ 0.000  & 16.01 $\pm$ 0.01 & \\
B & 1.668 $\pm$ 0.001 & $-$1.122 $\pm$ 0.001 & 16.35 $\pm$ 0.01 & \\
G1 & 1.357 $\pm$ 0.003 & $-$0.848 $\pm$ 0.003 & 17.29 $\pm$ 0.02 \\
G2 & 13.49 & $-$8.30 & 17.41 $\pm$ 0.02$^d$ \\
\hline \multicolumn{5}{c}{SDSS J1620+1203 ($\theta=2.807''\pm 0.004''$, $z_s=1.158$, $z_l=0.398$)$^b$} \vspace*{1.5mm} \\ \hline \vspace*{-2.0mm} \\ 
A & 0.000 $\pm$ 0.000 & 0.000 $\pm$ 0.000 & 17.02 $\pm$ 0.01 & \\
B & $-$2.107 $\pm$ 0.001 & $-$1.856 $\pm$ 0.001 & 18.56 $\pm$ 0.02 & \\
G & $-$1.764 $\pm$ 0.001 & $-$1.458 $\pm$ 0.001 & 15.52 $\pm$ 0.02 & 15.72 $\pm$ 0.01 \\
\hline 
\label{tab:lensastrometry}
\end{longtable}
\begin{tablenotes}
\item{\footnotesize Magnitudes are in the Vega system. All magnitudes are corrected for atmospheric and Galactic extinction. The positive directions of X and Y are West and North, respectively. The objects without attached error bars on the positions are located at a relatively large distance from the target, where the residual distortion is expected to be large. Here $n\equiv4$ refers to the Sersic index, and $z_l$, $z_s$ are the redshifts of the main lens galaxy and source quasar, respectively; where $z_l$ is approximate, its value is a photometric redshift estimate. For the objects where the quasar host galaxy has been explicitly modelled, relative photometry of the images includes the contribution of the point source as well as the host. The results for SDSS~J1405+0959, presented in a separate publication, are not included (see Appendix \ref{section:1405}). \\ 
$^a$ The fluxes are estimated with Hostlens for the point source, and reflect the uncertainty in the host galaxy - quasar disentangling, therefore attached errors should not be used to estimate the error on the flux ratio, which are smaller. \\ 
$^b$ The zero-point error is 0.14 mag for both SDSS~J0820+0812  and SDSS~J1620+1203. \\ 
$^c$ In addition to the quoted errors on photometry, a systematic error of 0.05 mag was obtained during simulations, which should be added to the absolute photometry, but does not affect the relative values. \\ 
$^d$ Aperture photometry is given, and the positions and their errors are measured with the PHOT task. \\ 
$^e$ The astrometry for SDSS~J1206+4332 should be considered less reliable (see Appendix \ref{section:1206}) \\ 
$^f$ Sersic index of G1a (not G1b) is fixed at 4. \\ 
$^g$ The astrometry represents the average over the $H$ and $K'$ bands, and the magnitudes of G1 and G2 were obtained from aperture photometry. \\ 
$^h$ Absolute photometry unreliable due to non-photometric conditions.} \\ 
\end{tablenotes}

\begin{longtable}{ccccc}
  \caption{Morphological parameters} \\ 
  \hline 
Object & 
$e$ &
$PA$ &
$r_e$ [$''$] &
$n$ \\ 
\hline
\hline \multicolumn{5}{c}{SDSS J0743+2457} \vspace*{1.5mm} \\ \hline \vspace*{-2.0mm} \\ 
G & 0.80 $\pm$ 0.12 & 23.8 $\pm$ 2.0 & 0.19 $\pm$ 0.05 & 4.13 $\pm$ 0.92 \\
G & 0.79 $\pm$ 0.12 & 23.8 $\pm$ 2.0 & 0.19 $\pm$ 0.05 & [4] \\
\hline \multicolumn{5}{c}{SDSS J0819+5356} \vspace*{1.5mm} \\ \hline \vspace*{-2.0mm} \\ 
G1 & 0.25 $\pm$ 0.03 & -35.5 $\pm$ 2.4 & 2.89 $\pm$ 0.45 & 6.28 $\pm$ 0.34 \\
G1 & 0.15 $\pm$ 0.01 & -31.8 $\pm$ 1.2  & 1.22 $\pm$ 0.05 & [4] \\
\hline \multicolumn{5}{c}{SDSS J0820+0812} \vspace*{1.5mm} \\ \hline \vspace*{-2.0mm} \\ 
G$^a$ & 0.25 $\pm$ 0.05 & $-$44.8 $\pm$ 5.9 & 1.58 $\pm$ 0.30 & 1.37 $\pm$ 0.75 \\
\hline \multicolumn{5}{c}{SDSS J0832+0404} \vspace*{1.5mm} \\ \hline \vspace*{-2.0mm} \\ 
G & 0.21 $\pm$ 0.02 & 9.0 $\pm$ 8.3 & 0.26 $\pm$ 0.05 & 3.85 $\pm$ 0.70 \\
\hline \multicolumn{5}{c}{SDSS J0904+1512} \vspace*{1.5mm} \\ \hline \vspace*{-2.0mm} \\ 
G1 & [0] & - & 0.32 $\pm$ 0.09 & 4.89 $\pm$ 0.71 \\
G1 & [0] & - & 0.29 $\pm$ 0.01 & [4] \\
G2 & 0.21 $\pm$ 0.12 & 43.0 $\pm$ 11.7 & 0.46 $\pm$ 0.05 & 0.56 $\pm$ 0.21 \\
\hline \multicolumn{5}{c}{SDSS J0946+1835} \vspace*{1.5mm} \\ \hline \vspace*{-2.0mm} \\ 
G1 & 0.09 $\pm$ 0.01 & -11.9 $\pm$ 3.9 & 1.37 $\pm$ 0.09 & 5.64 $\pm$ 0.30 \\
G1 & 0.09 $\pm$ 0.00 & -11.6 $\pm$ 3.6 & 0.79 $\pm$ 0.04 & [4] \\
G2 & 0.34 $\pm$ 0.04 & 68.7 $\pm$ 4.4 & 0.30 $\pm$ 0.01 & 0.78 $\pm$ 0.11 \\
G3 & 0.66 $\pm$ 0.09 & -9.6 $\pm$ 4.1 & 0.21 $\pm$ 0.07 & 0.27 $\pm \sim$ 0.5 \\
\hline \multicolumn{5}{c}{SDSS J1001+5027} \vspace*{1.5mm} \\ \hline \vspace*{-2.0mm} \\ 
G1 & 0.10 $\pm$ 0.08 & 8.7 $\pm$ 5.8 & 1.77 $\pm$ 0.36 & 5.74 $\pm$ 0.56 \\
G1 & 0.10 $\pm$ 0.01 & 1.0 $\pm$ 7.31 & 1.06 $\pm$ 0.04 & [4] \\
G2 & 0.16 $\pm$ 0.03 & $-$43.9 $\pm$ 10.1 & 0.41 $\pm$ 0.13 & 3.23 $\pm$ 1.80 \\
\hline \multicolumn{5}{c}{SDSS J1055+4628} \vspace*{1.5mm} \\ \hline \vspace*{-2.0mm} \\ 
G1 & 0.36 $\pm$ 0.04 & $-$1.7 $\pm$ 3.3 & 0.48 $\pm$ 0.30 & 5.83 $\pm$ 2.40 \\
G1 & 0.41 $\pm$ 0.05 & $-$0.5 $\pm$ 2.2 & 0.39 $\pm$ 0.05 & [4] \\
\hline \multicolumn{5}{c}{SDSS J1131+1915} \vspace*{1.5mm} \\ \hline \vspace*{-2.0mm} \\
G1 & 0.47 $\pm$ 0.07 & 87.5 $\pm$ 5.2 & 0.43 $\pm$ 0.11 & 6.58 $\pm$ 1.14 \\
G1 & 0.45 $\pm$ 0.07 & 87.8 $\pm$ 4.9 & 0.27 $\pm$ 0.02 & [4] \\
\hline \multicolumn{5}{c}{SDSS J1216+3529} \vspace*{1.5mm} \\ \hline \vspace*{-2.0mm} \\
G1 & 0.67 $\pm$ 0.11 & $-$14.60 $\pm$ 4.10 & 0.32 $\pm$ 0.10 & 5.02 $\pm$ 2.83 \\
G1 & 0.68 $\pm$ 0.11 & $-$15.20 $\pm$ 2.20 & 0.28 $\pm$ 0.05 & [4] \\
\hline \multicolumn{5}{c}{SDSS J1254+2235} \vspace*{1.5mm} \\ \hline \vspace*{-2.0mm} \\
G1a & 0.32 $\pm$ 0.05 & $-$61.3 $\pm$ 2.7 & 0.14 $\pm$ 0.03 & 2.6 $\pm$ 1.6 \\
G1b & 0.28 $\pm$ 0.05 & $-$26.2 $\pm$ 3.9 & 1.09 $\pm$ 0.06 & 0.23 $\pm$ 0.07 \\
G1a & 0.31 $\pm$ 0.05 & $-$60.8 $\pm$ 2.6 & 0.72 $\pm$ 0.19 & [4] \\
G1b & 0.30 $\pm$ 0.05 & $-$23.8 $\pm$ 3.1 & 1.13 $\pm$ 0.05 & 0.17 $\pm$ 0.09 \\
\hline \multicolumn{5}{c}{SDSS J1313+5151} \vspace*{1.5mm} \\ \hline \vspace*{-2.0mm} \\
G & 0.35 $\pm$ 0.01 & $-$68.7 $\pm$ 1.0 & 1.32 $\pm$ 0.09 & 7.0 $\pm$ 0.43 \\
G & 0.32 $\pm$ 0.01 & $-$68.8 $\pm$ 1.2 & 0.69 $\pm$ 0.03 & [4] \\
\hline \multicolumn{5}{c}{SDSS J1322+1052} \vspace*{1.5mm} \\ \hline \vspace*{-2.0mm} \\
G1 & 0.07 $\pm$ 0.03  & $-$52.3 $\pm$ 7.8 & 0.48 $\pm$ 0.02 & [4] \\
G3 & 0.59 $\pm$ 0.09 & 72.1 $\pm$ 5.7 & 0.59 $\pm$ 0.28 & 0.94 $\pm$ 0.30 \\
\hline \multicolumn{5}{c}{SDSS J1330+1810} \vspace*{1.5mm} \\ \hline \vspace*{-2.0mm} \\
G & 0.70 $\pm$ 0.03 & 25.5 $\pm$ 0.4 & 0.49 $\pm$ 0.03 & 2.23 $\pm$ 0.12 \\
\hline \multicolumn{5}{c}{SDSS J1334+3315} \vspace*{1.5mm} \\ \hline \vspace*{-2.0mm} \\
G ($J$) & unreliable & $\sim-$21 $\pm$ 22  & 0.25 $\pm$ 0.05 & 2.53 $\pm$ 1.36 \\
G ($H$) & 0.61 $\pm$ 0.29 & $\sim-$ 23 $\pm$ 5.4 & 0.22 $\pm$ 0.11 & [4] \\
G ($K'$) & 0.43 $\pm$ 0.23 & $-$36.5 $\pm$ 18.7 & 0.34 $\pm$ 0.17 & 3.4 $\pm$ 2.1 \\
G ($K'$ host) & 0.62 $\pm$ 0.23 & $-$5.0 $\pm$ 3.8  & 0.18 $\pm$ 0.05 & $\sim$2.6 $\pm$ 0.81 \\
G1 ($J$) & 0.44 $\pm$ 0.02 & $-$57.4 $\pm$ 1.4 & 0.78 $\pm$ 0.03 & 1.51 $\pm$ 0.12 \\
G1 ($H$) & 0.55 $\pm$ 0.03 & $-$60.5 $\pm$ 1.7  & 0.73 $\pm$ 0.16 & 2.58 $\pm$ 0.54 \\
G1 ($K'$) & 0.60 $\pm$ 0.02 & $-$59.5 $\pm$ 1.0 & 0.73 $\pm$ 0.13 & 3.40 $\pm$ 0.42 \\
\hline \multicolumn{5}{c}{SDSS J1353+1138} \vspace*{1.5mm} \\ \hline \vspace*{-2.0mm} \\
G1a & 0.50 $\pm$ 0.08 & $-$64.8 $\pm$ 0.9 & 0.47 $\pm$ 0.01 & 5.77 $\pm$ 0.19 \\
G1b & 0.69 $\pm$ 0.03 & $-$66.8 $\pm$ 0.9 & 1.04 $\pm$ 0.03 & 0.32 $\pm$ 0.08 \\
G1a & 0.52 $\pm$ 0.04 & $-$63.9 $\pm$ 1.4 & 0.20 $\pm$ 0.05 & [4] \\
G1b & 0.61 $\pm$ 0.03 & $-$66.6 $\pm$ 0.6 & 1.04 $\pm$ 0.07 & 0.88 $\pm$ 0.25 \\
\hline \multicolumn{5}{c}{SDSS J1400+3134} \vspace*{1.5mm} \\ \hline \vspace*{-2.0mm} \\
G1 & 0.32 $\pm$ 0.05 & $-$66.6 $\pm$ 7.2 & 0.31 $\pm$ 0.09 & [4] \\
\hline \multicolumn{5}{c}{SDSS J1406+6126} \vspace*{1.5mm} \\ \hline \vspace*{-2.0mm} \\
G1a & [0] & $-$ & 0.13 $\pm$ 0.04 & [4] \\
G1b & 0.69 $\pm$ 0.01 & $-$28.0 $\pm$ 0.60 & 1.14 $\pm$ 0.06 & [1] \\
\hline \multicolumn{5}{c}{SDSS J1455+1447} \vspace*{1.5mm} \\ \hline \vspace*{-2.0mm} \\
G1 & 0.23 $\pm$ 0.02 & 82.1 $\pm$ 5.3 & 0.87 $\pm$ 0.17 & 8.40 $\pm$ 1.25 \\
G1 & 0.26 $\pm$ 0.02 & 86.3 $\pm$ 2.5 & 0.45 $\pm$ 0.05 & [4] \\
\hline \multicolumn{5}{c}{SDSS J1515+1511} \vspace*{1.5mm} \\ \hline \vspace*{-2.0mm} \\
G1 & 0.81 $\pm$ 0.03 & -17.1 $\pm$ 0.3 & 0.66 $\pm$ 0.01 & 1.65 $\pm$ 0.15 \\
\hline \multicolumn{5}{c}{SDSS J1620+1203} \vspace*{1.5mm} \\ \hline \vspace*{-2.0mm} \\
G & 0.24 $\pm$ 0.02 & -27.2 $\pm$ 1.2 & 1.30 $\pm$ 0.07 & 4.88 $\pm$ 0.20 \\
G & 0.22 $\pm$ 0.01 & -24.7 $\pm$ 0.8 & 1.02 $\pm$ 0.01 & [4] \\ 
\hline 
\label{tab:lensmorphology}
\end{longtable}
{\footnotesize The position angle is positive from North towards East.
}



\onecolumn

\begingroup
\small
\begin{landscape}
\begin{longtable}{cccccc}
  \caption{Lensing models} \\ 
  \hline 
Model & 
$\sigma$ [km/s] (M [$h^{-1}  \mathrm{M}_{\sun}$]) / $\theta_\mathrm{Ein}$ [$''$] &
$e, \theta_e$  [deg] &
$\gamma, \theta_\gamma$  [deg] &
$\Delta t$ [days] &
$\mu$ \\ 
\hline
\hline \multicolumn{6}{c}{SDSS J0743+2457 ($z_s=2.165$, $z_l=0.381$)} \vspace*{1.5mm} \\ \hline \vspace*{-2.0mm} \\ 
SIS$+\gamma$ & $165.2^{+2.2}_{-2.2}/0.58^{+0.01}_{-0.03}$   & / & $0.063^{+0.02}_{-0.03}$, $-31.7^{+10}_{-12}$ & $23.1^{+0.4}_{-0.6}$ & $3.58^{+0.22}_{-0.45}$ \\
SIE & $163.3^{+1.2}_{-2.5}/0.56^{+0.01}_{-0.02}$  & $0.17^{+0.05}_{-0.07}$, $-31.9^{+12}_{-14}$ & / & $22.0^{+0.3}_{-0.2}$ & $3.55^{+0.18}_{-0.47}$ \\
SIE$+\gamma$ & $187.9^{-1.3}_{-13.6}/0.74^{+0.00}_{-0.10}$   & ($0.80^{+0.04}_{-0.17}, 23.8^{+2.0}_{-1.9}$) & $0.496^{-0.028}_{-0.187}$, $-63.4^{+2.8}_{-1.3}$ & $30.9^{+0.3}_{-2.7}$ & $4.45^{+0.27}_{-0.91}$ \\
\hline \multicolumn{6}{c}{SDSS J0819+5356 ($z_s=2.237$, $z_l=0.294$)} \vspace*{1.5mm} \\ \hline \vspace*{-2.0mm} \\ 
SIS$+\gamma$ & $304.5^{+0.6}_{-0.05}/2.12^{+0.00}_{-0.01}$ & / & $0.049^{+0.003}_{-0.003}$, $-60.0^{+0.3}_{-0.4}$ & $66.4^{+0.7}_{-0.6}$ & $16.95^{+0.95}_{-0.96}$ \\
SIE & $301.2^{+0.4}_{-0.3}/2.07^{+0.01}_{-0.00}$  & $0.13^{+0.01}_{-0.00}$, $-60.0^{+0.3}_{-0.4}$ & / & $63.4^{+0.7}_{-0.5}$ & $16.90^{+0.08}_{-1.00}$ \\
SIE$+\gamma$ & $300.2^{+1.1}_{-0.5}/2.06^{+0.01}_{-0.01}$ & ($0.25^{+0.02}_{-0.03}$, $-35.5^{+2.1}_{-2.5}$) & $0.079^{+0.009}_{-0.016}$, $69.9^{+3.9}_{-2.7}$ & $61.5^{+1.2}_{-0.6}$ & $17.22^{+1.08}_{-0.98}$ \\
\hline \multicolumn{6}{c}{SDSS~J0820+0812 ($z_s=2.024$, $z_l=0.803$)} \vspace*{1.5mm} \\ \hline \vspace*{-2.0mm} \\ 
SIS$+\gamma$ & $277.3^{+2.9}_{-2.0}/1.04^{+0.02}_{-0.02}$   & / & $0.029^{+0.009}_{-0.021}$, $25.9^{+13.7}_{-20.1}$ & $133.7^{+21.9}_{-10.6}$ & $5.39^{+0.44}_{-1.01}$ \\
SIE & $277.8^{+1.8}_{-1.6}/1.05^{+0.01}_{-0.02}$  & $0.08^{+0.02}_{-0.06}$, $24.2^{+22.0}_{-15.7}$ & / & $135.2^{+22.2}_{-11.9}$ & $5.35^{+0.39}_{-0.91}$ \\
SIE$+\gamma$ & $272.3^{+4.2}_{-2.6}/1.00^{+0.03}_{-0.02}$ &  ($0.25^{+0.03}_{-0.07}$, $-44.8^{+6.5}_{-5.5}$) & $0.122^{+0.009}_{-0.046}$, $40.4^{+8.6}_{-6.3}$ & $121.7^{+25.5}_{-7.3}$ & $5.56^{+0.30}_{-1.27}$ \\
\hline \multicolumn{6}{c}{SDSS J0832+0404 ($z_s=1.115$, $z_l=0.659$)} \vspace*{1.5mm} \\ \hline \vspace*{-2.0mm} \\ 
SIS$+\gamma$ & $332.6^{+2.4}_{-0.1}/1.08^{+0.01}_{-0.00}$ & / & $0.015^{+0.000}_{-0.013}$, $32.8^{+6.0}_{-12.0}$ & $233.0^{+3.6}_{-1.3}$ & $3.94^{+0.10}_{-0.02}$ \\
SIE & $333.9^{+1.2}_{-0.3}/1.08^{+0.01}_{-0.00}$  & $0.04^{+0.00}_{-0.04}$, $32.8^{+6.9}_{-5.0}$ & / & $236.4^{+1.2}_{-1.6}$ & $3.95^{+0.09}_{-0.04}$ \\
SIE$+\gamma$ & $337.2^{+1.7}_{-2.7}/1.10^{+0.02}_{-0.01}$ & ($0.21^{+0.02}_{-0.02}$, $9.0^{+5.4}_{-8.4}$) & $0.068^{+0.011}_{-0.010}$, $95.1^{+7.9}_{-9.2}$ & $244.2^{+1.1}_{-0.7}$ & $3.96^{+0.09}_{-0.13}$ \\
\hline \multicolumn{6}{c}{SDSS~J0904+1512 ($z_s=1.826$, $z_l\sim0.2-0.5$, $z_l=0.3$ used for modelling)} \vspace*{1.5mm} \\ \hline \vspace*{-2.0mm} \\ 
SIS$+\gamma$ & $161.3^{+0.5}_{-0.4}/0.57^{+0.0}_{-0.0}$ & / & $0.082^{+0.004}_{-0.006}$, $-85.6^{+2.2}_{-1.4}$ & $8.2^{+0.1}_{-0.1}$ & $7.90^{+0.46}_{-0.39}$ \\
\hline \multicolumn{6}{c}{SDSS~J0946+1835 ($z_s=4.799$, $z_l=0.388$)} \vspace*{1.5mm} \\ \hline \vspace*{-2.0mm} \\ 
SIS$+\gamma$ & $237.2^{+2.8}_{-1.6}/1.31^{+0.03}_{-0.02}$   & / & $0.152^{+0.011}_{-0.018}$, $-46.9^{+1.4}_{-1.0}$ & $121.6^{+2.7}_{-1.8}$ & $3.20^{+0.02}_{-0.06}$ \\
SIE & $249.4^{+0.8}_{-0.04}/1.45^{+0.01}_{-0.01}$  & $0.38^{+0.02}_{-0.03}$, $-46.7^{+1.3}_{-0.05}$ & / & $143.2^{+0.2}_{-0.3}$ & $3.13^{+0.01}_{-0.04}$ \\
SIE$+\gamma$ & $237.4^{+3.5}_{-1.3}/1.31^{+0.04}_{-0.01}$   & ($0.09^{+0.01}_{-0.01}$, $-11.9^{+3.3}_{-4.0}$) & $0.145^{+0.010}_{-0.023}$, $-51.9^{+1.0}_{-0.06}$ & $122.3^{+3.4}_{-1.5}$ & $3.18^{+0.02}_{-0.06}$ \\
\hline \multicolumn{6}{c}{SDSS J1001+5027 ($z_s=1.839$, $z_l=0.415$)} \vspace*{1.5mm} \\ \hline \vspace*{-2.0mm} \\ 
G1only; Power law$+\gamma$; $\gamma'$=$1.72^{+0.02}_{-0.05}$ & $/1.44^{+0.08}_{-0.01}$   & ($0.10^{+0.08}_{-0.00}$, $8.7^{+4.5}_{-5.4}$) & $0.050^{-0.008}_{-0.042}$, $18.2^{+15.8}_{-16.2}$ & $(119.3^{+2.1}_{-2.9})$ & $6.37^{+1.20}_{-0.04}$ \\
G1only; SIE$+\gamma+\kappa$; $\kappa=0.18^{+0.04}_{-0.03}$ & $225.1^{+4.9}_{-2.6}/1.00^{+0.03}_{-0.04}$   & ($0.10^{+0.07}_{-0.07}$, $8.7^{+3.9}_{-5.1}$) & $0.150^{+0.023}_{-0.034}$, $24.6^{+1.7}_{-2.6}$ & $(119.3^{+2.1}_{-2.2})$ & $4.69^{+0.38}_{-0.34}$ \\
G1only; SIE$+\gamma$ ($h$ free) $h=0.82^{+0.05}_{-0.02}$ & $247.9^{+4.6}_{-1.5}/1.21^{+0.03}_{-0.02}$   & ($0.10^{+0.08}_{-0.05}$, $8.7^{+6.5}_{-4.0}$) & $0.181^{+0.017}_{-0.040}$, $24.6^{+1.6}_{-2.3}$ & $(119.3^{+0.1}_{-0.1})$ & $3.19^{+0.04}_{-0.13}$ \\
G1/G2=ct. Power law$+\gamma$ (G1) + SIS (G2) & $/0.96^{+0.05}_{-0.15}$   & ($0.10^{+0.04}_{-0.07}$, $8.7^{+6.0}_{-5.2}$) & $0.170^{+0.176}_{-0.070}$, $28.8^{+5.8}_{-1.7}$ & $(119.3^{+3.3}_{-1.9})$ & $3.38^{+0.11}_{-0.80}$ \\
... $\gamma'$=$2.12^{+0.30}_{-0.16}$ & & & & \\
G1/G2=ct. SIE$+\gamma+\kappa$ (G1) + SIS (G2) & $231.8^{+4.0}_{-4.1}/1.05^{+0.05}_{-0.03}$ & ($0.10^{+0.06}_{-0.04}$, $8.7^{+6.9}_{-5.0}$) & $0.127^{+0.020}_{-0.037}$, $29.4^{+3.0}_{-3.0}$ & $(119.3^{+3.0}_{-2.6})$ & $3.56^{+0.32}_{-0.28}$ \\
... $\kappa=-0.06^{+0.04}_{-0.06}$ & & & & \\
G2 free; SIE$+\gamma$ (G1) + SIS (G2) & $230.1^{+3.3}_{-3.3}/1.04^{+0.03}_{-0.03}$ & ($0.10^{+0.08}_{-0.05}$, $8.7^{+5.3}_{-5.6}$) & $0.131^{+0.025}_{-0.037}$, $28.0^{+2.9}_{-5.6}$ & $(119.3^{+2.9}_{-2.9})$ & $3.78^{+0.16}_{-0.13}$ \\
... $\sigma_\mathrm{G2}/\sigma_\mathrm{G1}=0.59^{+0.09}_{-0.05}$  & & & & \\
G1only; Power law$+\gamma$; $\gamma'=1.76^{+0.02}_{-0.08}$ & $/1.38^{+0.05}_{-0.04}$ & / & $0.104^{+0.008}_{-0.033}$, $16.4^{+2.4}_{-3.1}$ & $(119.3^{+0.1}_{-4.3})$ & $5.68^{+1.26}_{-0.00}$ \\
G1only; SIS$+\gamma$ ($h$ free) $h=0.79^{+0.04}_{-0.02}$ & $245.7^{+3.6}_{-2.0}/1.19^{+0.03}_{-0.02}/$   & / & $0.210^{+0.016}_{-0.020}$, $22.6^{+1.0}_{-1.2}$ & $(119.3^{+0.1}_{-0.0})$ & $3.23^{+0.06}_{-0.12}$ \\
G1only; SIS$+\gamma+\kappa$; $\kappa=0.15^{+0.03}_{-0.03}$ & $226.9^{+3.8}_{-3.4}/1.01^{+0.4}_{-0.03}$   & / & $0.179^{+0.016}_{-0.029}$, $22.6^{+0.8}_{-1.0}$ & $(119.3^{+2.9}_{-2.5})$ & $4.44^{+0.32}_{-0.28}$ \\
G1/G2=ct. Power law$+\gamma$ (G1) + SIS (G2) & $/0.91^{+0.07}_{-0.05}$   & / & $0.234^{+0.022}_{-0.085}$, $27.2^{+0.7}_{-1.0}$ & $(119.3^{+2.4}_{-2.8})$ & $3.33^{+0.00}_{-0.64}$ \\
... $\gamma'=2.24^{+0.10}_{-0.09}$ & & & & \\
G1/G2=ct. 2SIS$+\gamma+\kappa$; $\kappa=-0.09^{+0.04}_{-0.04}$ & $233.8^{+4.0}_{-3.2}/1.07^{+0.04}_{-0.03}$ & / & $0.155^{+0.019}_{-0.023}$, $26.1^{+0.9}_{-1.1}$ & $(119.3^{+2.8}_{-2.3})$ & $3.35^{+0.19}_{-0.24}$ \\
G2 free; 2SIS$+\gamma$;  $\sigma_\mathrm{G2}/\sigma_\mathrm{G1}=0.54^{+0.08}_{-0.06}$ & $231.2^{+3.4}_{-2.5}/1.05^{+0.03}_{-0.03}$ & / & $0.163^{+0.020}_{-0.027}$, $24.6^{+1.0}_{-1.1}$ & $(119.3^{+2.7}_{-2.7})$ & $3.69^{+0.12}_{-0.10}$ \\
G1/G2=ct. 2SIS$+\gamma$ ($h$ free) $h=0.62^{+0.03}_{-0.03}$ & $/0.98^{+0.04}_{-0.02}$   & / & $0.142^{+0.010}_{-0.029}$, $26.1^{+1.1}_{-1.1}$ & $(119.3^{+0.0}_{-0.1})$ & $3.99^{+0.12}_{-0.17}$ \\
\hline \multicolumn{6}{c}{SDSS J1055+4628 ($z_s=1.249$, $z_l=0.388$)} \vspace*{1.5mm} \\ \hline \vspace*{-2.0mm} \\ 
SIS$+\gamma$ & $180.3^{+0.9}_{-0.9}/0.58^{+0.00}_{-0.01}$ & / & $0.020^{+0.005}_{-0.003}$, $-35.4^{+16.1}_{-9.8}$ & $22.2^{+0.3}_{-0.3}$ & $4.16^{+0.10}_{-0.10}$ \\
SIE & $180.1^{+0.5}_{-0.5}/0.58^{+0.00}_{-0.01}$  & $0.06^{+0.01}_{-0.01}$, $-35.0^{+13.8}_{-12.5}$ & / & $22.1^{+0.2}_{-0.1}$ & $4.15^{+0.08}_{-0.12}$ \\
SIE$+\gamma$ & $173.8^{+2.3}_{-1.1}/0.54^{+0.01}_{-0.01}$ & ($0.35^{+0.03}_{-0.05}$, $-1.7^{+2.4}_{-3.5}$) & $0.144^{+0.018}_{-0.035}$, $-87.5^{+3.0}_{-3.6}$ & $18.9^{+0.8}_{-0.4}$ & $4.24^{+0.12}_{-0.11}$ \\
\hline \multicolumn{6}{c}{SDSS J1131+1915 ($z_s=2.915$, $z_l\sim0.3$)} \vspace*{1.5mm} \\ \hline \vspace*{-2.0mm} \\ 
SIS$+\gamma$ & $178.1^{+0.6}_{-0.6}/0.74^{+0.01}_{-0.00}$ & / & $0.066^{+0.005}_{-0.007}$, $80.8^{+1.5}_{-1.5}$ & $15.5^{+0.2}_{-0.2}$ & $7.00^{+0.37}_{-0.36}$ \\
SIE & $176.0^{+0.4}_{-0.4}/0.73^{+0.00}_{-0.01}$  & $0.18^{+0.02}_{-0.02}$, $81.0^{+1.4}_{-1.7}$ & / & $14.7^{+0.1}_{-0.3}$ & $6.90^{+0.34}_{-0.31}$ \\
SIE$+\gamma$ & $170.5^{+3.0}_{-0.3}/0.68^{+0.03}_{-0.00}$ & ($0.47^{+0.03}_{-0.10}$, $87.5^{+4.2}_{-5.9}$) & $0.157^{+0.016}_{-0.067}$, $0.8^{+8.8}_{-8.4}$ & $12.5^{+1.2}_{-0.1}$ & $6.86^{+0.40}_{-0.38}$ \\
\hline \multicolumn{6}{c}{SDSS J1206+4332 ($z_s=1.789$, $z_l=0.748$)} \vspace*{1.5mm} \\ \hline \vspace*{-2.0mm} \\ 
G1only; SIS$+\gamma$ ($h$ free) $h=0.83$ & $341.5/1.55$   & / & $0.065$, $-57.2$ & $(111.3)$ & $9.14$ \\
G1only; Power law$+\gamma$; $\gamma'$=1.80 & $/1.57$ & / & $0.051$, $-55.3$ & $(111.3)$ & $14.10$ \\
G1only; SIS$+\gamma+\kappa$; $\kappa=0.19$ & $307.4$   & / & $0.053$, $-57.2$ & $(111.3)$ & $7.93$ \\
G1/G2=ct. 2SIS$+\gamma+\kappa$; $\kappa=0.11$ & $/1.22$   & / & $0.109$, $-89.4$ & $(111.3)$ & $10.17$ \\
G1/G2=ct. 2SIS$+\gamma$ ($h$ free) $h=0.76$ & $/1.38$   & / & $0.123$, $-89.4$ & $(111.3)$ & $8.79$ \\
G1/G2=ct. Power law$+\gamma$ (G1) + SIS (G2) & $/1.37$ & / & $0.116$, $-90.8$ & $(111.3)$ & $10.00$ \\
... $\gamma'$=1.88 & & & & & \\
\hline \multicolumn{6}{c}{SDSS J1216+3529 ($z_s=2.012$, $z_l\sim0.55$)} \vspace*{1.5mm} \\ \hline \vspace*{-2.0mm} \\ 
SIS$+\gamma$ & $200.6^{+2.8}_{-1.3}/0.71^{+0.02}_{-0.01}$ & / & $0.085^{+0.013}_{-0.014}$, $-21.7^{+4.2}_{-6.1}$ & $55.5^{+2.0}_{-0.8}$ & $3.28^{+0.16}_{-0.14}$ \\
SIE & $204.0^{+1.3}_{-0.9}/0.73^{+0.01}_{-0.00}$  & $0.23^{+0.03}_{-0.05}$, $-23.0^{+4.9}_{-5.1}$ & / & $59.3^{+1.6}_{-1.7}$ & $3.23^{+0.18}_{-0.14}$ \\
SIE$+\gamma$ & $221.5^{+0.9}_{-10.7}/0.87^{+0.00}_{-0.05}$ & ($0.67^{+0.03}_{-0.18}$, $-14.6^{+3.5}_{-4.1}$) & $0.256^{+0.012}_{-0.126}$, $78.3^{+6.1}_{-3.4}$ & $73.5^{+1.4}_{-7.1}$ & $3.52^{+0.07}_{-0.36}$ \\
\hline \multicolumn{6}{c}{SDSS J1254+2235 ($z_s=3.626$, $z_l\sim0.3$)} \vspace*{1.5mm} \\ \hline \vspace*{-2.0mm} \\ 
SIS$+\gamma$ & $182.2^{+0.5}_{-0.4}/0.80^{+0.00}_{-0.01}$ & / & $0.021^{+0.004}_{-0.005}$, $-56.0^{+2.2}_{-3.2}$ & $9.2^{+0.4}_{-0.0}$ & $11.54^{+0.43}_{-0.51}$ \\
SIE & $181.4^{+0.3}_{-0.2}/0.79^{+0.00}_{-0.00}$  & $0.06^{+0.01}_{-0.1}$, $-56.0^{+2.0}_{-2.9}$ & / & $9.2^{+0.3}_{-0.2}$ & $11.52^{+0.48}_{-0.43}$ \\
\hline \multicolumn{6}{c}{SDSS J1313+5151 ($z_s=1.875$, $z_l=0.194$)} \vspace*{1.5mm} \\ \hline \vspace*{-2.0mm} \\ 
SIS$+\gamma$ & $155.8^{+2.8}_{-0.0}/0.59^{+0.02}_{-0.00}$ & / & $0.033^{-0.008}_{-0.029}$, $-76.1^{+6.1}_{-9.1}$ & $12.4^{+0.5}_{-0.0}$ & $2.91^{+0.15}_{+0.03}$ \\
SIE & $157.2^{+1.3}_{+0.2}/0.60^{+0.01}_{-0.00}$  & $0.10^{-0.03}_{-0.09}$, $-76.1^{+8.1}_{-3.9}$ & / & $12.8^{+0.2}_{-0.1}$ & $2.91^{+0.15}_{+0.2}$ \\
SIE$+\gamma$ & $161.7^{+1.1}_{-1.3}/0.64^{+0.01}_{-0.01}$ & ($0.35^{+0.01}_{-0.01}$, $-68.7^{+0.9}_{-1.0}$) & $0.102^{+0.018}_{-0.016}$, $23.4^{+1.5}_{-2.1}$ & $14.1^{+0.2}_{-0.2}$ & $2.94^{+0.06}_{-0.09}$ \\
& & ... ($1.06, 7.00$) & & & \\
\hline \multicolumn{6}{c}{SDSS J1322+1052 ($z_s=1.716$, $z_l\sim0.55$)} \vspace*{1.5mm} \\ \hline \vspace*{-2.0mm} \\ 
SIS$+\gamma$ & $252.8^{+0.9}_{-1.0}/1.06^{+0.00}_{-0.01}$ & / & $0.057^{+0.007}_{-0.008}$, $-35.7^{+0.7}_{-0.8}$ & $85.1^{+0.7}_{-0.8}$ & $6.22^{+0.23}_{-0.34}$ \\
SIE & $249.6^{+0.5}_{-0.7}/1.03^{+0.00}_{-0.00}$  & $0.16^{+0.02}_{-0.02}$, $-35.7^{+0.9}_{-0.7}$ & / & $80.5^{+0.4}_{-0.3}$ & $6.20^{+0.19}_{-0.33}$ \\
SIE$+\gamma$ & $251.8^{+0.02}_{-0.02}/1.05^{+0.01}_{-0.01}$ & ($0.07^{+0.03}_{-0.02}$, $-52.3^{+11.1}_{-4.8}$) & $0.040^{+0.010}_{-0.012}$, $-25.3^{+5.8}_{-7.0}$ & $83.6^{+0.9}_{-1.3}$ & $6.20^{+0.17}_{-0.35}$ \\
\hline \multicolumn{6}{c}{SDSS J1330+1810 ($z_s=1.393$, $z_l=0.373$)} \vspace*{1.5mm} \\ \hline \vspace*{-2.0mm} \\ 
(no flux) Power law $ + \gamma$  & $/0.96^{+0.00}_{-0.00}$ & $0.64^{+0.02}_{-0.02}$, $21.8^{+0.1}_{-0.0}$ & $0.072^{+0.001}_{-0.001}$, $-111.2^{+1.2}_{-1.5}$ & $\Delta t_{AB}=0.4^{+0.0}_{-0.0}$ & $9.28^{+0.21}_{-0.33}$, (B/A=0.71) \\
... $\gamma'=2.49^{+0.00}_{-0.00}$ &  & & & $\Delta t_{AC}=-11.4^{+0.2}_{-0.6}$ & (C/A=0.55) \\
& & & & $\Delta t_{AD}=17.6^{+0.6}_{-0.1}$ & (D/A=0.20) \\
(flux) Power law $ + \gamma$ & $/0.95^{+0.00}_{-0.00}$  & $0.56^{+0.01}_{-0.01}, 21.7^{+0.0}_{-0.1}$ & $0.068^{+0.002}_{-0.001}$, $70.2^{+2.2}_{-0.08}$ & $\Delta t_{AB}=0.3^{+0.0}_{-0.0}$ & $11.78^{+0.24}_{-0.59}$ (B/A=0.80) \\
... $(\chi^2/\mathrm{d.o.f.}=11.2/3)  $ &   & & & $\Delta t_{AC}=-9.4^{+0.2}_{-0.5}$ & (C/A=0.51) \\
... $\gamma'=2.42^{+0.00}_{-0.00}$ & &  & & $\Delta t_{AD}=16.0^{+0.7}_{-0.0}$ & (D/A=0.20) \\
(flux) Power law $ + \gamma + \delta$  & $/0.95^{+0.01}_{-0.00}$  & $0.56^{+0.04}_{-0.02}, 23.3^{+1.1}_{-0.7}$ & $0.054^{+0.010}_{-0.005}$, $74.6^{+2.3}_{-21.5}$ & $\Delta t_{AB}=0.3^{+0.1}_{-0.0}$ & $12.34^{+1.0}_{-3.0}$ (B/A=0.79) \\
... $(\chi^2/\mathrm{d.o.f.}=0.8/1)$, $\gamma'$=$2.36^{+0.13}_{-0.02}$ &   & & & $\Delta t_{AC}=-9.9^{+1.0}_{-1.5}$ & (C/A=0.46) \\
... $(\delta, \theta_\delta)$=$(0.01^{+0.01}_{-0.00}, 7.1^{+28.7}_{+1.4})$ & &  & & $\Delta t_{AD}=16.2^{+1.9}_{-0.2}$ & (D/A=0.18) \\
(flux) SIE$+\gamma$  & $223.3^{+0.3}_{-0.1}/0.94^{+0.00}_{-0.00}$  & $0.35^{+0.02}_{-0.02}$, $19.6^{+1.3}_{-1.0}$ & $0.067^{+0.003}_{-0.003}$, $85.4^{+2.9}_{-2.5}$ & $\Delta t_{AB}=0.2^{+0.0}_{-0.0}$ & $32.0^{+1.7}_{-1.6}$ (B/A=0.97) \\
...$(\chi^2/\mathrm{d.o.f.}=46/4)$ & & &  & $\Delta t_{AC}=-4.9^{+0.3}_{-0.4}$ & (C/A=0.41) \\
& & &  & $\Delta t_{AD}=10.1^{+0.3}_{-0.3}$ & (D/A=0.21) \\
\hline \multicolumn{6}{c}{SDSS~J1334+3315 ($z_s=2.426$, $z_l=0.557$)} \vspace*{1.5mm} \\ \hline \vspace*{-2.0mm} \\ 
SIE & $146.8^{+0.8}_{-0.7}$  &   $0.29^{+0.05}_{-0.06}$, $-49.4^{+4.3}_{-3.3}$ & / & $5.6^{+0.7}_{-0.7}$ & $12.3^{+1.3}_{-1.7}$ \\
SIS+$\gamma$& $141.9^{+1.8}_{-2.5}$   & / & $0.11^{+0.02}_{-0.03}$, $-49.4^{+4.1}_{-4.0}$ & $5.6^{+0.8}_{-0.5}$ & $10.9^{+1.0}_{-1.2}$ \\
\hline \multicolumn{6}{c}{SDSS J1353+1138 ($z_s=1.629$, $z_l\sim0.25$)} \vspace*{1.5mm} \\ \hline \vspace*{-2.0mm} \\ 
SIS$+\gamma$ & $171.8^{+1.4}_{-1.2}/0.67^{+0.01}_{-0.01}$ & / & $0.080^{+0.005}_{-0.010}$, $-54.8^{+3.6}_{-4.2}$ & $17.7^{+0.3}_{-0.2}$ & $3.47^{+0.05}_{-0.04}$ \\
SIE & $174.4^{+0.7}_{-0.7}/0.69^{+0.00}_{-0.01}$  & $0.21^{+0.02}_{-0.02}$, $-53.6^{+4.6}_{-3.6}$ & / & $18.8^{+0.1}_{-0.0}$ & $3.42^{+0.04}_{-0.04}$ \\
SIE$+\gamma$ & $182.4^{+1.4}_{-3.6}/0.75^{+0.01}_{-0.03}$ & ($0.50^{+0.04}_{-0.12}$, $-64.8^{+0.9}_{-0.9}$) & $0.150^{+0.022}_{-0.064}$, $19.9^{+2.2}_{-2.2}$ & $21.6^{+0.5}_{-1.1}$ & $3.53^{+0.05}_{-1.0}$ \\
\hline \multicolumn{6}{c}{SDSS J1400+3134 ($z_s=3.317$, $z_l\sim0.8$)} \vspace*{1.5mm} \\ \hline \vspace*{-2.0mm} \\ 
SIS$+\gamma$ & $236.0^{+0.7}_{-0.6}/0.94^{+0.01}_{-0.00}$ & / & $0.059^{+0.002}_{-0.002}$, $-81.0^{+2.3}_{-2.5}$ & $29.9^{+0.9}_{-1.1}$ & $10.94^{+0.52}_{-0.26}$ \\
SIE & $236.5^{+0.5}_{-0.5}/0.95^{+0.00}_{-0.01}$  & $0.19^{+0.0}_{-0.01}$, $-80.7^{+2.5}_{-2.5}$ & / & $34.7^{+0.8}_{-1.1}$ & $9.30^{+0.30}_{-0.25}$ \\
SIE$+\gamma$ & $234.5^{+1.2}_{-0.4}/0.93^{+0.00}_{-0.02}$ & ($0.32^{+0.03}_{-0.06}$, $-66.6^{+4.2}_{-10.2}$) & $0.070^{+0.015}_{-0.034}$, $39.7^{+11.3}_{-13.9}$ & $35.1^{+2.9}_{-1.7}$ & $8.68^{+0.04}_{-0.92}$ \\
\hline \multicolumn{6}{c}{SDSS J1406+6126 ($z_s=2.134$, $z_l=0.271$)} \vspace*{1.5mm} \\ \hline \vspace*{-2.0mm} \\ 
SIS$+\gamma$ & $207.1^{+1.5}_{-0.7}/0.99^{+0.01}_{-0.00}$ & / & $0.024^{+0.004}_{-0.012}$, $-23.1^{+6.5}_{-4.8}$ & $27.7^{+0.0}_{-0.4}$ & $5.37^{+0.02}_{-0.07}$ \\
SIE & $208.3^{+0.8}_{-0.2}/1.00^{+0.01}_{-0.00}$  & $0.07^{+0.00}_{-0.04}$, $-23.1^{+7.4}_{-4.1}$ & / & $28.4^{+0.2}_{-0.1}$ & $5.36^{+0.02}_{-0.07}$ \\
\hline \multicolumn{6}{c}{SDSS J1455+1447 ($z_s=1.424$, $z_l\sim0.42$)} \vspace*{1.5mm} \\ \hline \vspace*{-2.0mm} \\ 
SIS$+\gamma$ & $225.4^{+0.7}_{-0.7}/0.91^{+0.01}_{-0.01}$ & / & $0.047^{+0.004}_{-0.003}$, $88.5^{+3.4}_{-2.5}$ & $33.8^{+0.7}_{-0.6}$ & $7.52^{+0.4}_{-0.2}$ \\
SIE & $224.0^{+0.4}_{-0.5}/0.90^{+0.0}_{-0.1}$  & $0.13^{+0.01}_{-0.01}$, $89.0^{+3.4}_{-3.1}$ & / & $33.3^{+0.7}_{-0.7}$ & $7.36^{+0.6}_{-0.6}$ \\
SIE$+\gamma$ & $224.1^{+1.3}_{-0.6}/0.90^{+0.1}_{-0.1}$ & ($0.23^{+0.01}_{-0.02}, 82.1^{+4.4}_{-4.8}$) & $0.043^{+0.009}_{-0.012}$, $-17.8^{+7.8}_{-23.0}$ & $33.6^{+1.4}_{-0.4}$ & $7.22^{+0.2}_{-0.6}$ \\
\hline \multicolumn{6}{c}{SDSS J1515+1511 ($z_s=2.054$, $z_l=0.742$)} \vspace*{1.5mm} \\ \hline \vspace*{-2.0mm} \\ 
SIS$+\gamma$ & $239.3^{+3.2}_{-1.6}/0.84^{+0.02}_{-0.02}$ & / & $0.175^{+0.009}_{-0.022}$, $-27.8^{+0.8}_{-0.7}$ & $145.2^{+3.4}_{-2.2}$ & $3.22^{+0.04}_{-0.09}$ \\
SIE & $254.5^{+0.5}_{-0.5}/0.95^{+0.00}_{-0.01}$  & $0.43^{+0.03}_{-0.04}$, $-27.6^{+0.9}_{-0.7}$ & / & $175.8^{+0.8}_{-0.08}$ & $3.13^{+0.03}_{-0.07}$ \\
SIE$+\gamma$ & $287.2^{+2.8}_{-2.1}/1.21^{+0.02}_{-0.02}$ & ($0.81^{+0.01}_{-0.01}$, $-17.1^{+0.3}_{-0.3}$) & $0.283^{+0.019}_{-0.011}$, $76^{+0.8}_{-0.9}$ & $216.4^{+4.1}_{-2.3}$ & $3.44^{+0.03}_{-0.05}$ \\
\hline \multicolumn{6}{c}{SDSS J1620+1203 ($z_s=1.158$, $z_l=0.398$)} \vspace*{1.5mm} \\ \hline \vspace*{-2.0mm} \\ 
SIS$+\gamma$ & $287.5^{+4.0}_{-0.8}/1.39^{+0.04}_{-0.01}$ & / & $0.054^{+0.006}_{-0.004}$, $2.9^{+4.2}_{-13.0}$ & $182.1^{+5.0}_{-1.0}$ & $3.11^{+0.16}_{-0.06}$ \\
SIE & $288.8^{+1.2}_{-1.1}/1.41^{+0.01}_{-0.02}$  & $0.15^{+0.01}_{-0.01}$, $1.6^{+6.3}_{-8.9}$ & / & $185.3^{+0.7}_{-0.3}$ & $3.09^{+0.08}_{-0.09}$ \\
SIE$+\gamma$ & $283.2^{+3.4}_{-1.5}/1.35^{+0.03}_{-0.01}$ & ($0.24^{+0.02}_{-0.02}$, $-27.8^{+1.3}_{-1.3}$) & $0.080^{+0.006}_{-0.025}$, $43.1^{+2.6}_{-3.0}$ & $170.6^{+0.8}_{-0.8}$ & $3.09^{+0.18}_{-0.09}$ \\
\hline 
\label{tab:massmodel}
\end{longtable}
\vspace*{-7.0mm}
{\scriptsize The position angle is positive from North towards East. Values inside parentheses are free to vary, but with a gaussian prior given by the observed uncertainties. For the lenses where a spectroscopic lens redshift is not available (SDSS~J0904+1512, SDSS~J1131+1915, SDSS~J1216+3529, SDSS~J1254+2235, SDSS~J1322+1052, SDSS~J1353+1138 and SDSS~J1400+3134), the effect of the redshift uncertainty on the estimated velocity dispersion and time delays is not included; this can be calculated by assuming typical redshift uncertainties of $\sim0.1$ in the discovery papers. Unless otherwise specified, all models have $(\chi^2/\mathrm{d.o.f.}=0/0)$. The nearby galaxy companions are ignored in these models. G1/G2 refers to the velocity dispersion ratio. Notation $\mu$ represents the total magnification factor; $\delta$ and $\gamma'$ represent the third order perturbation and the density slope. The results for SDSS~J1405+0959, presented in a separate publication, are not included (see Appendix \ref{section:1405}). For SDSS~J1330+1810, as well as all other systems with nearby galaxies, these models do not consider explicitly the nearby galaxies. For SDSS~J0904+1512, SDSS~J1206+4332, SDSS~J1254+2235, SDSS~J1334+3315 and SDSS~J1406+6126, where the ellipticity of the lensing galaxy is either negligible or unreliable, the SIE$+\gamma$ model was omitted. The systems where a close companion is identified can be inferred from Table \ref{tab:lensastrometry}, and those with a noticeable field overdensity in SDSS from Figure \ref{fig:environ}.}
\end{landscape}
\endgroup

\twocolumn


\section{Discussion of the technique}\label{sect:discutionoverall}

Here we take a look back on the morphological modelling technique we employed, by summarising several points:

\begin{itemize}
    \renewcommand{\labelitemi}{$\bullet$}
  \item For the majority of the systems, most of the error bars on the fitted parameters come from the simulations with either the PSF built on the bright quasar image or a separately observed PSF star. This is expected since these simulations are designed to test for systematic errors. These PSFs include components that cannot be modelled analytically, the noise frames used in the simulations may include faint background sources, and in the case that the PSF is built on the bright image, more noise is introduced, boosting the error estimate. In the cases where the systems are bright, it can be argued that the use of noise frame is not realistic because they do not contain Poisson noise, which is expected to be dominant. In regards to this, the use of noise maps produced by Hostlens/Galfit includes an estimate of Poisson noise, which was therefore taken into account. 
  \item While the separately observed PSF star that is most similar to the PSF of each system in terms of analytic parameters was used for simulations, we tested for SDSS~J1620+1203 that the use of a second PSF star (from SDSS~J0832+0404) would produce similar error bars, in case we model this system with an analytical PSF. Therefore the fact that we simulate our systems with different but similar AO PSFs does not appear to have a large effect.  
   \item The analytical PSF method can lead to spurious detections and mistaken interpretations, if it is not cross-checked. This has been seen in the case of SDSS~J1002+4449 and SDSS~J1353+1138. For SDSS~J1002+4449, a comparison of the $\chi^2$ values indicates that a lensing galaxy is present, whereas simulations indicate that noise is most likely mistaken for a galaxy. In the case of SDSS~J1353+1138, modelling with an analytical PSF suggests the detection of an additional stellar component, whereas modelling with a hybrid PSF shows that, once again, noise was mistaken for another component.
   \item Inconsistencies in astrometry are found in the multiple observations of extended objects in SDSS~J1405+0959 and SDSS~J1330+1810. While the scatter between observations has been taken into account in order to infer realistic error bars for these particular systems, the inconsistencies do not cast doubt on the astrometric precision estimated for the other objects in the current sample, because there are known sources of systematics for the two objects. For SDSS~J1405+0959, the low S/N and the uncertain morphology of GX, located unusually close to G2, is a source of systematics. For SDSS~J1330+1810, the residuals show that there are systematics in the modelling of the PSF (20 mas mode). However the astrometry of point sources in these two systems, as well as for SDSS~J1334+3315, with typical error bars of $\sim1$ mas, is consistent between observations, supporting the astrometric precision estimation. 
   \item As is expected because the final science frame has larger S/N and the residual astrometric distortion estimate was likely dominated by measurement errors, the systematic distortion estimate typically dominates the astrometry errors. These errors would be further reduced by using a more precise distortion map and applying corrections, not just error estimates, for differential refraction.
    \item Most of the galaxies are not consistent with a Sersic index $n=4$ (de Vaucouleurs profile). They show more pronounced residuals when the de Vaucouleurs profile is used. Instead, most galaxies are fitted with a higher value of $n=5-6$. This is in agreement with the mean index value for the lens galaxies found in \citet{bolton12}. 
        \item For SDSS~J1515+1511, the only object in the current sample for which a suitable PSF star exists in the FOV, masking the central $2\times2$ pixels (in order to reject pixels in both the PSF and the target that may be dominated by Poisson noise) has introduced significant changes in the parameters of the host in the sense of increasing the effective radius and Sersic index, and reducing its flux; on the other hand, for SDSS~J0819+5356 (modelled with an analytical PSF) the changes are insignificant. The effect could not be checked for SDSS~J0904+1512, which was modelled iteratively with a hybrid PSF built on the target. 
       \item A matter of concern is inferring the lens galaxy / host galaxy parameters in shallow images where the outer regions of the galaxies may be poorly detected. However, the simulations performed should in principle account for this effect. A more robust method is to compare deep and shallow exposures of the same object, which was done for the lens galaxies in SDSS~J0946+1835 and SDSS~J1330+1810, revealing that the morphological parameters match within $\sim1-2\sigma$. This however was not done for the host galaxies: for SDSS~J1330+1810, where both deep and shallow observations are available, the host is not fit satisfactorily in either observation. \citet{peng06-2} undertook this study and found differences in the estimated host magnitude and effective radius of $\sim$ 0.15 mag and 10\% - 20\%, respectively, comparable to the effects they found from using different PSFs.
  \end{itemize}

  \section{Quasar host galaxies in the AO sample}\label{section:hostfacts}

Here we discuss the properties of the detected quasar host galaxies, and carry out a study on whether their luminosities follow the relations found in the literature for quasars at similar redshifts. As shown in Table \ref{tab:host}, we measured the quasar host galaxy luminosity in four systems (SDSS~J0819+5356, SDSS~J0832+0404, SDSS~J0904+1512 and SDSS~J1515+1511), and found upper limits found for an additional system (SDSS~J1322+1052). 

When modelling the host galaxies, we considered a singular isothermal profile for the lensing galaxy. Since the real lenses may not be isothermal, this affects the inferred size and luminosity of the hosts. \citet{marshall07} translate the uncertainty in the mass profile slope $\sigma_{\gamma'}$ into additional uncertainty on the size and luminosity as $\sigma_r/r\propto\sigma_{\gamma'}$, where $r$ stands for the galaxy radius, and $\sigma_{m_{\mathrm{AB}}}\propto2.2\sigma_{\gamma'}$, respectively. Here $\sigma_{\gamma'}$ can be taken to be the intrinsic spread of power-law induced measured by \citet{koopmans09}, $\sigma_{\gamma'}\sim0.20$. This dominates the error budget for the majority of the systems.
    
    All four hosts with available measurements have small effective radii of 2.2, 3.2, 0.4 and 2.1 kpc for SDSS~J0819+5356, SDSS~J0832+0404, SDSS~J0904+1512 and SDSS~J1515+1511, respectively. It is a known fact that quiescent galaxies at high redshift $z\sim2$ are more compact than galaxies of comparable mass at $z\sim0$ \citep[e.g.,][]{daddi05,trujillo07}. Also, \citet{peng06-2} measured for their sample of quasar hosts effective radii $\lesssim3-5$ kpc, which are not much larger than obtained here (with the exception of SDSS~J0904+1512).
    
    All four hosts have small Sersic indices indicative of disk galaxies. At low redshifts, bulge-dominated hosts are more numerous, especially for the more luminous quasars \citep[e.g.,][]{dunlop03,guyon06}. However, at $z=1.5-2.5$, \citet{kocevski12} find a high fraction of disk hosts for moderately luminous AGN, \citet{schawinsky12} find heavily obscured quasar hosts to be mostly disks, and \citet{peng06-2} find signs that the hosts are of less concentrated morphology. On the other hand, it is plausible that, given the overall small effective radii, the even more compact and concentrated bulge component cannot be reliably distinguished and is integrated into the PSF by our modelling technique. As a result, we also performed fitting with de Vaucouleurs profile for the host galaxies. In particular, this needs to be used when checking if the same correlation between the luminosity of the host bulge and $M_\mathrm{BH}$ found in the literature holds in our AO sample. We also note that the S/N of all hosts is low, and visually the residuals after modelling with Sersic or de Vaucouleurs profiles look very similar. 

As the relations between the quasars and their host galaxies need to be determined in the quasar rest frame, we converted the observed magnitudes of the de-magnified hosts and quasars, recreated in the source plane, to absolute magnitudes in the rest-frame $R$-band. We used the equation 

\begin{equation}
M_R=m_K-K(z)-5\log d_L -25 \ \ ,
\label{eq:kcorrect}
\end{equation} 

\noindent where $K(z)$ is the $K$-correction between the $K_s$ and $R$ bands found by \citet{falomo08}, and $d_L$ is the luminosity distance (in Mpc). The choice of spectral template impacts the conversion at most by $\sim0.3$ mag for SDSS~J0832+0404. We did not apply an evolution correction. In addition, some extinction may be caused by light being deflected close to the lensing galaxy. However, no differential extinction is found for any of the five systems in the follow-up spectroscopy presented in the discovery papers, and \citet{peng06-2} conclude that such extinction is likely negligible. We present the results in Table \ref{tab:BH}.

We assumed that all five systems are radio quiet, as they are not found in any available catalogue of radio sources. Comparing the nuclear luminosities to the sample of high- and low-luminosity quasars from \citet{kotilainen07}, all systems contain low-luminosity quasars, except for SDSS~J1515+1511, which contains a high-luminosity quasar.

In order to study the relation between the host galaxies and their central black holes, $M_\mathrm{BH}$ needs to be estimated. An estimate exists in \citet{shen11}, where $M_\mathrm{BH}$ was computed for all SDSS quasars in Data Release 7. This estimate however cannot be used, as it does not account for the lensing nature of the systems, the multiple images and the spectral contamination from the lensing galaxies. 

In order to estimate $M_\mathrm{BH}$, as well as to facilitate comparison, we employed the same virial relation used by \citet{peng06-2}: 

\begin{equation}
M_\mathrm{BH}=A_\mathrm{line}\left[\frac{\mathrm{FWHM(line)}}{1000\ \mathrm{km\ s}^{-1}}\right]^2    \left[\frac{\lambda L_\lambda(\lambda_\mathrm{line})}{10^{44}\ \mathrm{erg\ s}^{-1}}\right]^{\gamma_\mathrm{line}} M_\odot        \ \ ,
\label{eq:einstringsourceplane}
\end{equation} 

\noindent where $L=\lambda L_\lambda$ is the continuum luminosity at $\lambda_\mathrm{line}\{\mathrm{CIV, MgII}\}=\{1350,3000\}$ \AA, $A_\mathrm{line}\{\mathrm{CIV,MgII}\}=\{4.5, 6.1\}\times{10^6}$, and $\gamma_\mathrm{line}\{\mathrm{CIV,MgII}\}=\{0.53, 0.47\}$ (\citet{vestergaard06} and \citet{mclure02}, respectively). At least one of the two emission lines, CIV and MgII, is found in the original SDSS spectrum of each system. 

The measured line FWHMs in \citet{shen11} are not modified by lensing, and can therefore be employed. We determined the continuum luminosities using the method of \citet{peng06-2}. First, by combining the discovery papers with the new AO data, all systems have individual resolved magnitudes for the quasar images and lensing galaxies in at least three bands. We converted the measured magnitudes of the bright quasar image in each system into flux densities\footnote{The conversions were performed using the tool at \url{http://www.gemini.edu/?q=node/11119}}, which we then fitted with a power law $F_\lambda\propto\lambda^{\alpha-2}$ and extrapolated to the observed wavelength $\lambda_\mathrm{line}(1+z)$. 
We then divided the extrapolated flux densities by the magnification ratio of the bright quasar image, resulting in the de-lensed $\mathrm{F}_\lambda$, which we converted to the desired luminosities $L=4\pi d_L^2\lambda_\mathrm{line}(1+z)\mathrm{F}_\lambda$. Although this approach is an approximation, and does not account, for example, for unreliable magnitudes in the low-resolution data or the effect of strong emission lines on the magnitudes, $M_\mathrm{BH}$ is only weakly dependent on the estimated continuum luminosity, roughly through its square root. We used the fitted power law to estimate the quasar nuclei $M_R$ in Table \ref{tab:BH}, with $K(z)=2.5\log(\alpha-1)$, and the final values in the Vega system.

We report the resulting $M_\mathrm{BH}$, together with the Eddington ratios, in Table \ref{tab:BH}. SDSS~J0819+535 and  SDSS~J1322+1052, where the CIV line is affected by partial absorption, show differences in $M_\mathrm{BH}$ estimated from the MgII and CIV lines, and therefore we discarded the result based on the CIV lines. For the four objects with reliable results, $M_\mathrm{BH}\simeq10^8-10^9 \mathrm{\ M_\odot}$. 

\begin{figure}
\begin{center}
\includegraphics[width=90mm]{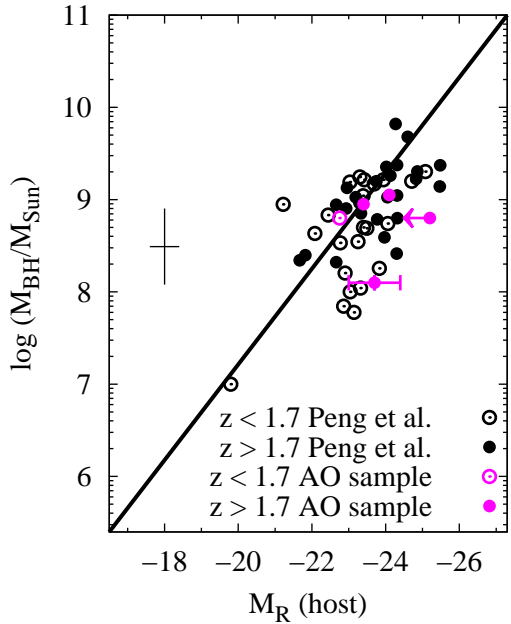}
\end{center}
\caption{The correlation between the black hole mass and the absolute $R-$band magnitude of the quasar hosts. The black points are from the sample of \citet{peng06-2}. The line shows the best-fit relation found for local AGNs (not plotted here; see \citet{peng06-2} for details). The error cross from \citet{peng06-2}, shown here, is fairly representative for the AO sample (magenta points) as well, with the exception of SDSS~J0819+5356. For SDSS~J0819+5356 and SDSS~J1322+1052, the $M_\mathrm{BH}$ estimates based on CIV are ignored. \it{A colour version of this figure is available in the online version of the paper.}
\label{fig:BH2}}
\end{figure}

In Figure \ref{fig:BH2} we show the 5 objects in the current AO sample compared to the sample from \citet{peng06-2}. They follow closely the correlation found between $M_\mathrm{BH}$ and the host galaxy luminosity.

\begin{figure}
\includegraphics[width=97mm]{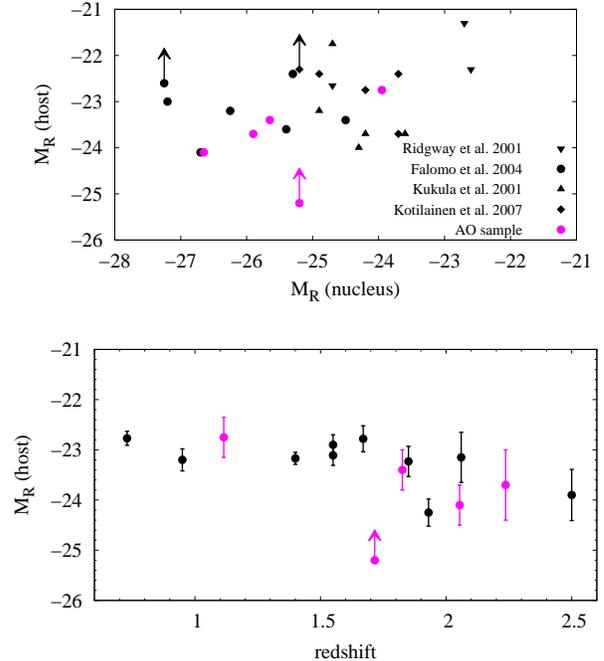}
\caption{\emph{Upper plot:} The absolute $R-$band magnitudes of the quasar hosts, assuming Sersic index $n=4$, and their nuclei, for the 5 objects with host galaxy estimates in the present AO sample (magenta) and various samples of radio quiet quasars at $1<z<2$. \emph{Lower plot:} The absolute $R-$band magnitudes of quasar hosts and their redshift. The black points are sample averages of radio quiet quasars, compiled in \citet{falomo08}. \it{A colour version of this figure is available in the online version of the paper.}
\label{fig:BH1}}
\end{figure}

Figure \ref{fig:BH1} plots the absolute $R$-band magnitudes of the quasar hosts versus those of their nuclei, as well as the $R$-band magnitudes as a function of redshift, for the AO sample as well as other quasars compiled by \citet{kotilainen07} and \citet{falomo08}. In both plots, the four systems with robust host luminosities occupy the loci found for radio quiet quasars. 

\begin{table*}
 \centering
 \begin{minipage}{155mm}
\caption{Quasar host galaxy properties}
  \begin{tabular}{@{}ccccccccc@{}}
\hline
Object &
$m_{K'}$ host &
$m_{K'}$ nucleus &
$e$ & 
$R_\mathrm{eff}$ [$''$] &  
$R_\mathrm{eff}$ [kpc] &  
Sersic index &  
host/qso \\ 
\hline
SDSS~J0819+5356 & $20.09\pm0.22$ & $19.26\pm0.11$ & $0.74\pm0.12$ & $0.26\pm0.07$ & 2.2 & $0.57\pm0.66$ & 0.46 \\
SDSS~J0819+5356 & $19.49\pm0.58$ & $19.13\pm0.13$ & $0.74\pm0.11$ & $0.33\pm0.30$ & $\sim2.8$ & [4] & $\sim0.77$ \\
SDSS~J0832+0404 & $19.21\pm0.10$ & $17.95\pm0.08$ & $0.67\pm0.15$ & $0.38\pm0.10$ & 3.2 & $0.25\pm0.32$ & 0.31 \\
SDSS~J0832+0404 & $18.46\pm0.14$ & $18.22\pm0.12$ & $0.65\pm0.13$ & $0.39\pm0.19$ & 3.2 & [4] & 0.80 \\
SDSS~J0904+1512 & $19.39\pm0.05$ & $17.74\pm0.02$ & $0.25\pm0.16$ & $0.04\pm0.00$ & 0.4 & $2.11\pm0.24$ & 0.22 \\
SDSS~J0904+1512 & $19.10\pm0.06$ & $17.73\pm0.06$ & $0.15\pm0.23$ & $0.04\pm0.01$ & 0.4 & [4] & 0.28 \\
SDSS~J1322+1052 & $\gtrsim17.15$ & $\sim18.02$  &  $\sim 0.3$ & - & - & - & $\gtrsim1.2$ \\
SDSS~J1515+1511 & $19.18\pm0.10$ & $16.68\pm0.01$ &  $0.30\pm0.05$ & $0.25\pm0.01$ & 2.1 & $1.71\pm0.80$ & 0.10 \\
SDSS~J1515+1511 & $18.85\pm0.04$ & $16.72\pm0.01$ &  $0.33\pm0.04$ & $0.18\pm0.02$ & 1.5 & [4] & 0.14 \\
\hline
\end{tabular}
{\footnotesize Error bars include those resulting from the simulations with a real PSF and noise, but not from the uncertainty in the radial slope of the lens mass profile. Here host/qso is the flux ratio in the source plane, and $e$ is the ellipticity of the host.}
\label{tab:host}
\end{minipage}
\end{table*}

\begin{table*}
 \centering
 \begin{minipage}{160mm}
\caption{Quasar host luminosity and black hole related properties}
  \begin{tabular}{@{}cccccccccc@{}}
  \hline
\multirow{2}{*}{Object} &
\multirow{2}{*}{$z_s$} &
$M_{R}$ &
\multirow{2}{*}{$L_\star$} & 
\multirow{2}{*}{$\alpha-2$} &
$M_{R}$ & 
$\log L_\mathrm{cont}  $ & 
FWHM &  
\multirow{2}{*}{$\log(\frac{M_\mathrm{BH}}{M_\odot})$} &
\multirow{2}{*}{$\frac{L_\mathrm{bol}}{L_\mathrm{Edd}}$} \\ 
 & & host & & & nucleus & [$\log{\mathrm{erg/s}}$] & [km/s] & & \\
\hline
 SDSS~J0819+5356 & 2.237 & $-23.7$ & 10.9 & $-0.71$ & $-25.9$ & 44.6, 44.7 & 1114, 3096 & 7.1, 8.1 & 1.04, 0.16 \\
 SDSS~J0832+0404 & 1.115 & $-22.75$ & 4.6 & $-1.75$ & $-23.95$ & -, $45.1$ & -, 5863 & -, 8.8 & -, 0.07 \\
 SDSS~J0904+1512 & 1.826 &  $-23.4$ & 8.6 & $-1.65$ & $-25.65$  & 45.7, 45.5 & 5754, 4364  & 9.1, 8.8 & 0.13, 0.22 \\
 SDSS~J1322+1052 & 1.716 & $>-25.2$ & $<44.5$& $-1.68$ & $\sim-25.2$  & 45.6, 45.4 & 1723, 4553 & 8.0, 8.8 & 1.26, 0.17  \\
 SDSS~J1515+1511 & 2.054 &  $-24.1$ & 15.9 & $-1.88$ & $-26.65$  & 46.4, 46.1 & 3860, 4267 & 9.1, 9.0 & 0.60, 0.47 \\
\hline
\end{tabular}
{\footnotesize \\ For all systems, the magnitudes and luminosities are calculated assuming Sersic index $n=4$ for the host galaxies. Here $\alpha$ is the spectral power-law index (see text). The two values in columns 7 (8) refer to 1350 \AA (CIV), and 3000 \AA (MgII), respectively; Corresponding values are given in columns 9 and 10.  $L_\star$ ($\mathrm{M}_{R,\star}$=21.2; \citet{gardner97}) is the characteristic luminosity of the Schechter luminosity function for elliptical galaxies. $L_\mathrm{cont}$ is the continuum luminosity in the proximity of the considered emission line. $L_\mathrm{bol}/L_\mathrm{Edd}$ is the Eddington ratio, where $L_\mathrm{bol}$ is the bolometric luminosity computed from   $L_\mathrm{3000}$ and $L_\mathrm{1350}$ with bolometric corrections 5.15 and 3.81, respectively, from the composite SED in \citet{richards06}; $L_\mathrm{Edd}=4\pi GM_\mathrm{BH}m_pc/\sigma_T\simeq1.3\times10^{38}{M_\mathrm{BH}/M_\odot}$ erg/s is the Eddington luminosity, in the usual units.}
\label{tab:BH}
\end{minipage}
\end{table*}

  \section{Lens galaxy light-mass profile and environment}\label{section:lightmass}
  
Although a dedicated study of the environment of the SQLS sample used in this paper is not possible, due to the lack of necessary spectroscopic data \citep[e.g.,][]{momcheva06,wong11}, in Appendix \ref{section:objectsdescript} we present considerations, whenever possible, based on the available AO data and the SDSS multi-band data. As mentioned in Section \ref{section:intro}, a motivating goal is the high-precision determination of the Hubble constant. This is possible in the case of strongly lensed quasars with measured time delays, however biases are introduced by degeneracies between the lens galaxy mass profile and the environment. It is therefore useful to identify and prioritise the modelling of lenses with a comparatively better understood environmental effect when planning time-consuming monitoring, for the purpose of measuring time delays. In this section, the goal is to study based on our AO sample whether or not the light ellipticity can be used as a proxy for the mass ellipticity of the lens galaxies, and then to obtain an overall classification of the present AO sample in terms of the degree in which the environmental influence is understood. 
  
Using an approach similar to that of \citet{chantry10} and \citet{sluse12}, we modelled each object (except for the quad lens SDSS~J1330+1810), whenever possible, with two independent lens models: SIS$+\gamma$ and SIE$+\gamma$, plus the SIE model (generally equivalent with SIS$+\gamma$), in order to estimate the effects of the lens ellipticity on the direction and strength of the predicted shear. Since for doubles the SIE$+\gamma$ models have a negative number of degrees of freedom, we used a prior, which assumes that the observed luminous lens ellipticity and orientation also hold for the mass profile. The considerations in this section are based on the assumption that we measure accurately the ellipticity of the luminous profile of the lensing galaxies. The simulations we performed for each system, as described in Section \ref{section:simulate}, were indeed designed to provide an accurate estimate of the uncertainties in these measurements, even when we approximate with an analytical or hybrid PSF the true, non-analytical PSF. In addition, many of our lenses have large ellipticity, which is less likely to be affected by inaccuracies in the PSF.  

  \subsection{Lens galaxy light-mass profile}\label{subsection:lightmass}

We first perform an insightful comparison to the sample of \citet{sluse12}. They model a sample of 10 doubles and 14 quads from the CASTLES survey, further processed with image deconvolution. The only quad in the current AO sample, SDSS~J1330+1810, is likely a disk-like galaxy with a significantly larger observed ellipticity, $\sim 0.7$. In addition, the model ellipticity is highly dependent on the power-law model, with a value of $\sim0.35$ for slope $\gamma'=2$, and $\sim0.6$ for slope $\gamma'\sim2.4$ (the best-fit model assuming no perturbers); \citet{sluse12} assume a uniform slope $\gamma'=2$ (isothermal).

 Among the 10 \citet{sluse12} doubles, 7 have shear orientations in agreement between the SIS$+\gamma$ and SIE$+\gamma$. All of these have observed light ellipticiy $e<0.22$ and in addition, the SIS$+\gamma$ model shear values are such that $e<3\gamma$ \citep[][]{keeton97}, therefore the lens ellipticity is not the main source of shear. There are two exceptions, in which case the light orientation and shear are nearly orthogonal, such that the intrinsic and extrinsic sources of shear are hard to disentangle. For the remaining objects, the SIS$+\gamma$ and SIE$+\gamma$ shear orientations are different, the observed ellipticity is larger, and the shear in the SIE$+\gamma$ model is nearly orthogonal to the lens. For these doubles (HE 2149-2745, CLASS B1600+434 and FBQS 0951+2635), the interpretation attempted in \citet{sluse12} of shear in terms of environment is not secure, if the assumption that the light and mass ellipticity are the same. On the other hand, all three objects are consistent with the observed orientation and SIS$+\gamma$ shear being in agreement to within $\sim10$ deg.

In the subsequent analysis of our sample of confirmed doubles with robust modelling, we remove 9 systems which may bias our analysis. These systems have unreliable light ellipticity/orientation, or are too complicated to be reliably modelled with simple single SIE or SIE$+\gamma$ models. They are SDSS~J0904+1512 (ellipticity measurement not available), SDSS~J1001+5027 (very close galaxy companion), SDSS~J1055+4628 (morphology uncertain due to possible quasar host contamination), SDSS~J1206+4332 (complicated system of lenses), SDSS~J1216+3529 (very close galaxy companion), SDSS~J1254+2235 (multiple Sersic components), SDSS~J1334+3315 (close galaxy companion responsible for most of the shear), SDSS~J1405+0959 (binary lens system) and SDSS~J1406+6126 (multiple Sersic components). We keep however SDSS~J1353+1138, which was modelled with two Sersic components of similar orientation and ellipticity. 
 
From the remaining sample of 13 doubles with robust SIS$+\gamma$ (SIE) and SIE$+\gamma$ mass models, two objects, SDSS~J0946+1835 and SDSS~J1322+1052, have the shear orientation in the SIS$+\gamma$ and SIE$+\gamma$ models consistent with each other. These have observed light $e\lesssim0.10$ consistent with $e<3\gamma$, resulting in the conclusion that the match between shear model orientations is due to the fact that the internal shear created by the lens ellipticity is too small to account, and in fact significantly modify, the model shear.  Qualitatively, these doubles are counterparts of the 7 objects in \citet{sluse12}.

Eleven objects in the current sample (SDSS~J0743+2457, SDSS~J0819+5356, SDSS~J0820+0812, SDSS~J0832+0404, SDSS~J1131+1915, SDSS~J1313+5151, SDSS~J1353+1138, SDSS~J1400+3134, SDSS~J1455+1447, SDSS~J1515+1511 and SDSS~J1620+1203) have observed orientations and SIE$+\gamma$ shear that are nearly orthogonal. This means that the 11/13 systems belong to this category are qualitative counterparts of the three systems in \citet{sluse12}. Most of these have ellipticity $\gtrsim0.25$. Lens modelling shows that once the ellipticity in the SIE$+\gamma$ model is larger than $e<3\gamma$, where $\gamma$ is the shear strength of the SIS$+\gamma$ model, the models require a nearly orthogonal shear in order to reproduce the observed configuration. As this ellipticity range lies outside the correlation range in \citet{sluse12}, the most likely explanation is that for these systems the overall mass ellipticity, dominated by the dark matter halo, is less elliptical than observed. Statistically, this is more likely than for such a large number of systems to be serendipitously located in environments where the main source of external shear acts orthogonal to the observed light orientation. 

There is nonetheless a theoretically predicted tendency for doubles (as opposed to quads), to have shears preferentially oriented perpendicular to the major axis of the lensing galaxies, but this is at a level of $\sim15\%$ (Figure \ref{fig:SIEshearhist} and reference therein), certainly not enough to explain the large fraction of objects in the AO sample. Statistically, the observed and theoretical histograms in Figure \ref{fig:SIEshearhist} have only a $\sim$ 3\% probability of representing the same distribution, following a Kolmogorov-Smirnov test. 

Therefore, this statistical analysis shows that in our sample of lensed quasars with an intermediate range of $R_\mathrm{Ein}/R_\mathrm{eff}$ a few times larger than $\sim0.5$ (with mean $R_\mathrm{Ein}/R_\mathrm{eff}\sim1.5$, larger than the mean value for the SLACS and SL2S lenses), and at large observed light ellipticity $\gtrsim0.25$, the light profile tends to be less elliptical than the mass. A similar conclusion was suggested by \citet{dye14} from a much smaller sample of individual lenses with extended source modelling. 

In Figure \ref{fig:PAellipt} we compile the mass versus light orientation and ellipticity for the robust doubles in our sample, using the alternative SIE (n.b. not the SIE$+\gamma$) model. 
The error bars on the mass ellipticity and orientation are boosted to include deviations from isothermality consistent with the lens sample of \citet{koopmans09}, i.e. with slope $1.8\lesssim\gamma'\lesssim2.2$. That is, the values of the mass ellipticity, and hence orientation, depend on the radial mass profile \citep[e.g.,][]{keeton98}, and we add in quadrature the error bars in Table \ref{tab:massmodel} for an isothermal profile ($\gamma'=2.0$) to the variation in ellipticity/orientation for the corresponding models with $\gamma'=1.8$ and $\gamma'=2.2$. The ellipticity plot shows that the great majority of lenses with observed ellipticity $\gtrsim 0.25$ require mass models with smaller ellipticity than the light, even after we allow for freedom in the radial slope. This supports the conclusion above. The orientation (PA) plot shows scatter due to the influence of environment. The scatter is larger than for the sample of quads in \citet{sluse12}, where the additional observational constraints can more reliably separate external and internal sources of shear.  
  
 \subsection{Sample classification based on environment}\label{subsection:environment}  
  
We have seen above that the use of the SIE$+\gamma$ model with a prior on the observed light ellipticity/orientation is misleading in breaking the degeneracy between the external shear and lens shape for our sample, as in most cases the light shape is not a good proxy for mass shape. Below we classify our sample based solely on the SIS$+\gamma$ model and clues from the environment. We identify three types of objects:
 \begin{itemize}
    \renewcommand{\labelitemi}{$\bullet$}
    \item For 6 objects (SDSS~J1131+1915, SDSS~J1254+2235, SDSS~J1313+5151, SDSS~J1353+1138, SDSS~J1406+6126 and SDSS~J1515+1511), the measured light orientation of the main lensing galaxy and the direction of the shear (in the SIS$+\gamma$ model), or alternatively the orientation of the SIE mass model (without shear), match to within $\sim$10 deg. As no prior on the ellipticity or orientation was used, the alignment of the orientation predicted by the mass models and the observed light orientation is therefore real. As expected, the SIS$+\gamma$ and SIE models are equivalent, the intrinsic lens ellipticity acting as internal shear, with $e\sim3\gamma$. For these objects, the environmental influence is not necessary in order to model the system, and is likely unimportant. 
    
    One caveat needs mentioning. SDSS~J1254+2235 is modelled by a superposition of two concentric Sersic profiles, whose orientations differ by $\sim35$ deg. The SIE mass model orientation corresponds to the orientation of the brighter profile. If the additional newly-discovered object G2 were close to the redshift of the lens, its effect on the model would be drastic. Also, for SDSS~J1406+6126, which was modelled with a superposition of two Sersic profiles, the orientation of the extended profile was considered, the other being spherical.
     
    \item For 12 objects (SDSS~J0819+5356, SDSS~J0820+0812, SDSS~J0832+0404, SDSS~J0946+1835, SDSS~J1216+3529, SDSS~J1322+1052, SDSS~J1330+1810, SDSS~J1334+3315, SDSS~J1400+3134, SDSS~J1405+0959, SDSS~J1455+1447 and SDSS~J1620+1203), the environment is likely to play an important role, as the observed light orientation and required shear (the SIS$+\gamma$ model) are different (see below for two exceptions), and there is some evidence as to what causes the external shear. 
    
\begin{figure}
\includegraphics[width=90mm]{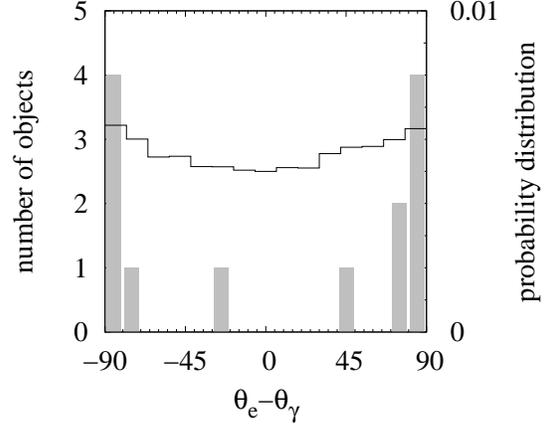}
\caption{Histogram of the difference in orientation between the observed lens and the shear (assuming the SIE$+\gamma$ model) for the doubles in the AO sample. As noted in the text, 10 objects which may bias our analyses were removed. Also plotted is the theoretically predicted alignment for doubles (solid line) from \citet{oguri10-1}; the non-biased (non-lens) population is flat. 
\label{fig:SIEshearhist}}
\end{figure}
    
    For SDSS~J1322+1052, the difference in orientations is just $\sim$17 deg, and there are several nearby galaxies, at least one of which could affect the system. For SDSS~J0819+5356, SDSS~J0832+0404, SDSS~J0946+1835 and SDSS~J1620+1203, the shears point toward nearby groups of galaxies, the likely source of influence. SDSS~J1330+1810, the only 4-image lens in the present sample, has enough observational  constraints to disentangle the internal and external shear, the latter of which is due to one of two nearby galaxy groups. SDSS~J0820+0812 is located behind an elongated galaxy group and a larger structure, whose orientation corresponds to the required shear. For SDSS~J1400+3134, the two orientations differ by $\sim14$ deg, but are brought in agreement if the effect of a nearby galaxy is considered. For SDSS~J1334+3315, a bright nearby galaxy has been shown to be responsible for the majority of the shear \citep[see][]{rusu11}. SDSS~J1405+0959 \citep[see][]{rusu14} is a complicated system with a binary lens, and likely an additional contribution from the observed nearby galaxies located in the foreground.
    
    There are two exceptions, where the observed light orientation and required shear are in fact consistent at $\sim$10 deg, but there is likely an environmental contribution.  For both SDSS~J1216+3529 and SDSS~J1455+1447, a single nearby galaxy, which may affect each system, is also located close to the direction of the shear. For the latter system, explicitly accounting for the companion galaxy in the lens model by scaling the strength of the two galaxies further improves the agreement between the shear orientation and the light orientation.
   
    \item For 5 objects (SDSS~J0743+2457, SDSS~J0904+1512, SDSS~J1001+5027, SDSS~J1055+4628, and SDSS~J1206+4332), the environment is likely to play an important role, but there is no clue as to what causes the measured external shear. For SDSS~J0743+2457 and SDSS~J1055+4628, the shear direction is significantly offset from the light orientation, and there is no obvious galaxy or galaxy group to account for the discrepancy. However the observed light orientation may be biased for SDSS~J1055+4628 due to possible unaccounted host contribution during the morphological modelling. SDSS~J0904+1512 has an ellipticity consistent with zero, therefore the source of the required shear is not known. SDSS~J1206+4332 has several nearby galaxies, and it is unclear if they are physically associated, and what their exact contributions to the model are. SDSS~J1001+5027 requires a large external shear whose direction does not match the nearby foreground group of galaxies. 
\end{itemize}

The doubles that are most promising in terms of reliably disentangling internal and external sources of shear, and would therefore be most useful for a determination of the Hubble constant based on time delay measurements, are SDSS~J1131+1915, SDSS~J1313+5151, SDSS~J1334+3315, SDSS~J1353+1138, SDSS~J1400+3134, SDSS~J1406+6126, SDSS~J1455+1447 and SDSS~J1515+1511.

\begin{figure*}
\begin{center}
\includegraphics[width=175mm]{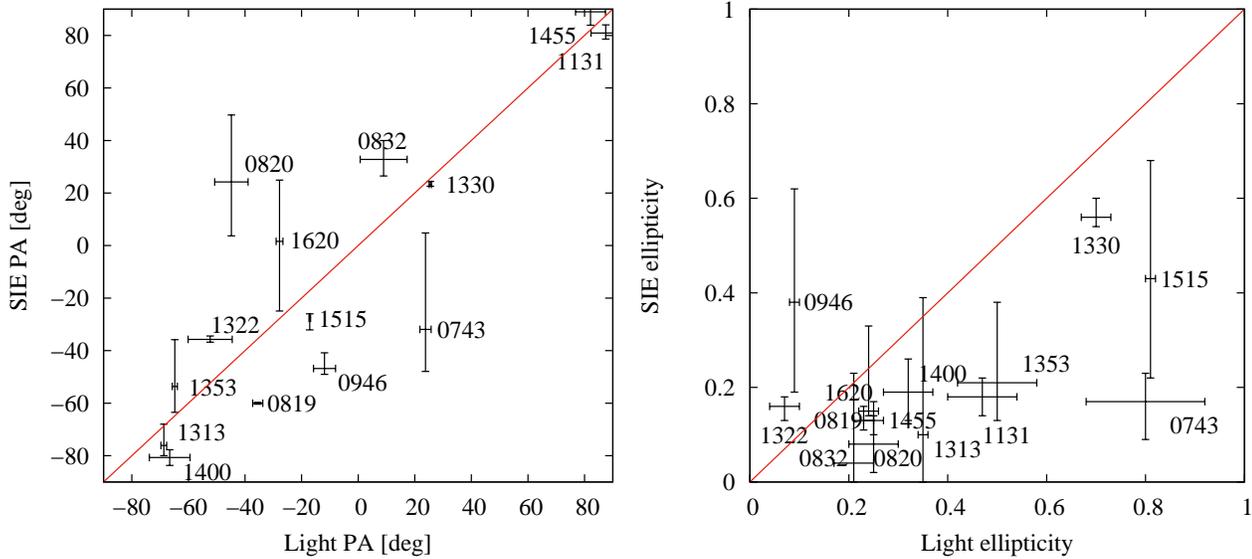}
\end{center}
\caption{Light and mass orientation (\emph{left}), as well as ellipticity (\emph{right}) for 14 objects without obvious sources of bias or systematics in the AO sample. For the observed (light) profiles, the values and error bars are taken from Table \ref{tab:lensmorphology}. However for the SIE (n.b. not the SIE$+\gamma$) mass models, the error bars from Table \ref{tab:massmodel} are boosted to account for variations in the power-law slope, by considering variations in mass orientation and ellipticity for $1.8\lesssim\gamma'\lesssim2.2$. For the quad SDSS~J1330+1810, only the power law $+\gamma+\delta$ model is considered, without boosting the errors. For SDSS~J1353+1138, the light profile of the flatter Sersic component is considered.
\label{fig:PAellipt}}
\end{figure*}



\section{Conclusions and future prospects}\label{sect:concl}

In this paper we described the results of a large imaging campaign with the Subaru Telescope adaptive optics, aimed at obtaining high-resolution observations of gravitationally lensed quasars from SQLS. 
The purpose of the high resolution observations was multifold. First, it was to provide accurate astrometry and relative photometry in the near-infrared, which we used to obtain accurate lensing models and masses inside the Einstein radii. Second, it was to derive robust morphology for the lensing galaxies and compare it with the mass profile. Third, it was to study (if detected) the quasar host galaxies at high redshift $\sim2$, as well as other potential features previously hidden in low-resolution data.
              
            In this work we introduced a new method for fitting the morphology of gravitationally lensed quasars in AO data. This was necessary, as the separately observed PSF stars proved to be ill-suited representations of the actual PSF, due to varying atmospheric turbulence characteristics. The new method consists of approximating the PSF analytically, and fitting it simultaneously to the quasar images and to the lensing galaxy (in the latter case, convolved with a Sersic profile). For the low Strehl ratios  ($\lesssim0.10$) typically obtained, we determined empirically that the best simple analytical PSF approximation is a superposition of 2 Moffat profiles. For the systems with larger signal-to-noise, and a relatively isolated brighter image, we created a hybrid PSF by using the observed brightness profile of the core, and replacing the wings with analytical approximations, due to the high noise level. We used a variety of methods to obtain realistic error bars on the observed parameters, including  simulations which employed the separately observed stars.
            
            In addition, using the natural stretch of the quasar host galaxy into arc-like features, we characterized the quasar host galaxy, or placed limits on its brightness, for 5 systems. The hosts follow relations found in the literature with redshift, nuclear luminosity and black hole mass. This is the first time such study was attempted without an a-priori known PSF. While the small number of host detections in the present AO sample cannot at the present significantly improve the results of previous studies, by showing that these hosts follow a known correlation determined for $HST$ data, and have similar error bars, we have shown that studies of hosts with AO are possible to perform in the future on larger samples, in a way that is competitive and complementary to $HST$-based studies.
            
            The typical achieved relative astrometry precision was $\sim1-2$ mas for the quasar images, but less for the lensing galaxies. This is comparable to the $HST$ results. Based on the new astrometry, we have shown that for many systems the original low-resolution astrometry was not accurate within the given error bars, in many cases providing unreliable gravitational lensing mass models. Due to the necessity of modelling the PSF on the target, the morphological parameters of the lensing galaxy were typically less accurate than those provided by $HST$.   
                      
     By using the SIE mass model together with the observed light distribution in the lensing galaxies, and using the known correlation between the orientation of light and mass, we have classified our sample according to the degree to which we expect the lens environment to influence the observed configuration. We have pointed out systems which seem well understood and would benefit most from time delay measurements in order to infer the Hubble constant. On the other hand, we have shown that the tight correlation found for the elliptical galaxy lenses in the SLACS Survey between the mass and light ellipticity does not hold statistically for the present sample at ellipticity $e\gtrsim0.25$, and the mass profile is less elliptical than the light profile. This is likely due to the larger mean ratio of the Einstein radius to the effective radius encountered in the present sample, the mass profile therefore being more dark-matter dominated and rounder. 

     As a result of the high-resolution observations, we obtained the following noteworthy results on particular systems (see Appendix \ref{section:objectsdescript} for details):
     
      \begin{itemize}
    \renewcommand{\labelitemi}{$\bullet$}
    \item SDSS~J0819+5356 and SDSS~J0832+0404: we characterised the host galaxy with the use of an analytical PSF. SDSS~J0904+1512 and SDSS~J1322+1052: we used an iterative approach with a hybrid PSF to model the quasar host galaxy. SDSS~J1515+1511: this is the only system where a star serendipitously located in the field has provided a suitable PSF, resulting in the detection of the quasar host galaxy. 

\item SDSS~J1001+5027 and SDSS~J1206+4332: the systems have available time delay measurements and consist of multiple lensing galaxies. The secondary galaxy in SDSS~J1001+5027 is either located at different redshift, or is undergoing tidal stripping. The system also suggests the existence of a massive nearby galaxy cluster, which future field spectroscopy would help confirm. In SDSS~J1206+4332, we discovered a previously undetected galaxy in the non-AO data. In order to reliably model each system however, spectroscopy is required to estimate the redshift of the individual lensing galaxies.

\item SDSS~J1254+2235, SDSS~J1353+1138 and SDSS~J1406+6126: the two galaxies are not fitted well by a single Sersic profile, but instead require a combination of two Sersic profiles each, revealing more morphological detail than in the previously available low-resolution data. 

\item SDSS~J1330+1810: in opposition to the discovery paper, we showed that the lensing galaxy is a late-type galaxy. Also, we discovered astrometric anomalies when applying the customary SIE$+\gamma$ mass model. Although substructure in the form of lens galaxy satellites is not detected but cannot be ruled out either, a power law ($\gamma'\sim2.3$) model with a contribution from the complicated environment in the form of a third order perturbation fits the data well. 

\item SDSS~J1405+0959: we presented this system in a separate paper \citep{rusu14}; this is an interesting system where the AO observations have revealed dramatical new detail, with a host galaxy arc produced by two lensing galaxies.     
\end{itemize}

Through the observations and subsequent data analysis presented here we have shown that a large campaign to observe gravitationally lensed quasars with AO is feasible even with the lack of an a-priory known PSF. We used the data to obtain new results on individual systems, as well statistically, in combination with other samples of objects. Upcoming wide-field imaging surveys such as the Large Synodic Sky Survey and Subaru Hyper Suprime-Cam Survey will discover thousands of additional lensed quasars \citep{oguri10-1}, and LGS/AO is a viable alternative to $HST$ for the follow-up of some of these new systems with high resolution observations. 

The technique employed can be further enhanced by the availability of a more accurate distortion correction map, and may be further enhanced by deconvolution techniques.



\section*{Acknowledgments}

This work was supported in part by the Japan Society for the Promotion of Science (JSPS) KAKENHI Grant Numbers 26800093, 24740171 and World Premier International Research Center Initiative (WPI Initiative), Ministry of Education, Culture, Sports, Science and Technology (MEXT), Japan. C.E.R. acknowledges the support of the JSPS Research Fellowship. Data analysis were in part carried out on common use data analysis computer system at the Astronomy Data Center, ADC, of the National Astronomical Observatory of Japan.

The authors would like to thank Dominique Sluse for helpful discussions, as well as the anonymous referee for criticism and suggestions that helped improve the lengthy manuscript. 
 The authors recognise and acknowledge the very significant cultural role and reverence that the summit of Mauna Kea has always had within the indigenous Hawaiian community. We are most fortunate to have the opportunity to conduct observations from this superb mountain.



\appendix


\section{Tables of tip-tilt stars, PSF stars, analytical PSF parameters, and astrometric distortion}\label{sec:app1}

\begin{table*}
 \centering
 \begin{minipage}{114mm}
  \caption{Tip-tilt and PSF stars}
  \begin{tabular}{@{}lllr@{}}
  \hline 
Object &
Description &
Magnitude &
Separation [\arcsec] \\ 
 \hline
SDSS~J074355.71+245745.62 & TT J0743+2457 & R=15.00 & $\sim$42.0\\
 \hline
SDSS~J082006.11+535620.42 & TT J0819+5356 & R=11.18 & $\sim$55.5\\
 \hline
SDSS~J083219.00+040505.03 & TT J0832+0404 & R=13.84 & $\sim$66.8\\
SDSS~J083427.71+031632.82 & TT  (PSF) J0832+0404 & R=13.9 & $71.2$\\
SDSS~J083425.13+031533.49 & PSF J0832+0404 & K=12.4 & \\
 \hline
SDSS~J090406.54+151229.94 & TT J0904+1512 & R=16.36 & $\sim$42.4; 39.2\\
SDSS~J090409.04+151214.85 & PSF J0904+1512 & K=13.87 & \\
 \hline
SDSS~J092635.53+310037.66 & TT J0926+3059 & R=12.4 & $\sim53.2$\\ 
 \hline
SDSS~J094605.79+183553.34 & TT J0946+1835 & R=14.43 & $\sim$20\\
SDSS~J094333.27+182720.20 & TT  (PSF) J0946+1835 & R=14.60 & $20.6$\\
SDSS~J094332.33+182735.79 & PSF J0946+1835 & K=12.36 & \\
 \hline
SDSS~J100135.04+502740.84 & TT J1001+5027 & R=18.02 & $\sim$63.3\\
SDSS~J100216.17+501059.07 & TT (PSF) J1001+5027 & R=18.04 & 65.7\\
SDSS~J100222.74+501117.49 & PSF J1001+5027 & K=12.24 & \\
\hline
SDSS~J100234.48+445036.47 & TT J1002+4449 & R=13.47 & $\sim$75.6\\
SDSS~J100229.61+445002.77 & PSF J1002+4449 & K=15.16 & 61.8\\
\hline
SDSS~J105440.61+273222.93 & TT J1054+2733 & R=16.89 & $\sim$43.4\\
\hline
SDSS~J105538.55+462804.93 & TT J1055+4628 & R=13.7 & $\sim$79.2\\
SDSS~J105208.39+463816.93 & TT (PSF) J1055+4628 & R=13.9 & 79.3\\
SDSS~J105200.79+463830.8 & PSF J1055+4628 & K=11.85 &\\
\hline
SDSS~J112821.12 +240253.1 & TT J1128+2402 & R=16.84 & $\sim$50.9\\
\hline
SDSS~J113201.76+191518.63 & TT J1131+1915 & R=15.33 & $\sim$58.3\\
SDSS~J112825.17+194704.17 & TT (PSF) J1131+1915 & R=14.73 & 58.6\\
SDSS~J112827.97+194747.46 & PSF J1131+1915 & K=10.90 &\\
\hline
SDSS~J121640+352949.51 & TT J1216+3529 & R=15.6 & $\sim$72.9\\
SDSS~J121505.47+330437.83 & TT (PSF) J1216+3529 & R=15.5 & 69.5\\
SDSS~J121507.13+330544.54 & PSF J1216+3529 & K=11.7 &\\
\hline
SDSS~J125414.9+223530.99 & TT J1254+2235 & R=17.59 & $\sim$56.0\\
SDSS~J125144.95+222445.79 & TT (PSF) J1254+2235 & R=17.37 & 59.7\\
SDSS~J125141.47+222521.42 & PSF J1254+2235 & K=10.48 &\\
\hline
SDSS~J131343.88+515214.61 & TT J1313+5151 & R=16.32 & $\sim$58.8\\
SDSS~J131618.12+512544.78 & TT (PSF) J1313+5151 & R=16.16 & 57.1\\
SDSS~J131619.66+512449.66 & PSF J1313+5151 & K=11.51 &\\
\hline
SDSS~J132058.29+164459.07 & TT J1320+1644 & R=16.7 & $\sim$58.0\\
\hline
SDSS~J132238.43+105237.71 & TT J1322+1052 & R=16.53 & $\sim$29.7\\
\hline
SDSS~J133023.35 +180957.3 & TT J1330+1810 & R=14.77 & $\sim$75.5\\
\hline
SDSS~J133402.02+331437.75 & TT J1334+3315 & R=13.64 & $\sim$57.0\\
\hline
SDSS~J135303.52+113752.67 & TT J1353+1138 & R=15.05 & $\sim$42.8\\
\hline
SDSS~J140017.13+313511.05 & TT J1400+3134 & R=14.1 & $\sim$58.8\\
\hline
SDSS~J140514.96+095820.10 & TT J1405+0958 & R=13.35 & $\sim$71.5\\
SDSS~J140517.09+103007.16 & TT (PSF) J1405+0958 & R=13.1 & 70.8\\
SDSS~J140521.31+102933.31 & PSF J1405+0958 & K=12.0 &\\
\hline
SDSS~J140626.66+612556.94 & TT J1406+6126 & R=14.37 & $\sim$46.0\\
\hline
SDSS~J145459.41+144838.21 & TT J1455+1447 & R=16.54 & $\sim$73.0\\
\hline
SDSS~J151538.33+151232.87 & TT J1515+1511 & R=13.37 & $\sim$57.2\\
SDSS~J151537.14+151114.41 & PSF J1515+1511 & K=14.79 & 80.3\\
\hline
SDSS~J162022.49+120343.81 & TT J1620+1203 & R=13.98 & $\sim$52.0\\
\hline
\end{tabular}
\\ 
{\footnotesize The magnitudes are taken from the the Naval Observatory Merged Astrometric Dataset \citep[NOMAD;][]{zacharias05} catalogue. ``Separation'' specifies the distance between the tip-tilt star and the object, and between the tip-tilt and the PSF star, respectively.}
\label{tab:AO}
\end{minipage}
\end{table*}

\begin{table*}
 \begin{minipage}{167mm}
  \caption{Analytical parameters of the best-fitted PSFs}
  \begin{tabular}{@{}lcccccccccc@{}}
  \hline 
\multirow{2}{*}{Object} &
\multirow{2}{*}{FWHM1} &
\multirow{2}{*}{e1} &
\multirow{2}{*}{PA1} &
\multirow{2}{*}{$\beta$1} &
\multirow{2}{*}{FWHM2} &
\multirow{2}{*}{e2} &
\multirow{2}{*}{PA2} &
\multirow{2}{*}{$\beta$2} &
flux1/ &
Strehl \\ 
& & & & & & & & & (flux1 + flux2) & ratio ($\%$) \\
 \hline
SDSS~J0743+2457 & $0.29$ & $0.30$ & $76.5$ & $3.1$ & $0.71$ & 0.10 & 75.0 & 2.0 & 0.34 & 5 \\
SDSS~J0743+2457 star & $0.21$ & $0.20$ & $-88.5$ & $2.8$ & $0.69$ & 0.04 & 69.3 & 2.3 & 0.44 &  \\
 \hline
SDSS~J0819+5356 & $0.25$ & $0.17$ & $-70.8$ & $1.5$ & $0.92$ & 0.09 & 61.0 & 1.9 & 0.36 & 2 \\
SDSS~J0819+5356 star & $0.23$ & $0.25$ & $19.8$ & $2.6$ & $0.80$ & 0.04 & 18.6 & 1.8 & 0.21 &  \\
 \hline
SDSS~J0832+0404 & $0.24$ & $0.16$ & $-70.3$ & $1.4$ & $1.20$ & 0.23 & 4.2 & 2.8 & 0.38 &  \\
SDSS~J0832+0404 PSF & $0.19$ & $0.04$ & $47.0$ & $9.0$ & $0.52$ & 0.12 & -49.2 & 2.2 & 0.22 & 5 \\
 \hline
SDSS~J0904+1512 & $0.21$ & $0.31$ & $81.6$ & $4.5$ & $0.77$ & 0.04 & $35.5$ & 2.8 & 0.32 & 5 \\
SDSS~J0904+1512 PSF & $0.21$ & $0.16$ & $33.2$ & $3.3$ & $0.72$ & 0.11 & $-63.6$ & 3.2 & 0.32 & 4 \\
 \hline
SDSS~J0926+3059 & $0.21$ & $0.29$ & $78.8$ & $21.1$ & $0.62$ & 0.10 & -81 & 1.8 & 0.09 & 2 \\
 \hline
SDSS~J0946+1835 (2011) & $0.26$ & $0.23$ & $78.4$ & $3.8$ & $0.81$ & 0.12 & $-47.7$ & 2.9 & 0.57 & 5\\
SDSS~J0946+1835 (2012) & $0.16$ & $0.14$ & $-86.1$ & $2.7$ & $0.98$ & 0.02 & 82.1 & 5.5 & 0.55 & 9 \\
SDSS~J0946+1835 PSF & $0.14$ & $0.07$ & $-35.9$ & $3.0$ & $0.90$ & $0.04$ & $-28.4$ & 4.5 & 0.57 & 11 \\
 \hline
SDSS~J1001+5027 & $0.20$ & $0.11$ & $-73.9$ & $5.3$ & $0.67$ & 0.02 & 31.1 & 2.0 & 0.22 & 3 \\
SDSS~J1001+5027 PSF & $0.28$ & $0.18$ & $66.8$ & $7.3$ & $0.63$ & $0.04$ & $79.8$ & 2.3 & 0.26 & 2 \\
 \hline
SDSS~J1002+4449 & $0.21$ & $0.22$ & $29.4$ & $12.5$ & $0.48$ & $0.27$ & 19.9 & 2.6 & 0.20 & 3 \\
SDSS~J1002+4449 PSF & $0.21$ & $0.11$ & $-11.3$ & $4.4$ & $0.58$ & $0.14$ & 2.9 & 2.5 & 0.30 & 5 \\
 \hline
SDSS~J1054+2733 & $0.34$ & $0.13$ & $69.9$ & $8.6$ & $0.83$ & 0.05 & 53.4 & 2.3 & 0.09 &  \\
 \hline
SDSS~J1055+4628 & $0.17$ & $0.12$ & $74.4$ & $3.8$ & $0.84$ & 0.25 & $-89.0$ & 3.2 & 0.26 & 5 \\
 \hline
SDSS~J1131+1915 & $0.15$ & $0.09$ & $-84.2$ & $3.7$ & $0.65$ & 0.05 & $-28.1$ & 2.4 & 0.39 & 12 \\
SDSS~J1131+1915 PSF & $0.14$ & $0.08$ & $-88.5$ & $4.0$ & $0.58$ & 0.22 & $5.2$ & 2.4 & 0.39 & 12 \\
 \hline
SDSS~J1216+3529 & $0.46$ & $0.20$ & $21.3$ & $[100]$ & $0.83$ & 0.13 & $-56.0$ & 2.3 & 0.48 & \\
 \hline
SDSS~J1254+2235 & $0.19$ & $0.20$ & $88.2$ & $8.1$ & $0.56$ & 0.15 & $-52.7$ & 2.0 & 0.31 & 6 \\
SDSS~J1254+2235 PSF & $0.21$ & $0.09$ & $-86.6$ & $18.2$ & $0.54$ & 0.09 & -85.3 & 2.1 & 0.41 & 6 \\
 \hline
SDSS~J1313+5151 & $0.18$ & $0.15$ & $72.4$ & $3.9$ & $0.63$ & 0.12 & $-79.1$ & 2.0 & 0.25 & 5 \\
SDSS~J1313+5151 PSF & $0.20$ & $0.11$ & $-84.5$ & $2.1$ & $0.72$ & 0.05 & 32.0 & 2.3 & 0.50 & 6 \\
 \hline
SDSS~J1322+1052 & $0.22$ & $0.24$ & $60.1$ & $7.0$ & $0.61$ & 0.03 & $48.0$ & 2.3 & 0.35 & 5 \\
 \hline
SDSS~J1330+1810 (2012) & $0.16$ & $0.06$ & $-52.8$ & $5.6$ & $0.61$ & 0.10 & $-56.4$ & 2.1 & 0.35 & 8 \\
SDSS~J1330+1810 (2013) & $0.18$ & $0.37$ & $72.9$ & $3.9$ & $0.68$ & 0.02 & $-83.9$ & 1.7 & 0.30 &  \\
 \hline
SDSS~J1334+3315 $J$ & $0.14$ & $0.08$ & $-48.1$ & $15.3$ & $0.32$ & 0.07 & $-67.3$ & 2.1 & 0.29 & 8 \\
SDSS~J1334+3315 $H$ & $0.14$ & $0.10$ & $-34.5$ & $4.7$ & $0.50$ & 0.08 & $-66.5$ & 3.2 & 0.51 & 11 \\
SDSS~J1334+3315 $K'$ & $0.15$ & $0.07$ & $-35.8$ & $7.1$ & $0.42$ & 0.19 & $-81.2$ & 2.5 & 0.42 & 12 \\
 \hline
SDSS~J1353+1138 & $0.21$ & $0.20$ & $62.7$ & $7.5$ & $0.61$ & 0.08 & $-30.8$ & 2.4 & 0.37 & 5 \\
 \hline
SDSS~J1400+3134 & $0.24$ & $0.28$ & $34.5$ & $6.6$ & $0.66$ & 0.07 & $78.4$ & 3.0 & 0.35 & 4 \\
 \hline
SDSS~J1405+0959 $J$ & $0.18$ & $0.07$ & $21.7$ & $6.0$ & $0.55$ & 0.16 & $-75.9$ & 2.5 & 0.15 & 2 \\
SDSS~J1405+0959 $H$ & $0.17$ & $0.07$ & $-48.0$ & $8.3$ & $0.51$ & 0.27 & $-63.2$ & 2.2 & 0.16 & 4 \\
SDSS~J1405+0959 $H$ PSF & $0.17$ & $0.09$ & $12.8$ & $43.3$ & $0.46$ & 0.14 & $-42.5$ & 2.3 & 0.27 & 6 \\
SDSS~J1405+0959 $K'$ & $0.18$ & $0.11$ & $27.9$ & $6.3$ & $0.54$ & 0.07 & $-60.4$ & 1.7 & 0.21 & 7 \\
 \hline
SDSS~J1406+6126 & $0.69$ & $0.12$ & $-70.3$ & $2.0$ & $-$ & $-$ & $-$ & $-$ & $-$ & $-$ \\
 \hline
SDSS~J1455+1477 & $0.19$ & $0.26$ & $-63.0$ & $2.4$ & $0.67$ & 0.07 & -52.0 & 2.1 & 0.22 & 3 \\
 \hline
SDSS~J1515+1511 & $0.21$ & $0.08$ & $-40.2$ & $8.2$ & $0.64$ & 0.03 & $-46.7$ & 2.0 & 0.17 & 3 \\
SDSS~J1515+1511 PSF & $0.21$ & $0.07$ & $-40.0$ & $10.3$ & $0.63$ & 0.04 & $-45.4$ & 2.0 & 0.19 & 6 \\
 \hline
SDSS~J1620+1203 (2011) & $0.16$ & $0.16$ & $-42.3$ & $3.0$ & $0.55$ & 0.07 & 41.4 & 2.0 & 0.50 & 9 \\
SDSS~J1620+1203 (2013) & $0.17$ & $0.13$ & $50.4$ & $4.0$ & $0.50$ & 0.01 & -63.8 & 1.8 & 0.33 & 7 \\
 \hline
\end{tabular}
\\ 
{\footnotesize Analytical parameters for all best-fitted object and star PSFs, determined with Hostlens. The fitting was performed on the coadded, final frames. Affix $\mathit{1}$ refers to the core, and affix $\mathit{2}$ refers to the wings, both modelled as Moffat profiles. Here $\mathit{FWHM}$ ($\arcsec$), $\mathit{e}$, $\mathit{PA}$ (deg) and $\beta$ are the profile full width at half maximum, ellipticity, position angle (positive from North towards East), and $\beta$ parameter, respectively. Values inside square brackets are kept fixed. The Strehl ratios were computed by comparing the peak flux of the PSF with that of a diffraction limited PSF model.}
\label{tab:analpsf}
\end{minipage}
\end{table*}


\begin{table*}
 \centering
 \begin{minipage}{170mm}
\caption{Residual distortion error}
  \begin{tabular}{@{}lccccccc@{}}
  \hline
Object (dith. pos.) &
X $\sigma_{\mathrm{pair\ sep.}}$ & 
Y $\sigma_{\mathrm{pair\ sep.}}$ &  
$\sigma_{\mathrm{atm.\ refract.}}$ & 
X $\sigma_{\mathrm{pix.\ scale}}$ &
Y $\sigma_{\mathrm{pix.\ scale}}$ &  
X $\sigma_{\mathrm{total}}$ &
Y $\sigma_{\mathrm{total}}$ \\ 
\hline
SDSS~J0743+2457 & [1.5] & [1.5] & $\ll$ 1 & 0.2 & 0.2 & 1.1 & 1.1 \\
SDSS~J0819+5356 & [1.5] & [1.5] & $\sim$1.5 & 0.3 & 0.3 & 1.5 & 1.5 \\
SDSS~J0832+0404 (5/6) & 5.5 & 5.5 & $\ll$1.0 & 0.1 & 0.1 & 3.9 & 3.9 \\ 
SDSS~J0904+1512 (5/5) & 1.6 & 0.7 & $\ll 1$ & 0.1 & 0.2 & 1.1 & 0.5 \\
SDSS~J0946+1835 (52mas) & [1.5] & [1.5] & $\sim$1.0 & 0.3 & 0.2 & 1.3 & 1.3 \\
SDSS~J0946+1835 (20mas) (9/9) & 1.7 & 1.4 & $\sim$1.0 & 0.4 & 0.4 & 1.4 & 1.2 \\
SDSS~J1001+5027 (7/9) & 0.8 & 0.6 & $\sim$1.0 & 0.2 & 0.3 & 0.9 & 0.9 \\
SDSS~J1055+4628 & [1.5] & [1.5] & $\ll$ 1 & 0.2 & 0 & 1.1 & 1.1 \\
SDSS~J1131+1915 (7/7) & 0.9 & 2.1 & $\ll$ 1 & 0.1 & 0.2 & 0.6 & 1.5 \\
SDSS~J1206+4332 & [1.5] & [1.5] & $\ll$ 1 & 0.3 & 0 & 1.1 & 1.1 \\
SDSS~J1216+3529 & [1.5] & [1.5] & $\ll$ 1 & 0.2 & 0.1 & 1.1 & 1.1 \\
SDSS~J1254+2235 (9/9) & 1.6 & 2.5 & $\ll$ 1 & 0.1 & 0.2 & 1.1 & 1.8 \\
SDSS~J1313+5151 (9/9) & 1.0 & 0.7 & $\ll$ 1 & 0.2 & 0.1 & 0.7 & 0.5 \\
SDSS~J1322+1052 (5/5) & 2.1 & 1.7 & $\ll$ 1 & 0.2 & 0.2 & 1.5 & 1.2 \\
SDSS~J1330+1810 (52mas) & [1.5] & [1.5] & $\ll$ 1 & $< 0.2$ & $< 0.2$ & 1.1 & 1.1 \\
SDSS~J1330+1810 (20mas) & [1.5] & [1.5] & $\ll$ 1 & $< 0.3$ & $< 0.4$ & 1.2 & 1.0 \\
SDSS~J1334+3315 (4/5) ($J$) & 2.2 & 1.7 & $\ll$ 1 & 0.1 & 0.1 & 1.6 & 1.2 \\
SDSS~J1353+1138 (5/5) & 1.2 & 1.6 & $\ll$ 1 & 0.2 & 0.1 & 0.9 & 1.1 \\
SDSS~J1400+3134 (4/5) & 1.6 & 2.0 & $\sim$1.5 & 0.2 & 0.2 & 1.6 & 1.8 \\
SDSS~J1405+0959 ($J,H,K'$) & [1.5] & [1.5] & $\ll$ 1 & 0.2 & 0.1 & 1.1 & 1.1 \\ 
SDSS~J1455+1447 (8/8) & 2.2 & 2.1 & $\ll$1.0 & $<0.1$ & 0.1 & 1.5 & 1.5 \\ 
SDSS~J1515+1511 (8/9) & 1.7 & 1.0 & $\ll$ 1 & 0.2 & 0.2 & 1.2 & 0.7 \\
SDSS~J1620+1203 (2013) (4/5) & 3.8 & 3.6 & $\sim$1.0 & 0.2 & 0.3 & 2.8 & 2.6 \\
\hline
\end{tabular}
{\footnotesize The detector X axis is typically aligned to the north-south direction, and therefore corresponds to the Y axis in Table \ref{tab:lensastrometry}. In the Objects column, the values in parenthesis refer to the number of dither positions used to calculate $\sigma_{\mathrm{pair\ sep.}}$ Here $\sigma_{\mathrm{pair\ sep.}}$ is the standard deviation of the quasar image separation between different dither positions. $\sigma_{\mathrm{atm.\ refract.}}$ is the expected stretch towards zenith due to the atmospheric distortion. $\sigma_{\mathrm{pix.\ scale}}$ is the uncertainty in the image separation due to the uncertainty in the pixel scale ($\sqrt{\mathrm{No. pixels}}\times \mathrm{average\ pixel\ size\ error}$). $\sigma_{\mathrm{total}}$ is calculated for the coordinates of an individual image on each axis as $\sqrt{\sigma^2_{\mathrm{pair\ sep.}}+\sigma^2_{\mathrm{atm.\ refract.}}+\sigma^2_{\mathrm{pix\ scale}}}/\sqrt{2}$. Values inside square brackets are for objects for which $\sigma_{\mathrm{pair\ sep.}}$ could not be calculated, and the average value from the objects with available calculations was used.}
\label{tab:distort-data}
\end{minipage}
\end{table*}

\section{Description of results on individual objects}\label{section:objectsdescript}

A discussion is given below for the objects observed during the present campaign, summarising the available data, analysis method used, as well as conclusions revealed from the lens mass models. The relative astrometry and photometry of each system, the morphological parameters of the galaxies, as well as the gravitational lens mass models are compiled in Tables \ref{tab:lensastrometry}, \ref{tab:lensmorphology} and \ref{tab:massmodel}, respectively. The objects for which the interpretation of the data is problematic are discussed in Appendix \ref{section:appendobjects}.

\subsection{SDSS~J0743+2457 }\label{section:0743}

SDSS~J0743+2457 is a small-separation ($\theta\sim1''$) double discovered independently in the MUSCLES survey \citep{jackson12} and in SQLS \citep{inada14}. The source redshift is $z_s=2.165$, and the spectroscopic lens redshift is $z_s=0.381$ \citep{inada12}. Follow-up low-resolution observations were obtained in the $VRI$ bands \citep{inada14}, identifying the lens galaxy.

\begin{figure*}
\includegraphics[width=165mm]{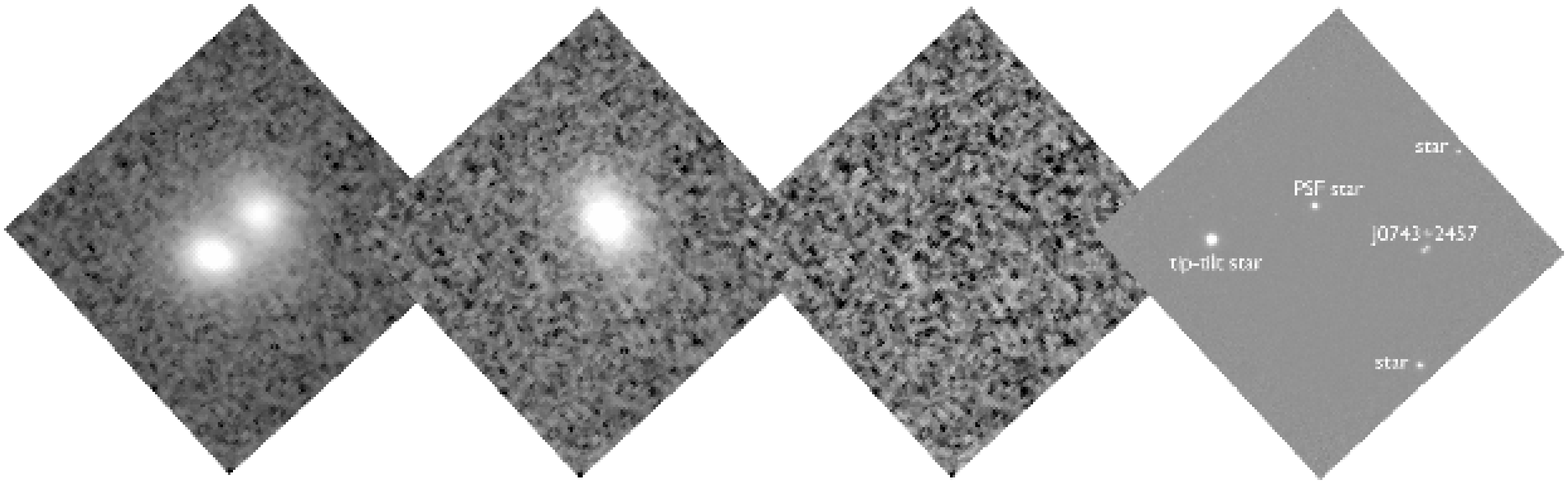}
\caption{Modelling of SDSS~J0743+2457. North is up and East is to the left. \emph{left}: original frame ($5''\times5''$), \emph{left center}: lens galaxy revealed after the subtraction of the quasar images. \emph{Right center}: residuals after the subtraction of all objects. \emph{Right}: FOV for SDSS~J0743+2457 ($66''\times66''$).
\label{fig:0743resid}}
\end{figure*}

The AO observations were performed in NGS mode, and there is a bright star in the FOV, closer to the NGS in relation to the system (Figure \ref{fig:0743resid}). This star was used as a PSF to perform simulations of the system. The system itself was modelled using an analytical PSF, which resulted in clean residuals. Image B is particularly close to the lensing galaxy, but the two can nonetheless be disentangled.


As the orientation of shear in the SIS$+\gamma$ model (Table \ref{tab:massmodel}) is significantly different from the light orientation of the lensing galaxy (Table \ref{tab:lensmorphology}), and there are no nearby perturbing galaxies in the FOV, the origin of the shear, although relatively small, cannot be determined. The shear orientation is vastly different in the discovery paper (by $\sim60$ deg), where no error bars are provided. The SIE$+\gamma$ model requires an unreasonably large amount of shear. We rule out that an inaccurate estimate of the galaxy morphology could bring the galaxy and shear orientation to match, as the $\sim55$ deg difference is very large, and the very large lens ellipticity would also facilitate the measurement of its orientation. 

\subsection{SDSS~J0819+5356}\label{section:0819}

The discovery of SDSS~J0819+5356 is reported in \citet{inada09}, where it was followed-up at low resolution in the $VRI$ bands. It is a large-separation ($\sim4''$) galaxy-scale lensed quasar at $z_s=2.237$, with spectroscopically estimated $z_l=0.294$. Even in the low-resolution data, \citet{inada09} notice an extended component C near A, and surmise it may be the quasar host galaxy.

\begin{figure*}
\includegraphics[width=145mm]{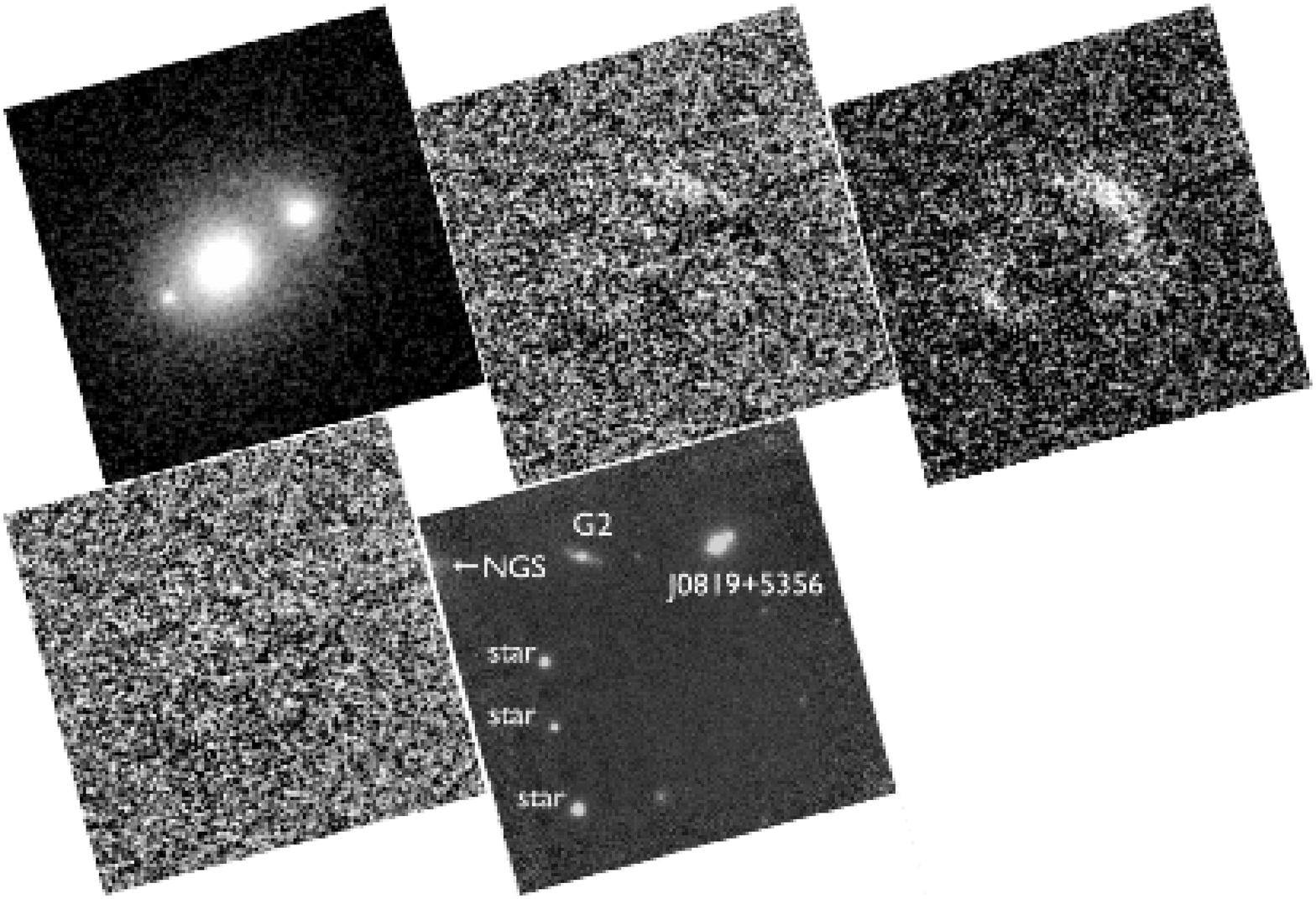}
\caption{Hostlens modelling for SDSS~J0819+5356. North is up and East is to the left. \emph{Top left}: original frame ($10''\times10''$), \emph{top centre}: residuals after modelling with an analytical PSF, without considering a host. \emph{Top right}: residuals after modelling with host galaxy, which is not subtracted.  \emph{Bottom left}: residuals after modelling with a host galaxy and subtracting all components.   \emph{Bottom right}: original $66''\times66''$ FOV. 
\label{fig:0819resid}}
\end{figure*}

Due to non-photometric observing conditions, the zero-point was calculated in the new NGS AO data using the stars in the FOV. Two of the stars have catalogue magnitudes with which the observed relative photometry is in agreement. The resulting photometric accuracy is dominated by the catalogue error bar of 0.1 mag. The system was modelled with an entirely analytical PSF, including the quasar host galaxy, which resulted in clean residuals. Simulations were performed using the star towards south in Figure \ref{fig:0819resid}, which is located at about the same distance from the NGS. The large error bars on the magnitudes of A and B (Table \ref{tab:lensastrometry}) reflect the uncertainty in the host-quasar decomposition. Systematics due to large Poisson noise were not found when the system was remodelled after masking the central 4 pixels of image A. The inferred quasar host properties are shown in Table \ref{tab:host}, where the fit was done for a host galaxy of unconstrained Sersic index, as well as $n=4$. Residuals are shown in Figure \ref{fig:0819resid}. As a note, if the system is modelled without a host, the morphology of the lensing galaxy changes dramatically (effective radius $\sim1.2''$, Sersic index $\sim5$). The companion galaxy G2 (Figure \ref{fig:0819resid}) is disk-like, according to the SDSS classification, and requires two Sersic components to model. 

The morphological parameters of the lens galaxy G1 are generally similar to the ones reported in \citet{inada09}, with the light orientation differing by 5 deg and the Sersic index slightly smaller. However the effective radius is half the value in the discovery paper. A possible reason could be that for this particularly extended object the sky background estimate is less reliable in the AO data. The wings of a steep Sersic profile are known to be particularly sensitive to the sky level. 

\begin{figure*}
\includegraphics[width=175mm]{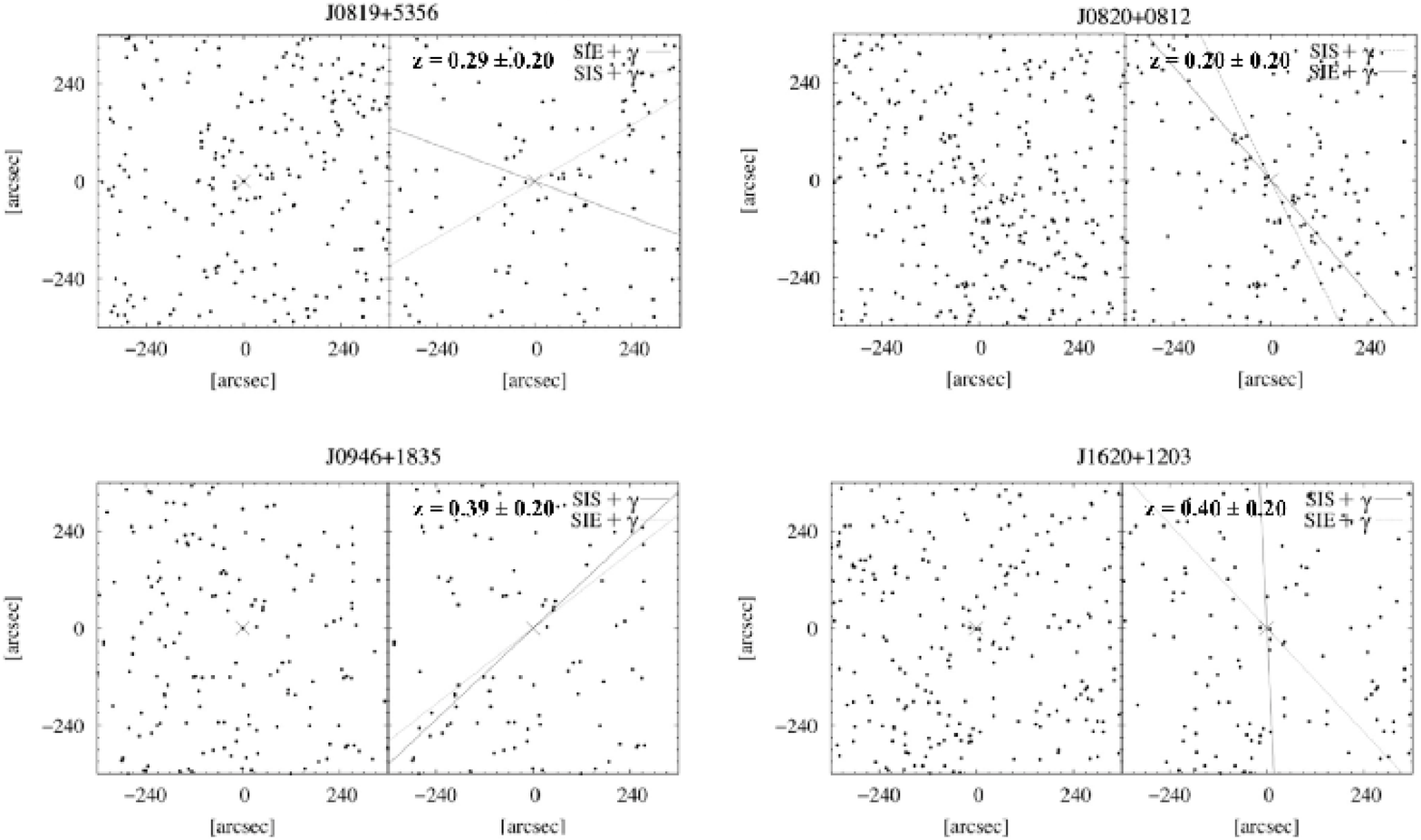}
\caption{SDSS environment $(left)$ and redshift cuts $(right)$ for selected systems. All galaxies kept after the cut have $i<21$ and photometric redshift error $\Delta z < 0.2$, ensuring accurate photometric redshifts \citep{csabai03}. The cuts satisfy at least one of the two redshift estimates in the SDSS database, PHOTOZ1 and PHOTOZ2 (see \protect\url{http://www.sdss3.org/dr10/algorithms/photo-z.php}).
\label{fig:environ}}
\end{figure*}

The shear direction in the SIS$+\gamma$ model is offset by $\sim$ 25 deg from the light orientation, suggesting external influence. The model parameters are very similar to the ones in the discovery paper, due to the large image separation which allows more robust astrometry. If accounting for G2 in the lens modelling, the shear direction changes by $\sim$ 16 deg, making it more aligned with the observed light orientation. Although the SDSS photometric redshift of G2 is consistent the redshift of G1, it is a disk-like galaxy, therefore its effect is expected to be smaller than inferred from the Faber-Jackson law. The value of the shear is not very large, particularly in the SIS$+\gamma$ model, and G1 is the brightest among the nearby galaxies, suggesting that it may be located close to the centre of a group. There is a group of galaxies (Figure \ref{fig:environ}) consistent with the redshift of G1, surrounding the system but concentrated towards west. This is consistent with the rough direction of the shear in the mass models. In addition, the large image separation/velocity dispersion of the lens may be boosted by the environment.

\subsection{SDSS~J0820+0812}\label{section:0820}

SDSS~J0820+0812 was observed without AO. The object was originally reported in \citet{jackson09}, as a two-image lensed quasar with separation $\sim2.3''$, with $z_s=2.024$ and $z_l=0.803$. The observations in that paper were so unreliable that no astrometry/lens modelling is published. 

\begin{figure*}
\includegraphics[width=175mm]{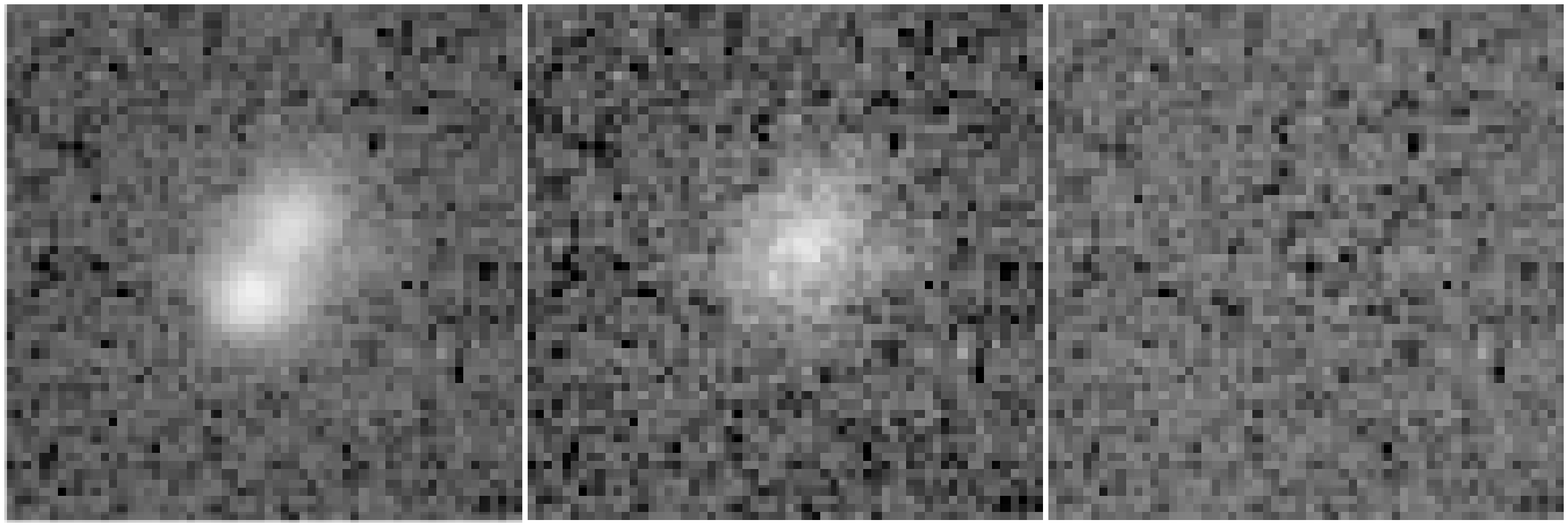}
\caption{Galfit/Hostlens modelling for SDSS~J0820+0812. \emph{Left}: original frame ($12.6''\times12.6''$, $4\times4$ binning); \emph{centre}: lens galaxy revealed after the subtraction of the quasar images. \emph{Right}: residuals after the subtraction of all objects.
\label{fig:0820model}}
\end{figure*}

In our new data, with the exception of the target, there are two galaxies and a star in the FOV. Due to the presence of the star in the FOV, the morphology of the system was fitted with Galfit in the traditional way. Because the original PSF FWHM ($\sim30$ pixels $= 1.35''$) is extremely oversampled, $2\times$ and $4\times$ pixel binning was performed. The photometry of all components and the effective radius of the galaxy are the same, but the positions, Sersic index, ellipticity and position angle (PA) differ for different binning. The results from $4\times$ pixel binning (final pixel scale $0.210''$, Figure \ref{fig:0820model}) were used, because the PSF for $2\times$ pixels binning is still largely oversampled, the S/N is lower, and the best-fit Sersic index $\sim0.25$ is unreasonably small. The sky was estimated in four different empty regions surrounding the target, the dispersion sigma of the values was calculated, and the sky was held fix at $\pm1\sigma$ when using Galfit. For each parameter, the statistical error reported by Galfit and the dispersion due to the sky was added in quadrature. The latter is dominant in most cases, including for the astrometry.

The 2MASS catalogue was used to calibrate the photometry. The $K-$band magnitude of the star is $15.63\pm0.21$, and $15.53\pm0.14$ for the galaxy. Since the error bar on the galaxy is smaller, and there is also the possibility that the star is variable, the galaxy aperture photometry was used for calibration. The magnitude of the star then becomes 15.87, marginally inconsistent with the 2MASS value. 

 A small Sersic value $\sim1$ is preferred for the lensing galaxy. However, if the Sersic index is fixed to 4, the difference is statistically small ($\Delta\chi^2/\mathrm{d.o.f.}=0.002$), and the residuals look the same. 
The flux ratio $2.03 \pm 0.13$ is significantly lower than reported in the discovery paper ($\sim6$). Even if the Sersic index 4 is used, it is no larger than $\sim3$. This is likely due to the poor quality of the data used in the discovery article. In addition, the flux ratio determined here is obtained in a redder band, and therefore less prone to quasar intrinsic variability, microlensing and extinction effects.

The large lens velocity dispersion $\sim 270$ km/s may be boosted by the lens environment, and indeed there are seven nearby bright galaxies in SDSS photometric redshift consistent with $\sim0.19$ (therefore in the foreground of the lens), surrounding the system. Most of the group members are located on the NE-SW direction. In addition, many more galaxies consistent with this redshift are found, particularly towards SW (Figure \ref{fig:environ}). This is consistent with the direction of the shear in both SIS$+\gamma$ and SIE$+\gamma$ models. Since the differences in shear position angle are small in the two models (however the error bars are very large), this suggests that the environmental contribution is the main source of shear, dominating over the lens internal ellipticity and exact mass profile.

\subsection{SDSS~J0832+0404}\label{sect:0832}

SDSS~J0832+0404 was reported in \citet{oguri08-3} as a double with $z_s=1.115$ and $z_l=0.659$, with available $V, I, K$ band images. 

\begin{figure*}
\includegraphics[width=175mm]{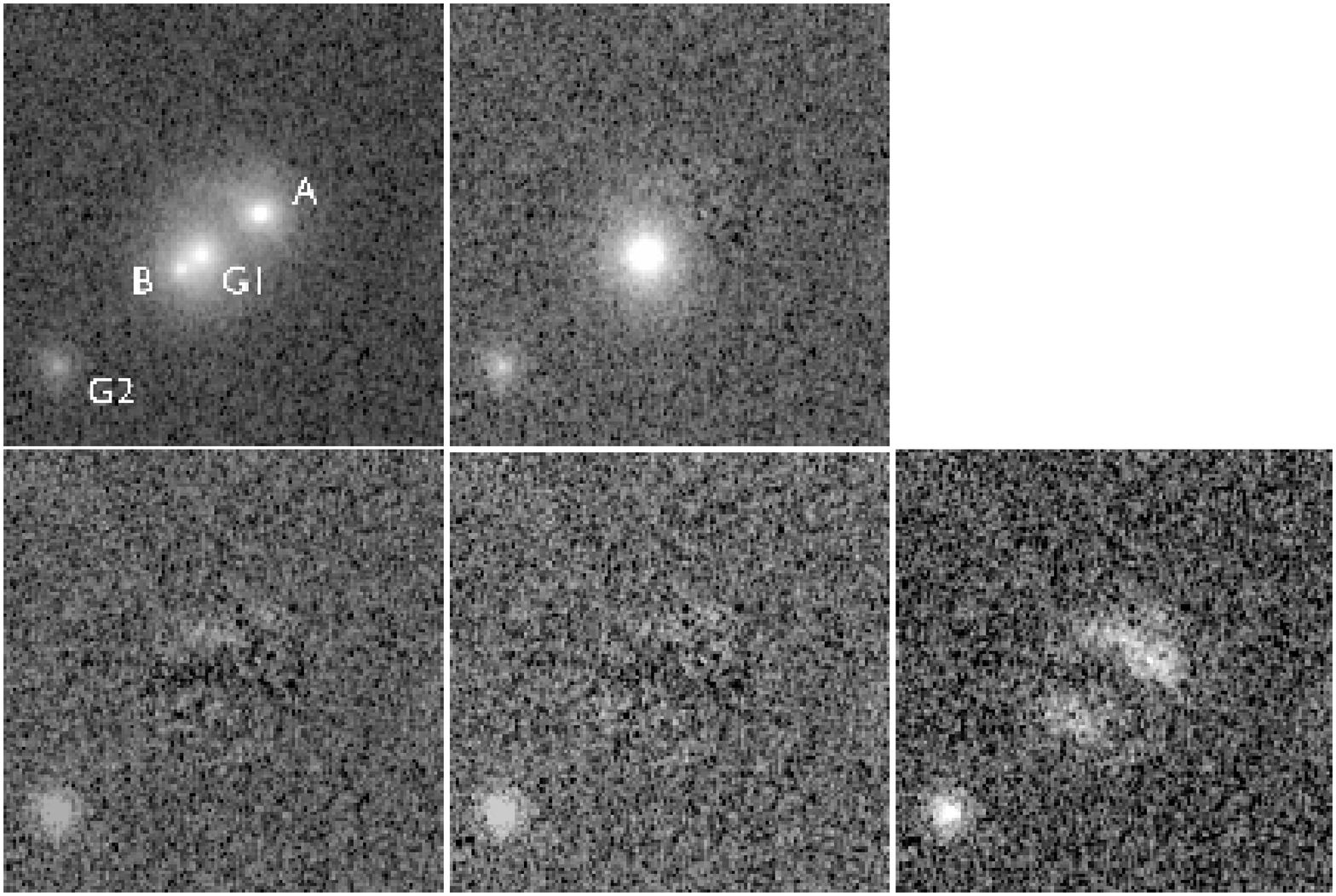}
\caption{Hostlens modelling for the AO observation of SDSS~J0832+0404. North is up and East is to the left. \emph{Top left}: original frame cut ($10''\times10''$). \emph{Top centre}: residuals after modelling the system, but without subtracting the lensing galaxy.  \emph{Bottom left}: residuals after modelling with an analytical PSF without fitting the host galaxy. \emph{Bottom centre}: residuals after subtracting the quasar images and host galaxy, using an analytical PSF.  \emph{Bottom right}: residuals after subtracting the lens galaxy and quasar point sources, but without subtracting the host galaxy.
\label{fig:0832AOresid}}
\end{figure*}

In the AO data, the presence of a faint arc-like feature towards East (Figure \ref{fig:0832AOresid}), which is significantly reduced if the quasar host galaxy is considered, indicates a host detection. As few residuals remain if the PSF is modelled analytically, a hybrid PSF was not considered, and therefore the system was modelled in a manner similar to SDSS~J0819+5356. In addition, since the overall AO correction is similar to that of the bright star in the FOV of SDSS~J0819+5356 (Table \ref{tab:analpsf}), that star was used to perform simulations. The parameters of the host galaxy are given in Table \ref{tab:host}. 

\begin{figure*}
\includegraphics[width=175mm]{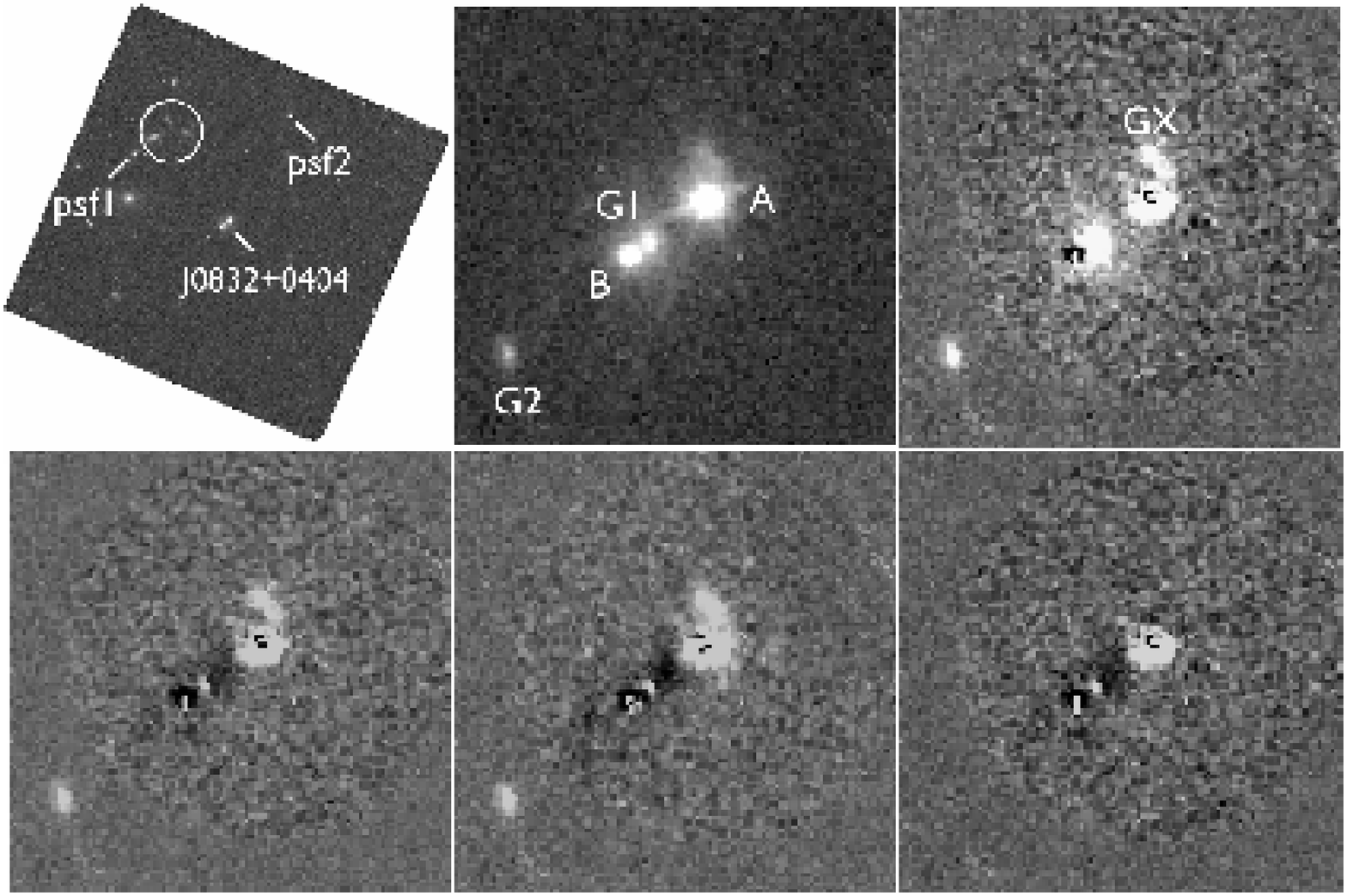}
\caption{Galfit/Hostlens modelling of the {\it HST} observation for SDSS~J0832+0404. North is up and East is to the left. \emph{Top left}: original frame cut ($75''\times75''$), showing the location of the two PSF stars, and a group of nearby galaxies ({\it circle}). \emph{Top centre}: original frame cut ($10''\times10''$). \emph{Top right}: residuals after modelling the system, but without subtracting the lensing galaxy. \emph{Bottom left \& bottom centre}: residuals after modelling the system with PSF1 \& PSF2, respectively. \emph{Bottom right}: residuals after subtracting G2 and GX as well.
\label{fig:0832HSTresid}}
\end{figure*}

This object is also available in archival HST observations (see Section \ref{sect:hst}). The $3\times400$ s observations were obtained with the Wide-Field Planetary Camera 2 \citep{holtzman95} on 2009 July 8. We performed standard reduction using Multidrizzle\footnote{Multidrizzle, part of the STSDAS package, is a product of the Space Telescope Science Institute, which is operated by AURA for NASA.} for distortion correction, cosmic ray removal, frame alignment and rotation. The reduced science frame is shown in Figure \ref{fig:0832HSTresid}. We used two stars in the FOV as PSF templates (Figure \ref{fig:0832HSTresid}), and modelled the system using Galfit. Compact residuals remain after the subtraction of all objects, particularly at the location of the lens and of image A. These may be due either to variations in the PSF across the field, or due to Poisson noise. We checked that compact residuals remain if, for example, the star PSF1 is modelled with the PSF2 template, in support of the first hypothesis. The fact that the residuals are compact and are not visible in image B suggest that they are not due to the quasar host galaxy, which appears to be undetected. This is not surprising even though the $HST$ observations are deeper, since the F606W filter falls shortward of the 4000 \AA\ rest-frame break of the host (assuming it is an early-type galaxy).

In addition to A, B and G1, an extended object GX is visible very close to image A, stretched tangentially to the location of the lens. 
This object, if indeed detected at all in the AO data, is much fainter, therefore has a blue colour. Its elongated morphology and position close to A suggest it is a lensed background object, although it cannot be ruled out that it is associated with the lens, without additional colour information. In the SIS$+\gamma$ model, the shear direction does not match the position of this object, making it unlikely to be associated with the lens. If the redshift of GX it smaller than $\sim1.2$, it is singly imaged, whereas if not, a counter image should be produced, given its proximity to the lens, of flux $\sim1/3$. Such image should likely be observable in the $HST$ data in the proximity of B, but is undetected.

There is a companion galaxy G2 at $\sim4.1\arcsec$ from the main lens, towards south-east. Assuming the Faber-Jackson law and that G1 and G2 are at the same redshift, it would be expected that G2 produces a shear of $\sim0.05$, however the measured shear assuming the SIS$+\gamma$ model is just $\lesssim0.02$, directed towards the foreground group of galaxies detected in SDSS and the $HST$ data towards north-east (Section \ref{sect:hst}). For this double, the most plausible explanation is that the there is an interplay between the effect of the intrinsic lens ellipticity, as well as the shear from the nearby companion and the nearby group, with the later being dominant. 

\subsection{SDSS~J0904+1512}\label{section:0904}

The discovery of SDSS~J0904+1512 was reported in \citet{kayo10}. It is a small-separation ($\sim1.1''$) double at $z_s=1.826$. The redshift of the lensing galaxy has not been spectroscopically determined, and estimates are discrepant. \citet{kayo10} find $z_l\sim0.2$ based on the Faber-Jackson method \citep{rusin03}, but $z_l\sim0.5$ based on the $V-I$ colour.  

\begin{figure*}
\includegraphics[width=165mm]{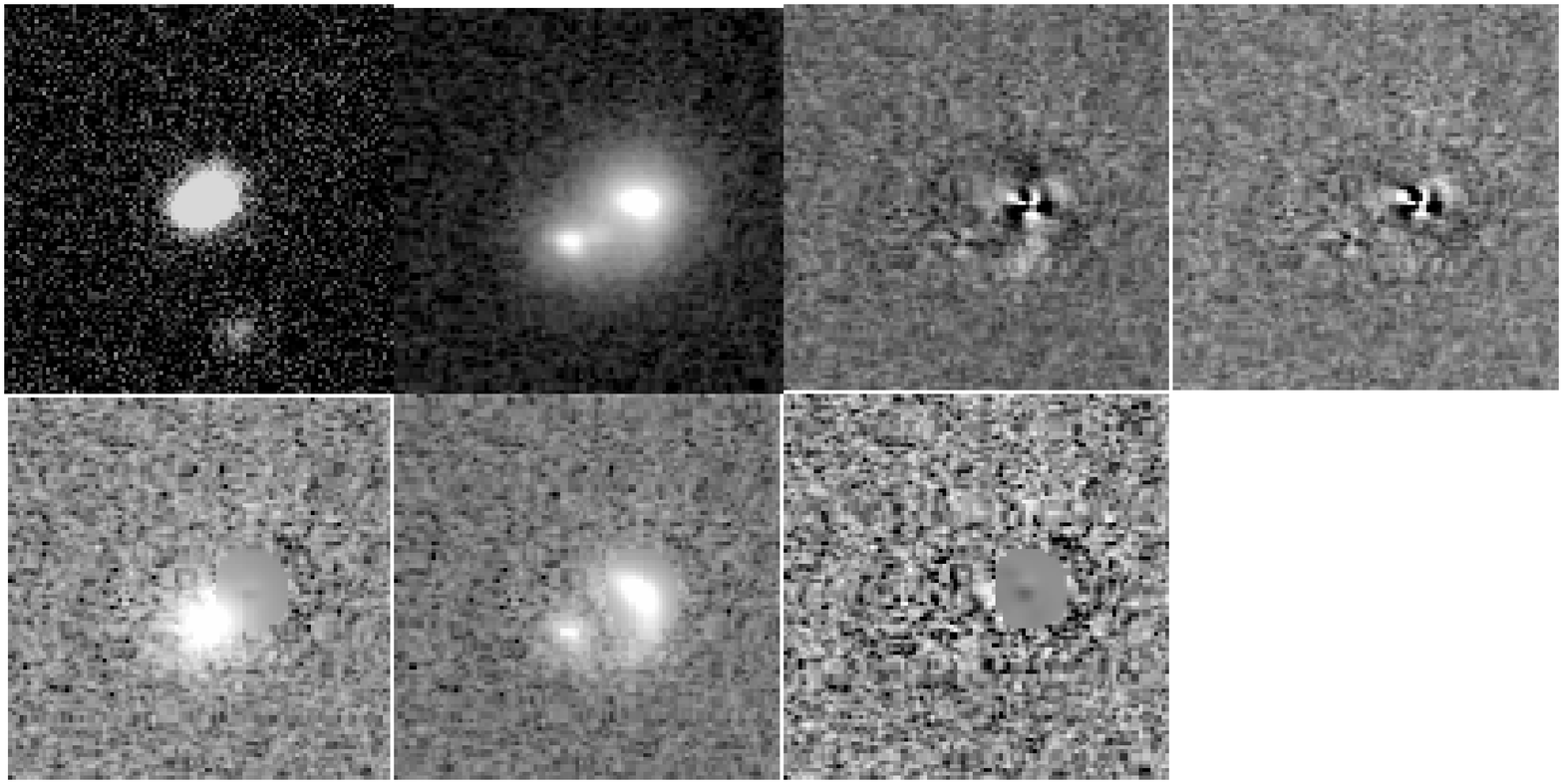}
\caption{Hostlens modelling for SDSS~J0904+1512. North is up and East is to the left. \emph{Top left}: original frame cut ($13''\times13''$), showing the location of G2 to the South. \emph{Top center left}: original frame cut ($5.1''\times5.1''$). \emph{Top center right}: residual after modelling with an analytical PSF, without fitting a host galaxy. \emph{Top right}: residual after modelling with an analytical PSF and fitting the host galaxy.  \emph{Bottom left}: residuals after subtracting the quasar images and host galaxy, using a hybrid PSF.  \emph{Bottom center}: residuals after subtracting the lens galaxy and quasar point sources. \emph{Bottom right}: residuals after subtracting all components. 
\label{fig:0904resid}}
\end{figure*}

Figure \ref{fig:0904resid} shows asymmetric residuals in the form of an arc (over-subtraction on one side, under-subtraction on another) at the location of A, oriented  tangentially to the critical curves. These are accounted for if a quasar host galaxy is considered. Since Figure \ref{fig:0904resid} shows non-analytical PSF residuals at the location of image A, these may affect the properties of the host galaxy fitted with Hostlens. Therefore an iterative approach was taken to construct a hybrid PSF which accounts for the non-analytical components, as described in Section \ref{section:hostfit}.  The hybrid PSF core consists of the observed pixel values (after subtracting the host contribution) in a circle of 10 pixel radius, centred on image A. This was chosen to be large enough to include the non-analytical PSF components, but as small as possible to leave outside a visible contribution from the host galaxy. The iteration was stopped once the $\chi^2$ stopped decreasing, in the third iteration. A similar iteration was also performed by fixing the Sersic index of the host galaxy at $n=4$. The resulting host parameters are written in Table \ref{tab:host}. Although it is very compact ($\sim1$ pixel effective radius in the source plane, but larger in the observed, magnified image plane), it is considered a true detection because of the visually confirmed arcs. In order to perform simulations, the image A cut produced in the last step of the iteration was used, with the host previously subtracted. This introduces additional noise in the simulations, however the image is quite bright. The same iterative procedure was applied to the simulations. The ellipticity of the lens galaxy converges towards zero and was therefore ignored. The nearby companion G2 (Figure \ref{fig:0904resid}, top left) was modelled with Galfit using the hybrid PSF.

This object shows the largest differences compared to the discovery paper. It is also one of the smallest separation systems in the AO sample, which might explain the discrepancies. In the discovery paper, A and B are only $\sim4$ and $\sim2.5$ pixels apart, on each axis, with FWHM $\sim 3.5$ pixels. The lens galaxy astrometry differs by $\sim$ 1.5 pixels on each axis from the AO data. In the AO data, the lens G1 appears clearly to be located closer to image B, as opposed to the discovery paper. While the effective radius and Sersic index are compatible, the discovery paper fits an ellipticity of $\sim0.5$, while in the AO data it is consistent with 0. 

Lens modelling was performed by assuming $z_l=0.3$. Converting the SIE model result in the discovery paper with the $e\sim3\gamma$ relation gives a shear of $\sim0.24$, the largest in this sample for a SIS$+\gamma$ model. The value using the AO astrometry is $\sim 0.08$, more in line with the other objects; the orientation of the shear is different from that in the discovery paper at $\sim60$ deg. This further supports the fact that the astrometry in the discovery paper is inaccurate. As the ellipticity was fixed at 0, mass models incorporating it are not used. The nearby galaxy G2, assuming it lies at the redshift of G1, would produce a shear of $\sim$ 0.03, or less than half of the value required by the SIS$+\gamma$ model. In addition, the shear direction is orthogonal to the direction towards G2. As a result, no conclusion can be drawn on the source of the shear, although the SDSS data shows that the system lies in a crowded environment.

\subsection{SDSS~J0946+1835}\label{section:0946}

The discovery of SDSS~J0946+1835 is reported in \citet{mcgreer10}. At $z_s=4.799$, it is the highest redshift lensed quasar known to date. The lens redshift has been determined spectroscopically to be $z_l=0.388$. The lens is a massive elliptical galaxy, causing a relatively large image separation $\sim3.1''$. Two faint components were discovered in the close proximity of the system, the bright one inconsistent with the colours of the images.

The object was observed twice in the $K'-$ band with AO, with the observation in 2012 being much deeper. This latter observation was performed in the 20 mas detector mode, with 9-point dither of separation $5''$. In order to increase S/N even more in the hope of the quasar host galaxy detection, $2\times2$ binning was performed. Due to the long exposure time, a significant number of cosmic rays were visible in each frame. These were removed automatically using the rejection-combine, however those in the immediate proximity of the target were manually removed from each individual frame in order to prevent any contamination. 

Figure \ref{fig:0946psf} shows that the separately observed PSF star cannot be used as a suitable PSF. Therefore, a hybrid PSF was used to model the system. The core component contains the observed light distribution within a circle of radius 12 pixels centered on image A. A larger hybrid core size was not considered, because it might be affected by inaccurate sky to the North of image A, which was located close to the edge of the field in several dither sequences. Simulations were done using the separate PSF star. The system was simulated with this PSF, but all objects were simulated 0.3 mag fainter, in order to produce about the same counts as the real system, since the separate PSF has slightly higher Strehl ratio.

\begin{figure*}
\includegraphics[width=175mm]{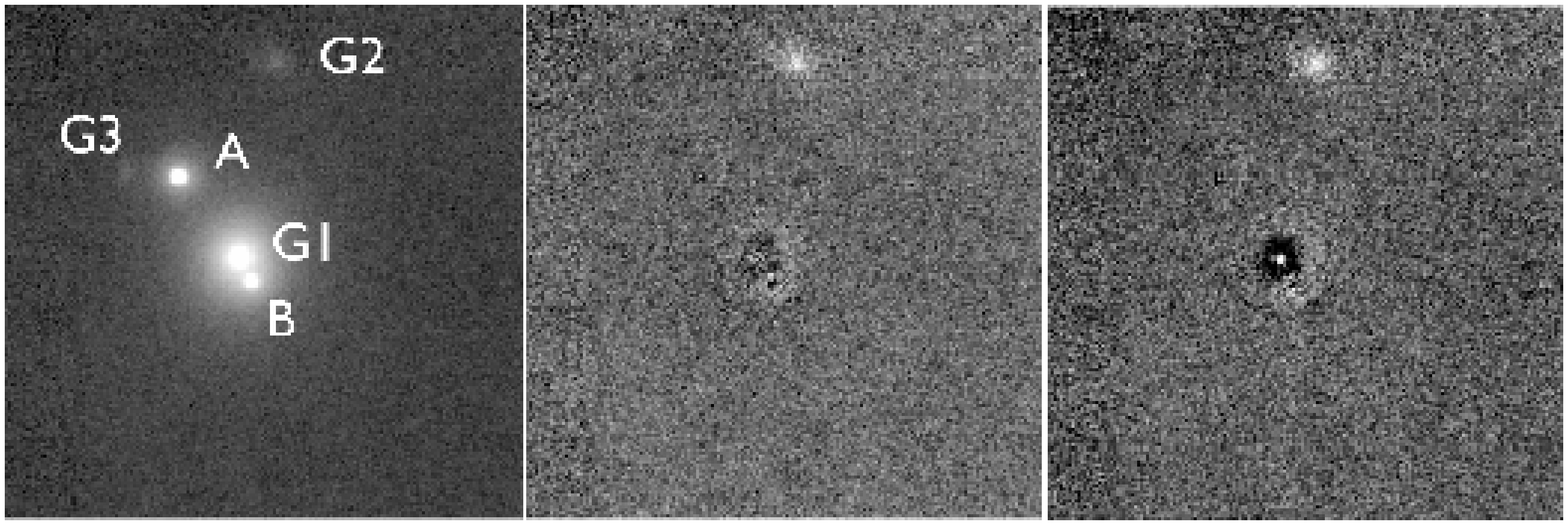}
\caption{Galfit/Hostlens modelling for SDSS~J0946+1835. North is up and East is to the left. \emph{Top left}: original frame ($12''\times12''$), \emph{top centre}: residuals after modelling with a free Sersic index and a hybrid PSF. \emph{Top right}: residuals after modelling with Sersic index 4. 
\label{fig:0946resid}}
\end{figure*}

The faint circular residuals seen in Figure \ref{fig:0946resid} around image B are due to masking when creating the sky frame, and not due to quasar host galaxy arcs. There are no visual signs of arc detection. A quasar host was fit using an analytical PSF to model the system, which resulted in a bright compact host $\sim$1/3 the flux of the point source, and the lack of visible arcs. It is likely that part of the quasar light is incorporated in the host, whose magnitude it therefore overestimated. This was tested by creating 100 simulated systems, using the separate PSF star, without a host, and fitted with an analytical PSF, both with and without a host. The difference between the reduced $\chi^2$ of the models with a host and without is $-0.0197\pm0.0014 \ (1\sigma$ of the scatter from the 100 values), whereas in the two corresponding models of the real system it is $\chi^2/\mathrm{d.o.f.}=63425.86/74147-64450.97/74152=-0.0138$. It is therefore very likely that the host detection is spurious. Since this is a very high-redshift quasar, the study of its host galaxy would be interesting. The fact that there is no clear detection is due to the cosmological surface brightness dimming and the low intrinsic magnification of this system.

Figure \ref{fig:0946resid} shows that the two faint components noted in the discovery paper are also detected in the AO data (G2 and G3). Both have elongated morphology, indicative of galaxies. Galaxy G2, which the discovery paper suggests to be similar in colour with G1, is 5 mag fainter than G1 in the new observations, and photometric redshift suggests that it is a faint local galaxy. For G3, there is no colour information to estimate a redshift.

The shear direction is consistent in the SIS$+\gamma$ and SIE$+\gamma$ models, but different at $\sim$ 35 deg from the orientation of the galaxy. The effect of G2 and G3 on the lens model is small, modifying the direction of the estimated shear by $\sim 3$ deg and $\sim 1$ deg, respectively. SDSS imaging (Figure \ref{fig:environ}) shows that there is a group of galaxies in the proximity of the system, towards NW, that are consistent with the lens galaxy redshift. The direction of the shear is consistent with this direction. In addition, the estimated shear values are quite large, consistent with the existence of nearby structure. In conclusion, the lensing models are well explained by the observed environment, which dominates over the contribution of the intrinsic ellipticity of the lensing galaxy.

\subsection{SDSS~J1001+5027}\label{section:1001}

SDSS~J1001+5027 was reported in \citet{oguri05}, based on observations in the $V, R$ and $I$ bands, as a lensed quasar with two images of intermediate separation $\sim2.9''$, and with a second galaxy in very close proximity to the main lens. The redshift of the main lens is determined spectroscopically in \citet{inada12} to be 0.415, and the source redshift is $z_s=1.839$. The time delay between the two images is measured by \citet{rathna13} to be $119.3\pm3.3$ days.

\begin{figure*}
\includegraphics[width=165mm]{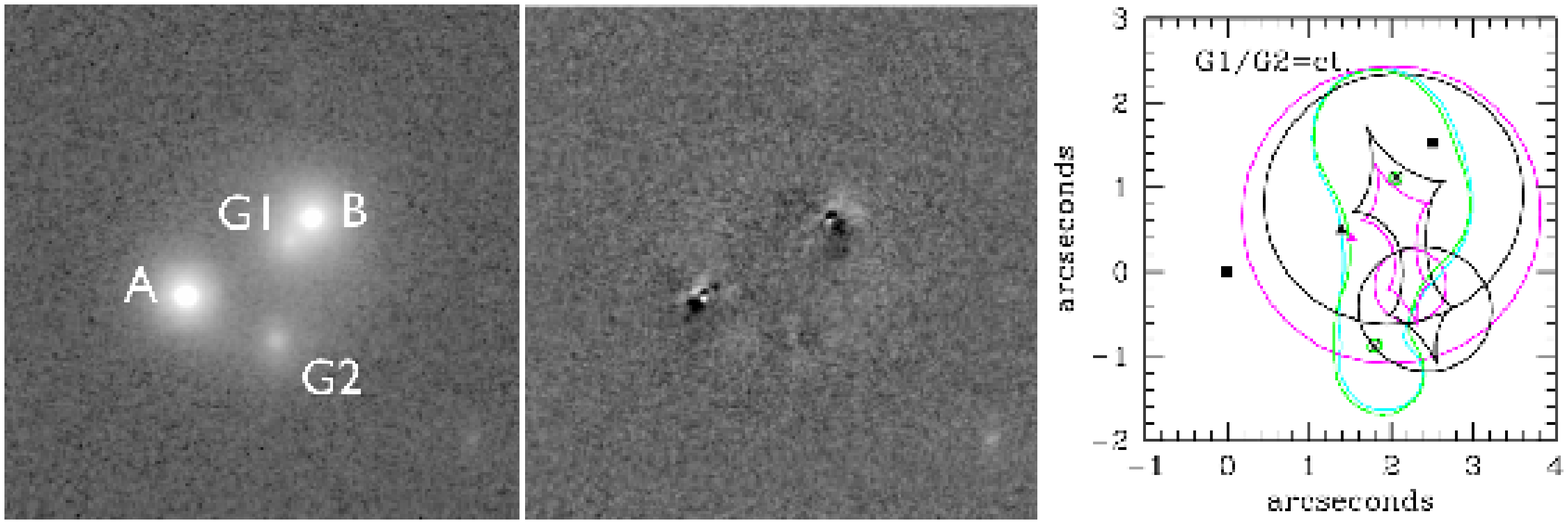}
\caption{Galfit/Hostlens modelling and lens model for SDSS~J1001+5027. North is up and East is to the left. \emph{Left}: original frame ($10''\times10''$), \emph{centre}: residuals after modelling with an analytical PSF.  \emph{Right}: Lens models: critical curves / caustics are shown in cyan and magenta for the 2SIS$+\gamma$, green and black for a 2Sersic$+\gamma$ model (which assumes mass fallows light), respectively. Squares mark the positions of the images, crosses mark the position of the lenses, and triangles mark the position of the source. \it{A colour version of this figure is available in the online version of the paper.}
\label{fig:1001resid}}
\end{figure*}

For this object, image A is isolated enough in the AO data to be used as a PSF. However, this leaves more residuals than usual at the location of B and G1, indicating that image A is not a very good approximation of the PSF at the location these components. Therefore the decision was made to model the system with the analytical PSF (Figure \ref{fig:1001resid}).

For the simulations, the separate PSF star observed with this system was a natural choice, however it has a smaller FWHM than the real system, producing objects that are too sharp. Therefore the system was simulated using an actual cut around image A as PSF. This likely overestimates the error bars, as noise is added to an already noisy PSF.

\citet{oguri05} noted that the two lensing galaxies of SDSS J1001+5027 have colours consistent with those of early-type galaxies at 0.2 $\leq$ z $\leq$ 0.5.  However, in the AO $K'$ band G2 appears significantly fainter compared to G1, signifying that the two galaxies (both ellipticals, as suggested by the value of the best-fit Sersic index) may not be located at the same redshift.  

As time delays are measured for this system, the mass models can have an extra parameter compared to the others used in the present paper (for d.o.f. = 0). Due to its proximity to the system, G2 is expected to have an important effect on the observed configuration. Therefore it was taken into consideration as an SIS when modelling the system. For G1, it was proceeded as follows:

 \begin{itemize}
    \renewcommand{\labelitemi}{$\bullet$}
    \item The fiducial SIS$+\gamma$, SIE, SIE$+\gamma$ models were used for G1, but the Hubble constant $h$ was allowed to vary as a free parameter.
    \item As above, but $h$ was fixed at the value measured by \citet{planck13}, and instead the relative masses of G1 and G2 were no longer constrained by the Faber-Jackson relation.
    \item $h$ and the relative strengths of G1 and G2 were fixed, but models were considered with one of the following two changes: instead of the singular isothermal profile, a power law profile of free slope ($\gamma'=2$ in the case of the singular isothermal profile) was considered for G1; in addition to shear, external convergence was also considered at the location of G1.
\end{itemize}

The critical curves and caustics for two of the models above are depicted in Figure \ref{fig:1001resid} (bottom right). Of course, the real model would be a combination of the models considered above, but such a model would be significantly under-constrained. Therefore these models were considered separately.

\begin{figure*}
\includegraphics[width=175mm]{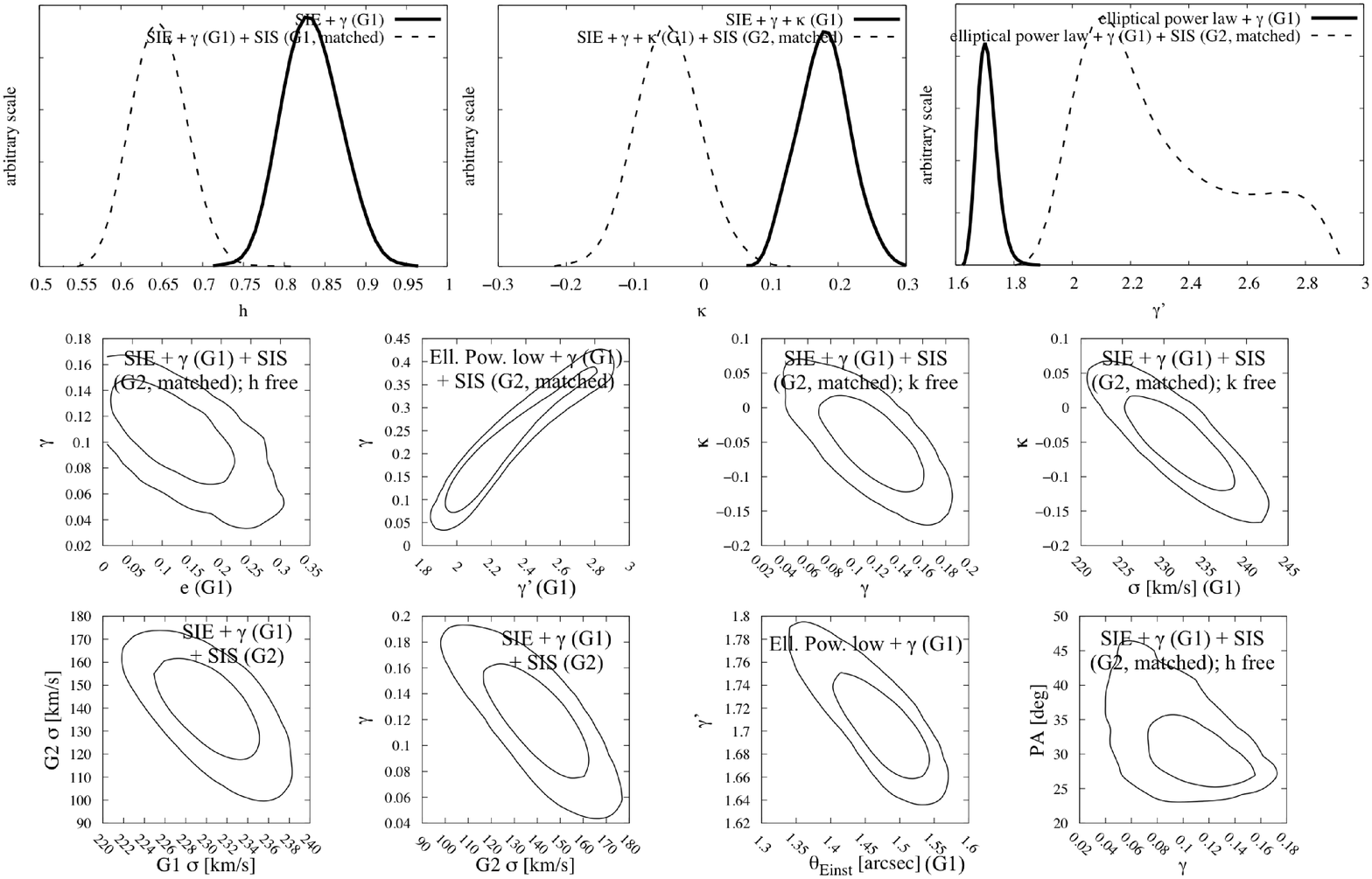}
\caption{1D and 2D selected lens model plots for SDSS~J1001+5027, created with Glafic and MCMC. The 2D contours represent the $1\sigma$ and $2\sigma$ regions.
\label{fig:1001degen}}
\end{figure*}

The one-dimensional probability distributions for the G1 density slope $\gamma'$ and for $h$, in Figure \ref{fig:1001degen}, show that models which account for the effect of G2 require a relatively larger $h$ and less steep slope. Also, from the two dimensional probability contours, among others, the following conclusions can be drawn, where all other parameters have been marginalised over:

 \begin{itemize}
    \renewcommand{\labelitemi}{$\bullet$}
    \item models with steeper mass profiles require less mass and more external shear;
     \item models with more external convergence require less mass and less external shear;
          \item models with more massive G2 require less massive G1 and less external shear;
\end{itemize}

The most interesting result is that all spherical mass models considered have similar shear orientation, regardless of whether G2 is modelled or not. The shear direction in the SIS$+\gamma$ model differs from that of the observed light profile of G1 by only $\sim20$ deg. However the light profile of this galaxy is not elliptical enough for the quadrupole to account for the large measured shear. This suggests strong environmental influence, dominant over both the internal shear generated by G1 and the influence of G2. \citet{oguri05} does report the existence of a group of galaxies \citep[Figure 8 in][]{oguri05} consistent with $z\sim0.2$ towards north-west, however this is roughly perpendicular to the direction of the shear. 

\begin{figure}
\includegraphics[width=80mm]{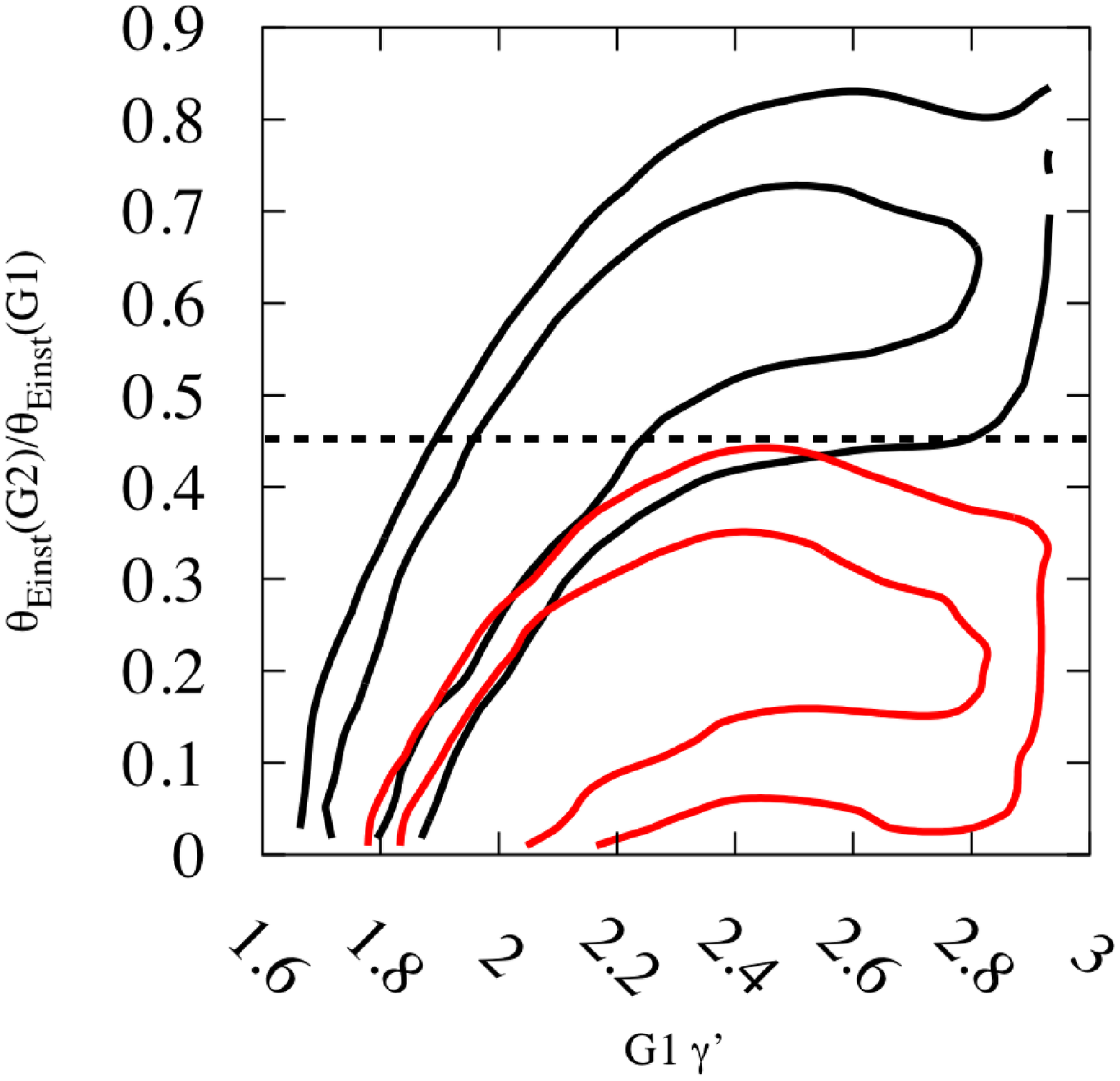}
\caption{Degeneracy for SDSS~J1001+5027 in the case convergence $\kappa=0$ (black) and $\kappa=0.15$ (red). An elliptical power low profile $\ +\gamma + \kappa$ (G1) + SIS (G2) mass model was used, with the relative contribution of G2 free to vary, but with fix $\kappa$ (d.o.f.$=-1$).   The contours represent the $1\sigma$ and $2\sigma$ regions. The horizontal line shows the relative ratio of the Einstein radii as inferred from the Faber-Jackson relation. \it{A colour version of this figure is available in the online version of the paper.}
\label{fig:1001k}}
\end{figure}

Another fact suggesting that the influence of G2 is not strong is that, if it is considered, not just small convergence, but negative convergence is required to reproduce the system in the SIS/SIE (G1) + $\gamma + \kappa$ + SIS (G2) case. This is peculiar, since the SIS profile produces $\gamma=\kappa$, and convergence is a positive scalar quantity, whereas shear is a headless vector (invariant under
rotation by 180 deg); therefore it is expected that $\gamma<\kappa$, assuming the shear is caused by the added contributions from multiple SIS perturbers. On the other hand, negative convergence does exist for regions of space that are relatively void. This however is not the case, as is shown by the existence of the nearby group of galaxies in the foreground. Comparing the one-dimensional probability distributions for convergence, it is seen that $\gamma\sim\kappa$ if the contribution of G2 is ignored. Also, a weaker G2 reduces the mass slope of G1, making it consistent with an isothermal profile (Figure \ref{fig:1001k}). 

Therefore, the models and the observed environment strongly suggest that G2 should have a small effect on the system, corresponding to either a different redshift from G1, or a mass much smaller than suggested from the relative brightness of G1 and G2, and the Faber-Jackson relation. Note that \citet{oguri05} measure even closer brightness for the two galaxies at shorter wavelengths, and therefore without the AO data the discrepancy would be even larger. There are two explanations for this result:
 \begin{itemize}
    \renewcommand{\labelitemi}{$\bullet$}
    \item The redshift of G2 is indeed different, as suggested from the different colours of the two galaxies in the visible and at $K'$ band. Therefore, its gravitational effect on the system is reduced.
     \item Alternatively, the proximity of G1 and G2 could suggest that G2 is tidally stripped by the larger galaxy, which reduces its mass without changing the overall luminosity of the galaxy \citep[e.g.,][]{treu09}.
\end{itemize}

This appears to be a complicated system, where the measured time delay cannot be straightforwardly converted into a measurement of $h$. A spectroscopic redshift of G2 is required for additional insight, as well as field spectroscopy to understand the origin of the large required shear.

\subsection{SDSS~J1055+4628}\label{section:1055}

 This object was reported in \citet{kayo10} as a small-separation ($\sim1.1''$) double at $z_s=1.249$. \citet{inada12} give a spectroscopic lens redshift of $z_l=0.388$. \citet{kayo10} also identifies two nearby galaxy companions and notes that the one to the North has similar colours to the main lens, but the one to the South is much bluer.

\begin{figure*}
\includegraphics[width=165mm]{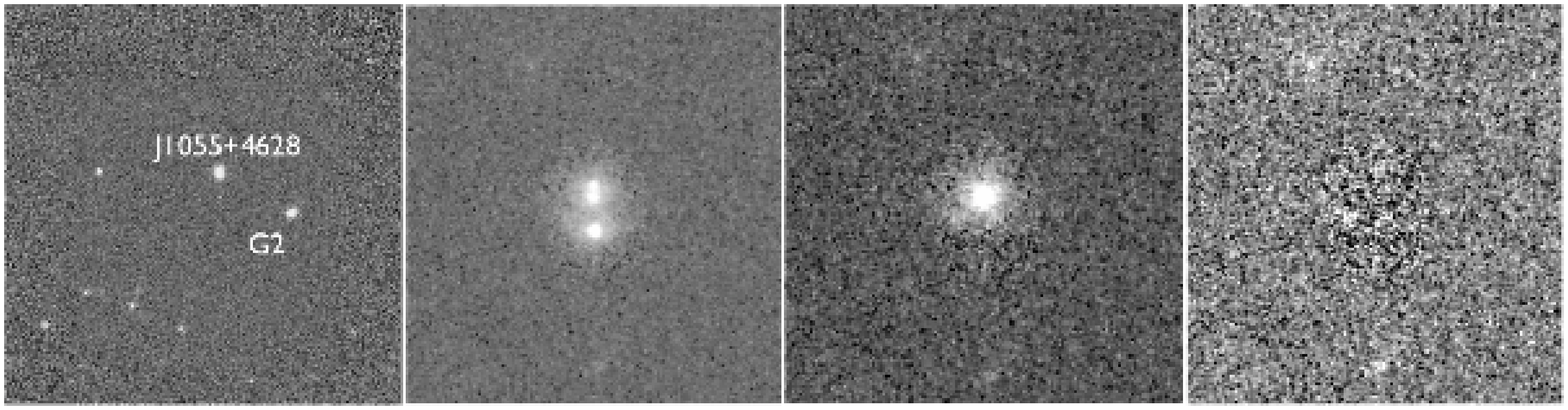}
\caption{Galfit/Hostlens modelling for SDSS~J1055+4628. North is up and East is to the left. \emph{Left}: original frame ($66.5''\times66.5''$). \emph{Centre left}: original frame ($10''\times10''$).  \emph{Centre right}: residuals after modelling with a free Sersic index and an analytical PSF; galaxy is not subtracted. \emph{Right}: residuals after subtracting all components. 
\label{fig:1055resid}}
\end{figure*}

In the AO data, an entirely analytical PSF was employed due to the low S/N, with only 8 frames of 60 s exposure obtained for this objects. Simulations were performed with the PSF star observed for SDSS~J1405+0959. There are features around image A that may be interpreted as arcs due to the lensed host galaxy (Figure \ref{fig:1055resid}, bottom right), however due to the low S/N and lack of improvement in the residuals, a host galaxy fit is not reported. The possible presence of arcs, which are unaccounted for, makes this a less reliable object to explore with the present technique. For comparison, the morphological parameters of the lensing galaxy are all consistent within $\sim1\sigma$ with the discovery paper, except for the lens ellipticity, which is significantly larger in the AO data. 

The companion galaxies from the discovery paper, although visible, appear very faint in these shallow observations. The magnitude of G2 (Figure \ref{fig:1055resid}, top left) was measured with aperture photometry, since it is too distant from the system to have a reliable PSF.

In the SIS$+\gamma$ model, the required shear is small, and its orientation is different at $\sim 34$ deg from the light orientation. The shear increases significantly for the SIE$+\gamma$ model, with a large change in the position angle. The effect of considering G2 is to change the position angle of the shear by $\sim 3$ deg. This effect is expected to be even smaller, since the SDSS photometric redshift of G2, $z\sim0.3$, is lower than the spectroscopic redshift of G1. The effect of the two nearby galaxies identified in the discovery paper is not considered. In addition to being very faint, they are significantly offset from the direction of the shear in all three models considered, therefore they are unlikely to be important in the mass models. No environmental clues for the origin of the shear are found in the larger-scale SDSS environment.


\subsection{SDSS~J1131+1915}\label{section:1131}

This object was reported in \citet{kayo10}. It is a small-separation ($\sim1.5''$) double at $z_s=2.915$. A spectroscopic lens redshift is not available, but the use of the Faber-Jackson relations \citep{rusin03} provides an estimate of $z_l\sim0.3$. 

\begin{figure*}
\includegraphics[width=165mm]{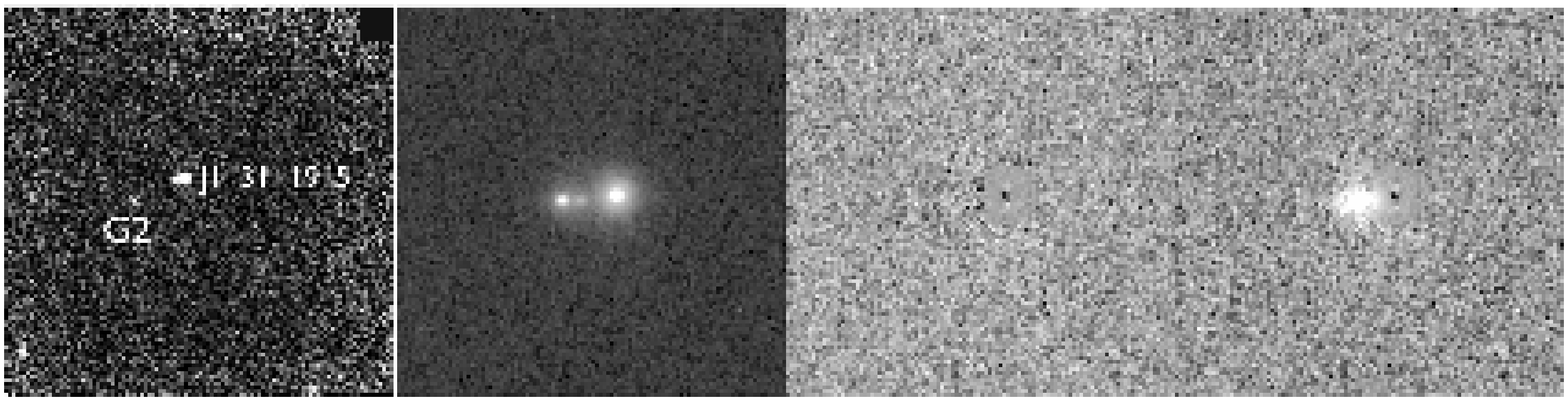}
\caption{Galfit/Hostlens modelling for SDSS~J1131+1915. North is up and East is to the left. \emph{Left}: complete FOV ($66''\times66''$).  \emph{Centre left}: original frame ($10''\times10''$), \emph{centre right}: residuals after modelling with an analytical PSF. \emph{Right}: the lens galaxy is kept, while the quasar images are subtracted.
\label{fig:1131resid}}
\end{figure*}

Using the AO data, a hybrid PSF was created from image A. Two different core radii, of 15 and 30 pixels, were used. As this is a bright system, simulations were produced with a cut around image A itself. For the nearby galaxy G2 (Figure \ref{fig:1131resid}), astrometry and aperture photometry were measured with the IRAF PHOT task. 

The astrometry is highly incompatible with the one in the discovery paper. In particular, the lens galaxy is closer to the brighter image in the original data, and to the fainter image in the AO data The problem lies with the original low-resolution, large seeing observations and not the AO data, because components B and G1 are clearly resolved in the AO imaging. The direction of shear in the SIS$+\gamma$ model is consistent with the observed light orientation to within 7 deg, suggesting that the intrinsic lens ellipticity is the main source of shear. The lens model in the discovery paper is highly inconsistent, due to the vastly different astrometry. Scaling its relative velocity dispersion by the Faber-Jackson relation, nearby galaxy G2 is expected to change the direction of the shear in the SIS$+\gamma$ model by 18 deg. However this effect would be smaller if G2 is located at a different redshift.


\subsection{SDSS~J1206+4332}\label{section:1206}

The discovery of SDSS~J1206+4332 is reported in \citet{oguri05}, as an intermediate-separation ($\sim 2.9''$) double at $z_s=1.789$. Time delay was measured in \citet{eulaers13} to be $111.3\pm3$ days. The system appears to have three lensing galaxies where the main lens, G1, has an associated absorber at redshift $z_l = 0.748$. The second galaxy, G2, is identified as a high redshift galaxy at $z \geq$ 0.7 and, the third galaxy G3, is blue and may be a chance superposition of a foreground galaxy.

\begin{figure*}
\includegraphics[width=150mm]{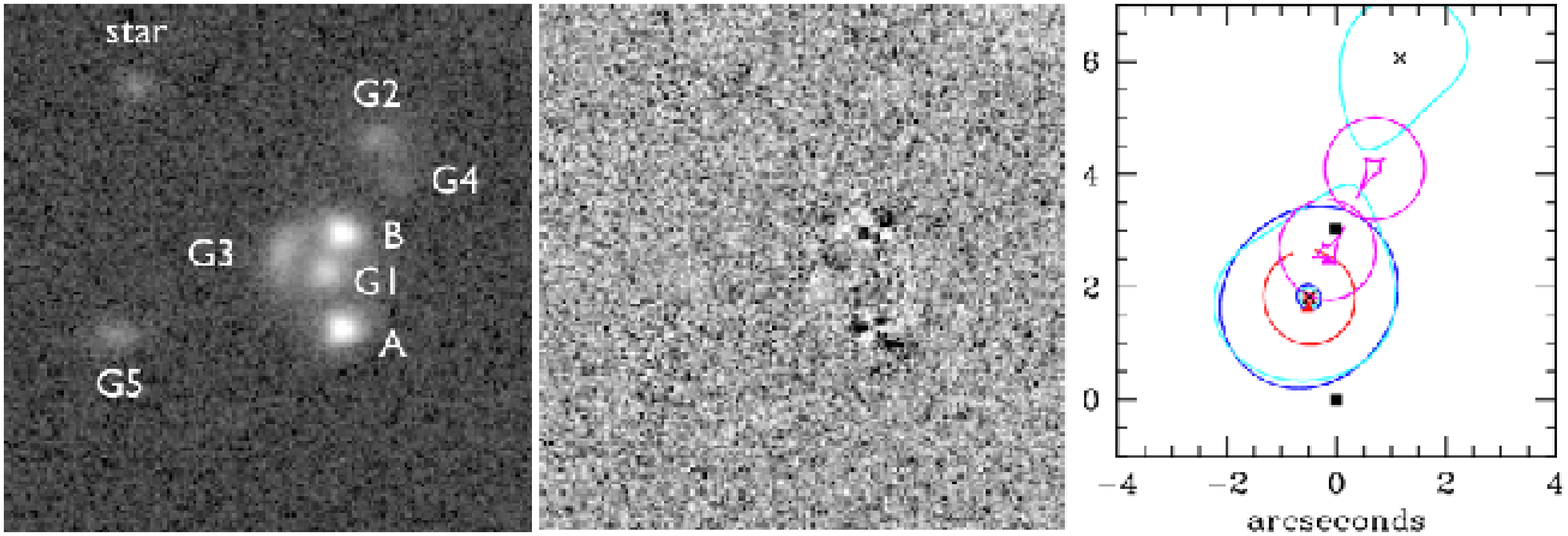}
\caption{Galfit/Hostlens modelling and lens model for SDSS~J1206+4332. North is up and East is to the left. \emph{Left}: original frame ($17''\times17''$); \emph{centre}: residuals after modelling with an analytical PSF. \emph{Right}: Lens models: critical curves are shown in blue, and caustics in red for the 1SIS$+\gamma$ model, cyan and magenta for the 2SIS$+\gamma$ model, respectively. Squares mark the positions of the images, a cross marks the position of the lens, and triangles mark the position of the source. \it{A colour version of this figure is available in the online version of the paper.}
\label{fig:1206resid}}
\end{figure*}

Since the PSF is severely oversampled in the non-AO observations, $2\times2$ pixel binning was performed. As there is no star in the FOV, the PSF was built analytically. When modelling the system with Hostlens, only A, B, G1 and G3 (Figure \ref{fig:1206resid}) were considered, and the other objects were masked. Despite the poor seeing, a cut around image A, after analytically removing the contribution of other objects, was successful. Using this cut as PSF however produces significantly more residuals at the location of image B, and subsequently it is discarded. Nonetheless, in the absence of other suitable non-AO objects to use for simulation, the cut around image A itself was used to simulate the system, and the resulting simulations were fitted in the same way as the real system.

In addition to the known galaxies, a new one (G4) was identified to the South of G2, which explains the peculiar morphology of G2 in the deconvolved images from \citet{eulaers13}. In addition, another galaxy G5 was modelled, further away from the system. Several of the morphological parameters of the model galaxies were marked by Galfit as unreliable, and therefore no morphological parameters are reported. However, G1 appears to have a large Sersic index $\sim10$, and G3 is also modelled with a large index, whereas the other galaxies appear to require low values. Also, G1 seems to have a large ellipticity of $\sim0.7$, with the major axis at $\sim-70$ deg. Given the quality of the data, we do not consider mass models incorporating the observed ellipticity, and we also omit error bars on the estimated mass models. 


This system is similar to SDSS~J1001+5027, in the sense that a time delay measurement is available, but it is even more complicated, due to the presence of more galaxies. If the morphology is to be trusted, G1 and G2 are early-type galaxies, and it is likely that they are physically associated, since they have similar colours. They are therefore expected to dominate the mass models. \citet{oguri05} obtain that G2 is indeed required in order to reproduce the observed configuration, but based on the new data G1 is enough to produce a perfect-fit SIS$+\gamma$ or SIE model. The counterpart model of the one in the discovery paper, based on the new data (G1 as SIE with G2 as SIS), produces similar results. 

According to the models summarised in Table \ref{tab:massmodel}, whether or not G2 is considered (scaled using the Faber-Jackson relation), the shear position angles are in relative agreement with the tentatively measured light orientation (within $\sim20$ deg), suggesting that the intrinsic mass ellipticity is likely to dominate the quadrupole. The strength of G3 is constrained by the number of observed images, and modelling it with SIS at the same redshift as G1 puts an upper limit of G3/G1 $\lesssim0.3$ (compared to $\lesssim0.7$ from the Faber-Jackson law) on its velocity dispersion in relation to G1 (in a model where G2 is also considered). This is in agreement with the different colours of G1 and G3, in the discovery paper. 

Analogous to the models considered for SDSS~J1001+5027, we obtain that G1 should have a slope of $\sim1.8$ assuming a power law mass model, or alternatively, for an SIS profile, convergence $\kappa\sim0.15$ or Hubble constant $h\sim0.8$. Models with or without G2 are more similar to each other in terms of these results than in the case of SDSS~J1001+5027, and in particular the required convergence is positive. The critical curves and caustics for two representative models are shown in Figure \ref{fig:1206resid} (bottom right).

The observed flux ratio in $K'-$band is $\sim1$, whereas it is $\sim1.5$ both in \citet{oguri05} (at smaller wavelength), and using the method which estimated a time delay in \citet{eulaers13}. This value has a small effect on the results in Table \ref{tab:massmodel}.

As a final remark, during the review process of this paper, we have become aware of the analysis presented for this system by \citet{agnello15}. Based on higher quality adaptive optics imaging and additional spectroscopy, they obtained improved morphology results and were able to show that G3 is in fact a lensed image of the quasar host galaxy. 

\subsection{SDSS~J1216+3529}\label{section:1216}

SDSS~J1216+3529 was reported in \citet{oguri08-3} as a small-separation ($\sim1.5''$) double at $z_s=2.012$. Based on the galaxy magnitude and colours, a redshift $z_l\sim0.55$ was estimated. A secondary galaxy was identified close to the system.

\begin{figure*}
\includegraphics[width=165mm]{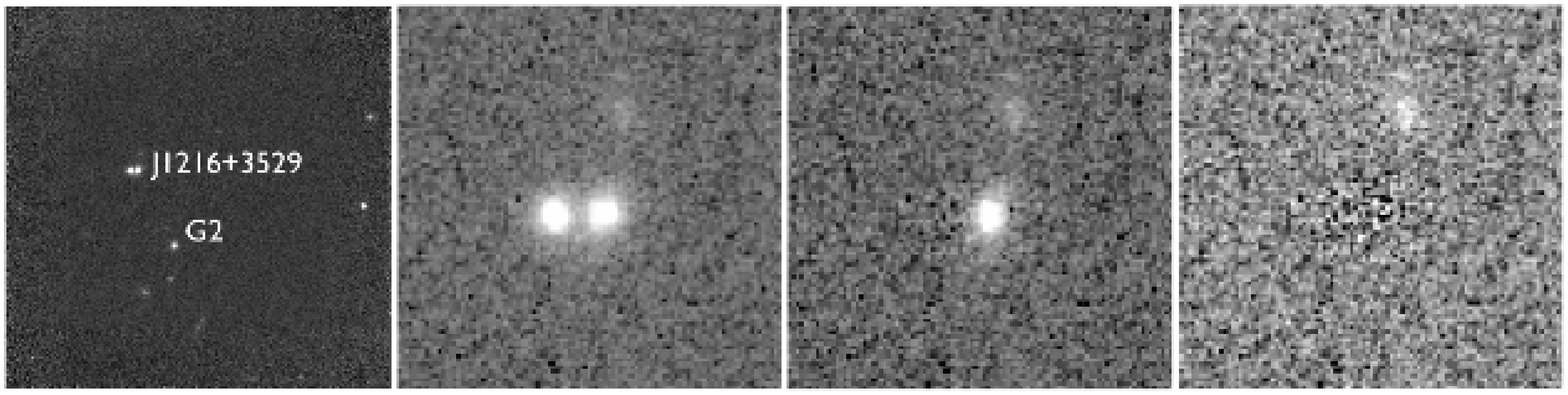}
\caption{Galfit/Hostlens modelling for SDSS~J1216+3529. North is up and East is to the left. \emph{Left}: complete FOV ($65''\times65''$). \emph{Centre left}: original frame ($10''\times10''$), \emph{centre right}: residuals after modelling with an analytical PSF; the lens galaxy is not subtracted. \emph{Right}: residuals after modelling with an analytical PSF and subtracting all components.
\label{fig:1216resid}}
\end{figure*}

This object has a very poor AO correction (PSF FWHM $\sim0.46\arcsec$), and only 9 frames of 60 s exposure were obtained. As a result, $2\times2$ pixel binning was used. The system was modelled with an analytical PSF, and simulated with the separate PSF star from SDSS J1001+5027, appropriately binned and convolved to match the observed FWHM of the system PSF. The magnitude and position of the more remote, brighter galaxy G2 (Figure \ref{fig:1216resid}, top left) was obtained with the IRAF PHOT task. For the closer, fainter component G3 measurements were made using the same analytical PSF.

Here, similar to the discovery paper, G3, whose magnitude is unreliable and photometric redshift is uncertain, was ignored in the lens models. If physically associated with G1, its effect on the lens model would be drastic however. modelling only G1, the position angle of the shear is in agreement within 7 deg with the light orientation of G1. In addition, the SIS$+\gamma$ shear direction is also close to the direction to G3, within 10 deg. It is therefore likely that there is a degenerate contribution of the effect of G3 and intrinsic ellipticity of G1 for this system. As a note, G2, whose photometric redshift is consistent with G1, would change the estimated shear direction by 6 deg.

\subsection{SDSS~J1254+2235}\label{section:1254}

This is a small-separation $\sim1.6''$ double \citep{inada09} at high redshift ($z_s=3.626$), with a disk-like lens at $z_l\sim0.3$, based on its $R-I$ colour.

In order to eliminate the residuals in the AO data, and obtain a suitable fit, two concentric Sersic profiles were required for the lensing galaxy. These are likely representative of the bulge and disc components of late-type galaxies. The orientations of these components are relatively misaligned by $\sim$ 35 deg. modelling was performed without using a hybrid PSF, because the proximity of the objects and the use of two Sersic profiles are likely to affect the construction of the hybrid PSF. The system was simulated using the separately observed PSF star. 

\begin{figure*}
\includegraphics[width=165mm]{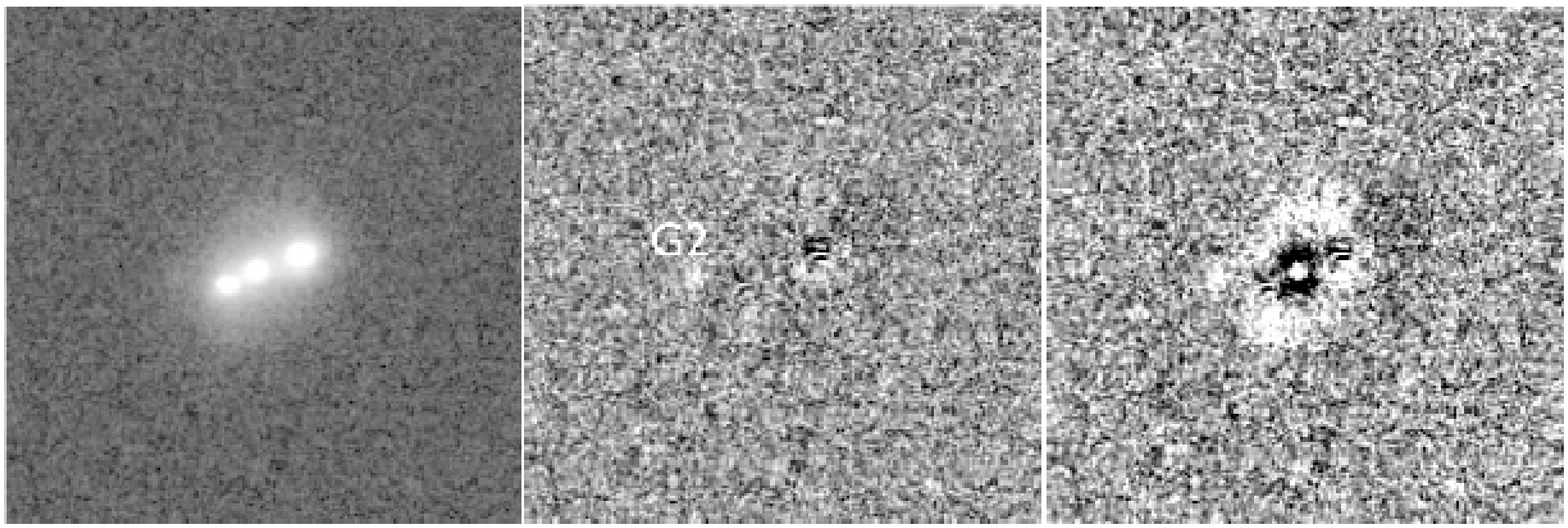}
\caption{Galfit/Hostlens modelling for SDSS~J1254+2235. North is up and East is to the left. \emph{Left}: original frame ($10''\times10''$), \emph{centre}: residuals after modelling with 2 Sersic profiles, using an analytical PSF. \emph{Right}: residuals after modelling with one Sersic profile. 
\label{fig:1254resid}}
\end{figure*}

A new, faint object well-modelled by a Sersic profile was found just East of image B (Figure \ref{fig:1254resid}, top right). Since colour information is unavailable, it is impossible to tell whether it is located in the foreground, background, or physically associated with the lensing galaxy. 

An SIE$+\gamma$ model is not employed, since the position angle to be used with this model is not obvious. It is worth noticing that in the SIS$+\gamma$ model, the position angle corresponds within $5$ deg to that of the brighter Sersic profile needed to fit the morphology. It is therefore likely that the dark matter halo of the galaxy is aligned to this brighter Sersic profile.

It must be noted that taking into account the component G2 significantly changes the mass models ($\gamma\sim0.044$ at $-12.4$ deg for the SIS$+\gamma$ with G2). This is because of its proximity to the system, despite its faintness. However, since the redshift of this component is unknown, its real effect on the system cannot be estimated.

\subsection{SDSS~J1313+5151}\label{section:1313}

The system is reported in \citet{ofek07} as a small-separation $\sim1.2''$ double at $z_s=1.875$, with spectroscopic lens redshift $z_l=0.194$. 

For this bright object, the standard deviation of pixels in a region containing B and G when fitting the object with a hybrid PSF and an analytical PSF is smaller in the case of the hybrid PSF, regardless of the PSF cut size. This means that the contribution from the non-analytical PSF structure is more dominant than the noise in the PSF wing. The object was modelled with a hybrid PSF built on the bright image A. The scatter between two different PSF core radii was included in the error budget: a small radius of 25 pixels, and another including the whole image A cut. The simulations were done with a cut of image A, after all other contributions were removed. This object is particularly bright enough that even accounting for additional noise introduced in the simulations, the resulting error bars are small. 

\begin{figure*}
\includegraphics[width=165mm]{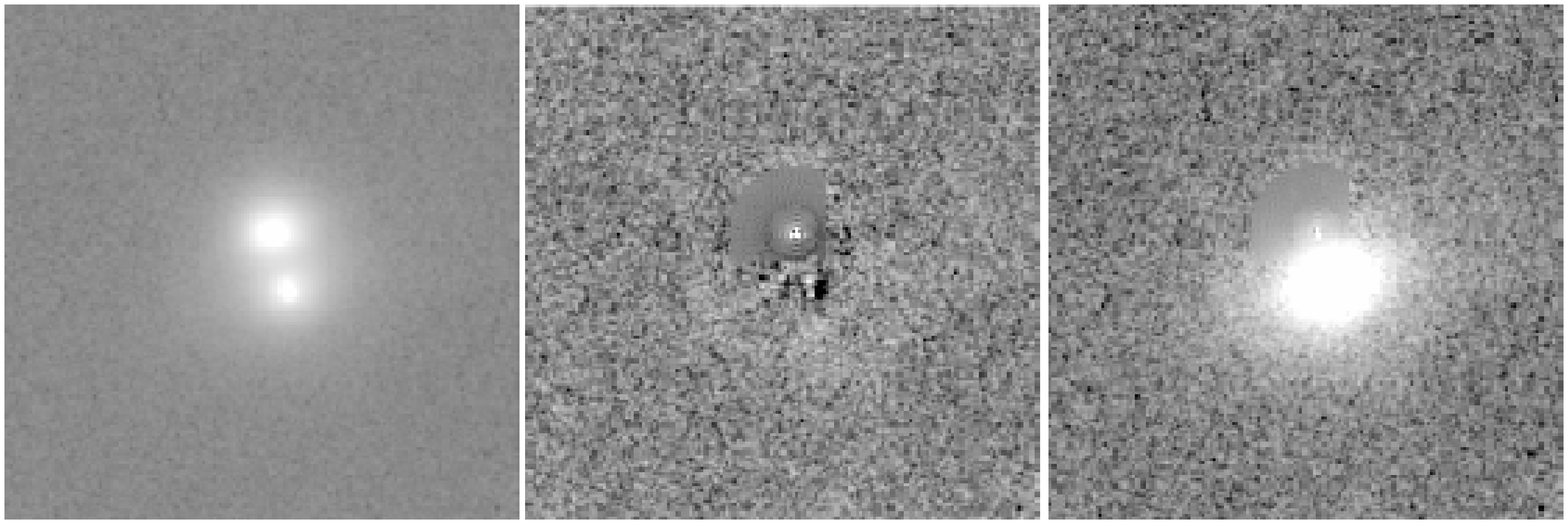}
\caption{Galfit/Hostlens modelling for SDSS~J1313+5151. North is up and East is to the left. \emph{Left}: original frame ($10''\times10''$), \emph{centre}: residuals after modelling with a free Sersic index and a hybrid PSF. \emph{Right}: residuals after subtracting only the quasar images.
\label{fig:1313resid}}
\end{figure*}

As a caveat however, this object has more significant residuals compared to the other in the current AO sample, amounting to as much as 10\% of the flux peak of G, at its location (Figure \ref{fig:1313resid}). Adding additional Sersic profiles would not improve the residuals, which are concentrated over a small area around the peaks of G and B. The reason for the residuals is likely the fact that Poisson noise is large at the peaks.

Compared to \citet{ofek07}, the image brightness rank has changed. The new brightness ranking, with the brightest image more distant from the lens, was predicted in that paper, with microlensing and extinction brought forward as explanations for the anomalous flux ratio and its observed strong dependence on wavelength. Indeed, both of these effects should be weaker at longer wavelengths.

The shear value in the SIS$+\gamma$ model is very small, and the model parameters are similar to the ones in \citet{ofek07}. The observed ellipticity is $\sim 3$ times larger than the equivalent ellipticity $e\sim3\gamma$, however the position angles are consistent to within $\sim8$ deg. 

\subsection{SDSS~J1322+1052}\label{section:1322}
\sectionmark{SDSS~J1322+1052}

SDSS~J1322+1052 was reported in \citet{oguri08-3}. It is a small-separation $\sim2.0''$ double at $z_s=1.716$, with the lens redshift estimate  $z_l\sim0.55$ based on the Faber-Jackson relation.

\begin{figure*}
\includegraphics[width=145mm]{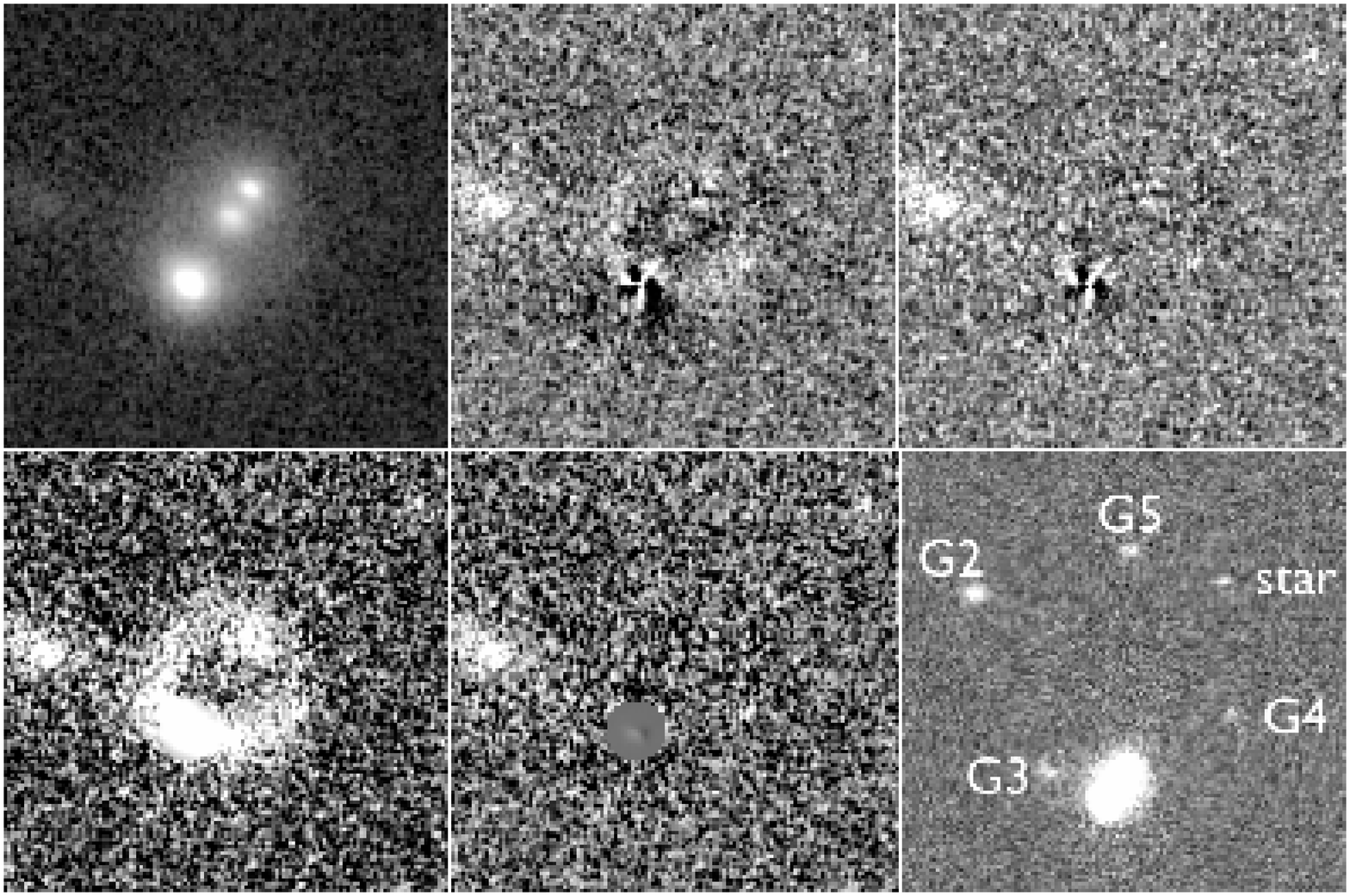}
\caption{Hostlens modelling for SDSS~J1322+1052. North is up and East is to the left. \emph{Top left}: original frame ($8''\times8''$), \emph{top centre}: residuals after modelling with an analytical PSF without considering the host galaxy. \emph{Top right}: residuals after modelling with an analytical PSF, and accounting for the host galaxy. \emph{Bottom left}: residuals after modelling the system with a hybrid PSF and subtracting all components except the host galaxy. \emph{Bottom centre}: residuals after modelling the system with a hybrid PSF and subtracting all components. Residuals when fixing the Sersic index of the lens at 4 or leaving it as a free parameter are visually indistinguishable. \emph{Bottom right}: $20\arcsec \times 20\arcsec$ region around the system.
\label{fig:1322resid}}
\end{figure*}

In the AO data, this object is similar to SDSS~J0904+1512, in the sense that a quasar host galaxy was found and modelled iteratively with a hybrid PSF (with a 10 pixel core radius). It is the brightest and most extended host galaxy in the current sample (Figure \ref{fig:1322resid}), with an almost complete Einstein ring. 

It became apparent during the modelling that the fit is problematic, with a very large Sersic index $>$ 10 for the lensing galaxy. This is the largest value encountered in the current sample. Therefore, another model was constructed in the same way, where the Sersic index is fixed at $n=4$. The residuals look remarkably similar in both models, and the final $\chi^2$/d.o.f values (obtained after the forth iteration) are similar. 
It is likely that the Einstein ring from the host galaxy biases the modelling of the lensing galaxy, introducing systematics. 
In view of the difficulty in estimating the correct morphology of the two galaxies, the host galaxy magnitude in Table \ref{tab:host}, obtained when the Sersic index of the lens is $n=4$, serves as an upper brightness estimate.

There are four galaxies in the close proximity of the system (Figure \ref{fig:1322resid}, bottom right). 
The direction of the shear in the SIS$+\gamma$ model is fairly close, within 17 deg, to the light orientation. The measured lens ellipticity is small, and including it in the SIE$+\gamma$ model has the effect of lowering the amount of required shear. Both models have similar shear orientations. Among the nearby galaxies, G3 the closest changes the estimated shear direction by 10 deg, making it less compatible with the orientation of G1, if modelled as an SIS scaled to the main lens through the Faber-Jackson law; the effect of the other galaxies is negligible. 

\subsection{SDSS~J1330+1810}\label{section:1330}

\subsubsection{Background and morphological modelling}

This object was reported and studied in \citet{oguri08-4}, based on visible and near-infrared observations. It is the only four-image lensed quasar in the current AO sample. It is a small-separation ($\sim1.8''$) system, with source redshift $z_s=1.393$ and spectroscopically measured lens redshift $z_l=0.373$. \citet{oguri08-4} reports no flux ratio anomaly in their modelling.

\begin{figure*}
\includegraphics[width=165mm]{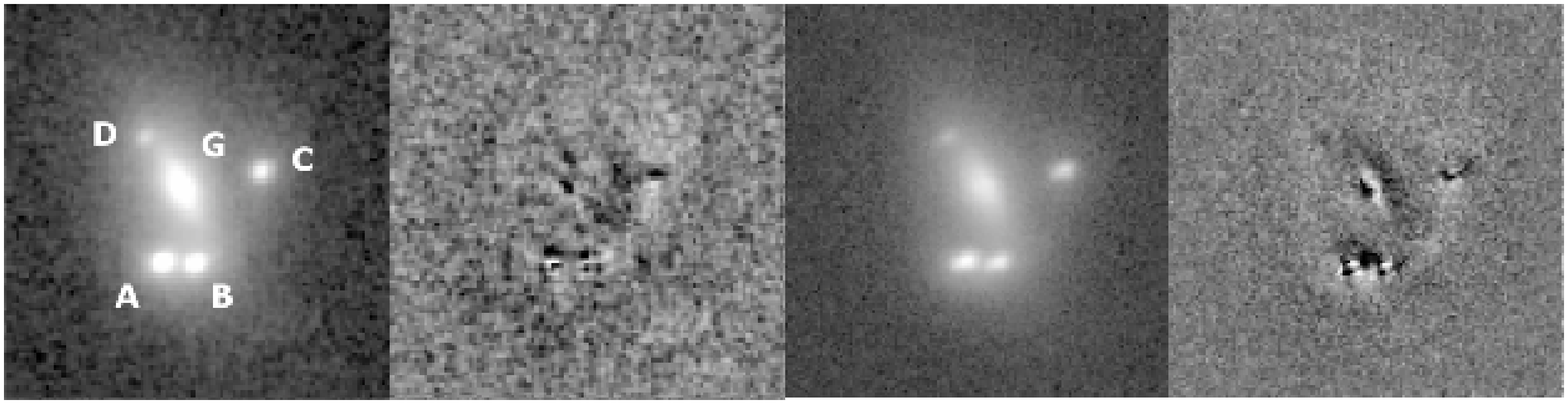}
\caption{SDSS~J1330+1810 imaging observations in 52 mas mode (2012; $\mathit{left, center left}$) and 20 mas mode ($2\times2$ binning, 2013; $\mathit{center right, right}$). Shown are original $5''\times5''$ frames and residuals after Hostlens/Galfit modelling. The lens is modelled as two concentric Sersic profiles in the 52 mas mode, and one Sersic profile in the 20 mas mode.
\label{fig:1330resid}}
\end{figure*}

This system was observed with AO twice. The deeper AO data from 2013, obtained with 20 mas pixel scale, was binned (2$\times$2 pixels), since the PSF was oversampled. The imaging data from both 2012 and 2013 was fitted with an analytical PSF, as four point-sources and a galaxy. The galaxy was modelled as two concentric Sersic profiles, resulting in a good overall subtraction (Figure \ref{fig:1330resid}). In the deeper 2013 data, two Sersic profiles were not found to be a significant improvement over a single profile. 
As seen in Figure \ref{fig:1330resid}, the residuals look worse in the 2013 data, as the objects seem to be overestimated on one side, and underestimated on another. 
The shallower 2012 data shows less signs of a host galaxy, which becomes slightly more prominent, in the form of arcs around images A, B and C, in the 2013 data. Unfortunately, due to the peculiar PSF shape in the data and the faintness of the arcs, the host galaxy could not be fitted convincingly, and it was ignored in the subsequent morphological modelling, with the exceptions of the simulations described below. 

When simulating the 2012 data, the separate PSF star in SDSS~J0946+1835 was used, as it is the most similar in terms of the analytical approximation. The simulations were only used to infer error bars for the astrometry and photometry. The morphology of the lens in Table \ref{tab:lensmorphology}, and photometry of all objects in Table \ref{tab:lensastrometry}, is taken from the deeper observations in 2013.
When simulating the 2013 data, the separate PSF star from SDSS~J0946+1835 was used, with a Gaussian convolution performed so that it matches the observed ellipticity and FWHM of the PSF estimated in the observed data. In order to check if the faint host galaxy introduces any biases, a model was created where the system was fitted with Hostlens as a lensed system with a lensing galaxy (SIE+$\gamma$) model and four lensed quasar images. Since this does not produce the observed flux ratios, a dark lens was used near image D in order to produce the observed ratio (see the section below on lens modelling). A host galaxy was then fitted, and its brightness was varied until it matched the brightness of the observed arcs near the three brighter images. This resulting brightness map of the host was added on top of each of the 100 simulations of the system, which were refit as five unrelated objects, without a lensing mass model.

The morphology of the lensing galaxy in the 2012 data, if a single Sersic component is fitted, is in good overall agreement with the morphology obtained in the 2013 data. The astrometry is quite consistent within error bars for A, B and C in the two data sets, in spite of the large residuals. Larger errors exist for the lensing galaxy, likely due to the less accurate PSF fitting of this more extended object. As a result, the average of the astrometry in the two observations was used, and the scatter between the two was used as the error bar on astrometry, calculated as the sample standard deviation. This method accounts for any possible systematics, as the data is of course independent between the two observations, conducted with different pixel size, with a distinctively different PSF, and different profiles for the galaxy (two Sersic and one Sersic component, respectively).

The photometry of the lensing galaxy is consistent between the AO data and the $K_s$-band value in \citet{oguri08-4}. Images A, B and C appear to have become brighter by about $0.15-0.3$ mag, while D is $\sim0.25$ mag fainter. Assuming that the error bars in \citet{oguri08-4}, which do not account for systematic errors coming from uncertainties in the PSF and the galaxy profile, are not significantly underestimated, this effect must be due to quasar intrinsic variability and microlensing. As remarked by \citet{oguri08-4}, microlensing likely plays the main role: while intrinsic variability would not change the observed flux ratios, particularly in view of the short estimated time delays (Table \ref{tab:massmodel}), the flux ratios A/B and A/C change (Figure \ref{fig:1330flux}), particularly in the shortest wavelength band, whereas A/C does not. 
In addition, B and D are located closer to the galactic disk, and therefore more likely to encounter stars.

The relative astrometry is consistent with the larger error bars in the discovery paper. The main difference in morphology is the lower Sersic index ($n=2.2$ compared to 3.6). This is closer to the fiducial value of $n=1$ for a disk-like galaxy. \citet{oguri08-4} notice that the galaxy is brighter than expected from the Faber-Jackson relation given its velocity dispersion (or alternatively should have a lower redshift), and surmise that this could be due to tidal stripping. However, they find no obvious companion galaxy near G, and no obvious signature of interaction. No such sign is observed in the AO data either. However, the new, lower Sersic index offers a more direct alternative. As shown in Figure \ref{fig:1330photoz}, an Sbc spectral template produces a photometric redshift that is in agreement with the spectroscopic redshift ($\chi^2$/d.o.f.=1.7 for 3 d.o.f.).

\subsubsection{Lens modelling}

\begin{figure}
\includegraphics[width=85mm]{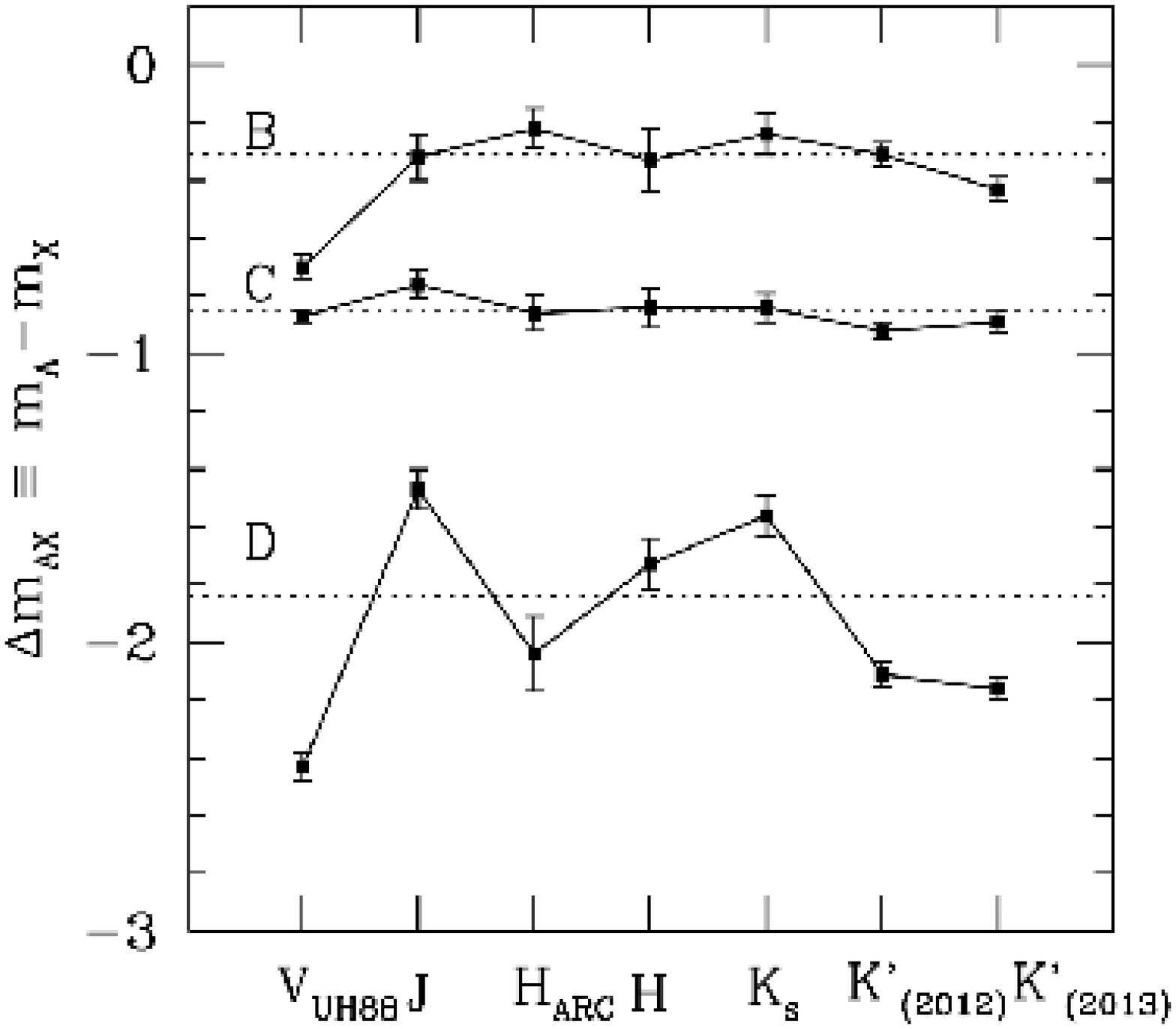}
\caption{Flux ratios of the SDSS~J1330+1810 quasar images from follow-up observations. Plotted are magnitude differences between images B-D and the brightest image A, $\Delta m_{AX} = m_A - m_X (X = B,C,D)$. Dotted horizontal lines indicate average values of individual ratios (excluding $V_{UH88}$), which are adopted in the mass modelling.
\label{fig:1330flux}}
\end{figure}

Instead of the 10\% error bars on the flux ratios, employed throughout this paper, since information is available in multiple bands, the average flux ratios from Figure \ref{fig:1330flux} (except for those in $V$-band) and their scatter were considered: B/A=0.75+0.05, C/A=0.46+0.02, D/A=0.19+0.05. In case of image D, this error is larger than 10\%. 

The SDSS environment is presented in Figure 7 from \citet{oguri08-4}. They identify two galaxy groups, one towards south-west, and a second to north-east, closers to the system, in which G3 \citep[using the original notation in ][]{oguri08-4} is the brightest galaxy. G3 is located at $z=0.311$, however for the majority of the other galaxies the photometric redshift is not accurate enough to distinguish which galaxies are associated with the redshift of G, or of G3. 

Here we consider two classes of mass models, one which accounts for the nearby galaxies (only those visible in the $K'$-band FOV), by modelling them as SIS lenses scaled relative to the main lens through the Faber-Jackson relation, and one which does not account for the environment. In both of these cases, models that account, and do not account for the measured flux ratios are considered separately. A summary of the models used, their corresponding $\chi^2$ and d.o.f. are shown in Table \ref{tab:1330chi}. Taking account of the environment is not a rigorous procedure, and is meant only for reference. This is mainly because it uses the untested assumption that the mass distribution is discrete, following individual galaxies rather than tracing the potential of a galaxy group. In addition, galaxies outside the $K'$-band FOV are not considered, and uncertainties in the scaling based on the Faber-Jackson relation are also ignored. We also assume that all galaxies are at the redshift of G. However this effect is not large: according to \citet{oguri08-4}, it is estimated that a weak external shear $\gamma$ at $z = 0.31$ is equivalent to an external shear of $\sim0.8\gamma$ at z = 0.373. 

As this system is a quad, there are enough d.o.f. to model both the external ellipticity and shear. The first model considered here, SIE$+\gamma$, was also considered in the discovery paper. Accounting or not for the measured flux ratios, \citet{oguri08-4} finds $\chi^2$/d.o.f.=1.14/4 and $0.33/1$, respectively. They therefore conclude that the object exhibits no sign of flux anomaly, which is typically encountered in the case of quads. Based on the order-of-magnitude more accurate astrometry, the updated values of 46/4 and 17/1 are found, respectively. The fits become relatively better it the environment is considered (20/4 and 14/1, respectively). These model are therefore significantly worse fits than suggested by the low resolution data, in terms of astrometry and flux ratios. 
 
A possibility is that the center of mass does not coincide with that of the light. However, the minimum offset from the observed lens position, necessary to obtain a good fit (when the flux ratios are discarded) is $\sim30$ mas, or $\sim150$ pc at $z=0.37$. This is five times the offset found by \citep{yoo06} based on the analysis of four lensed systems with Einstein rings, and is therefore unlikely. 

\begin{figure*}
\includegraphics[width=145mm]{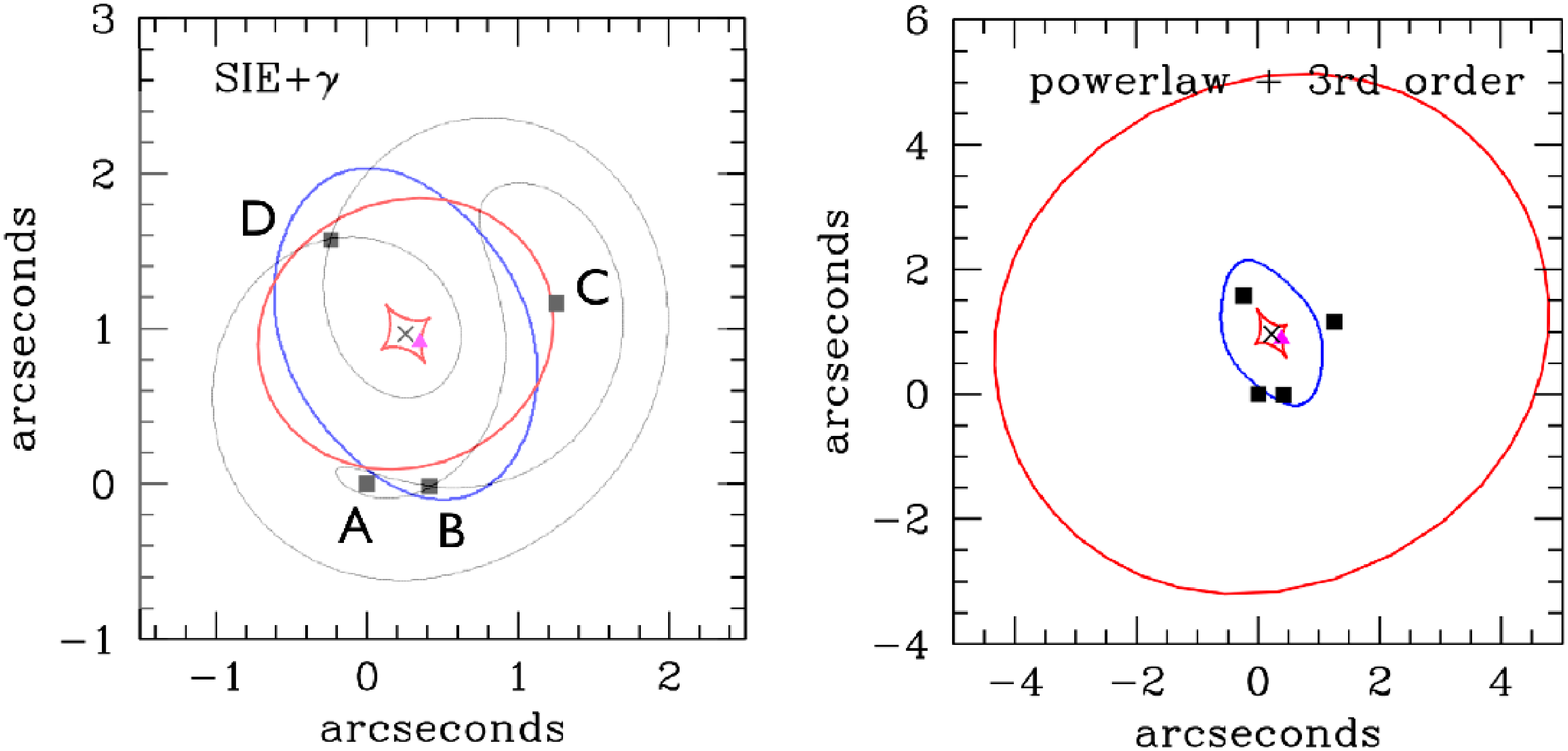}
\caption{ SIE$+\gamma$ $\mathit{(left)}$ and power law ellipsoid $+\gamma+\delta$ $\mathit{(right)}$ mass models for SDSS~J1330+1810. Critical lines $\mathit{(blue)}$, caustics $\mathit{(red)}$ and time delay contours $\mathit{(gray)}$ are shown. The cross shows the position of the lensing galaxy, and the squares show the positions of the images. \it{A colour version of this figure is available in the online version of the paper.}
\label{fig:1330model}}
\end{figure*}

\begin{figure}
\includegraphics[width=88mm]{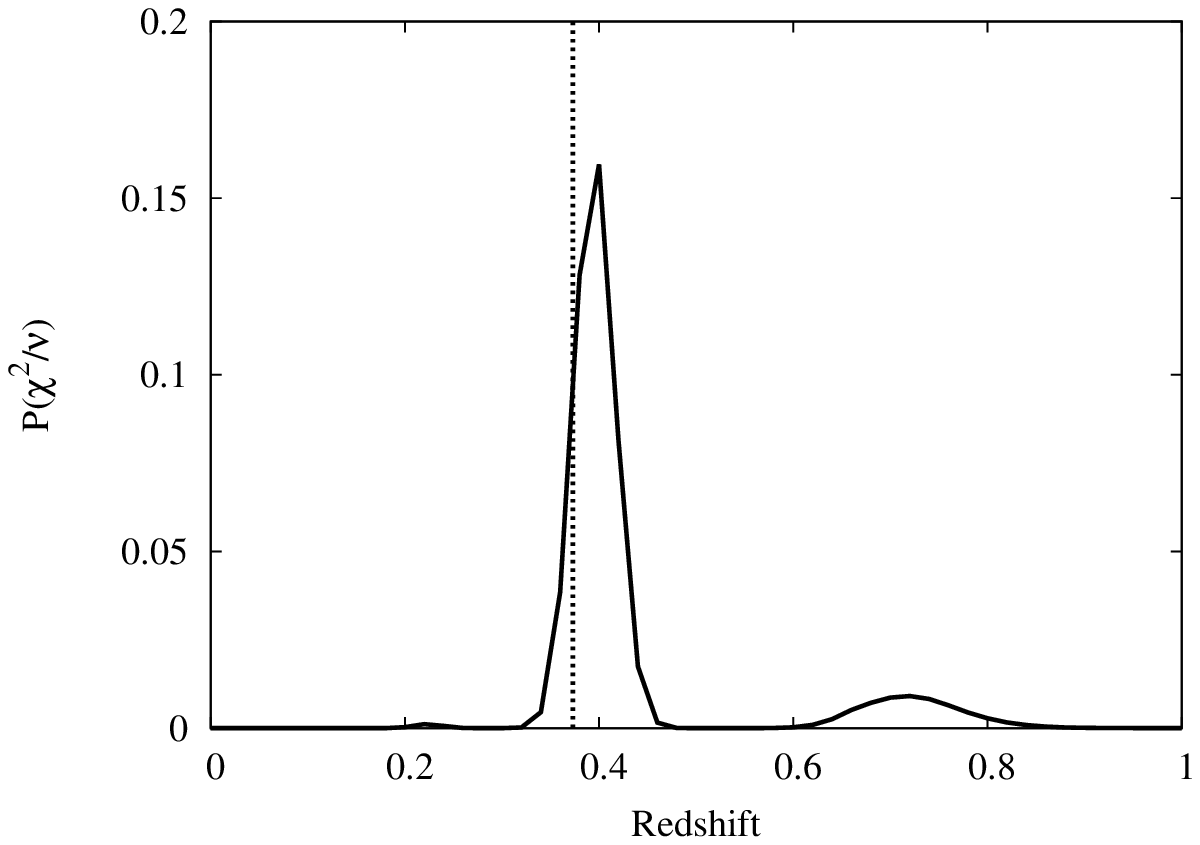}
\caption{Photometric redshift for the SDSS~J1330+1810 lensing galaxy, obtained with Hyperz for an Sbc template. The vertical line marks the spectroscopic redshift of the lens. The vertical axis represents the probability of exceeding the $\chi^2$ value obtained during the fit by chance, as a function of the numbers of d.o.f. (= 3).
\label{fig:1330photoz}}
\end{figure}

Next, more complicated but physically motivated mass models were fitted to explain the observed astrometry and photometry. These models are SIE + $\gamma + \delta$ (where $\delta$ refers to the third order perturbation component), Power law ellipsoid + $\gamma$, NFW + Sersic + $\gamma$, SIE + $\gamma$ + substructure and Power law ellipsoid + $\gamma + \delta$. Only two of the models, SIE + $\gamma + $substructure and Power law ellipsoid + $\gamma +\delta$, produce a formally good fit. The introduction of the $\delta$ component (Tables \ref{tab:massmodel} and \ref{tab:1330chi}, Figure \ref{fig:1330model}) is suggested in terms of the lens being located in a crowded environment, which contributes more than a shear term. But the compatibility (best-fit offset $\sim16$ deg and $\sim14$ deg w/o and w/ modelling the nearby galaxies, respectively) between the orientation of this component and the orientation of the lensing galaxy disk makes it likely that the component is physically associated with mass in the disk. The fitted mass slope $\sim2.4$ is larger than the average value of the SLACS lenses \citep{koopmans09}, but consistent with their intrinsic scatter. As a caveat however, the SLACS lenses are elliptical galaxies.  

\begin{figure*}
\includegraphics[width=120mm]{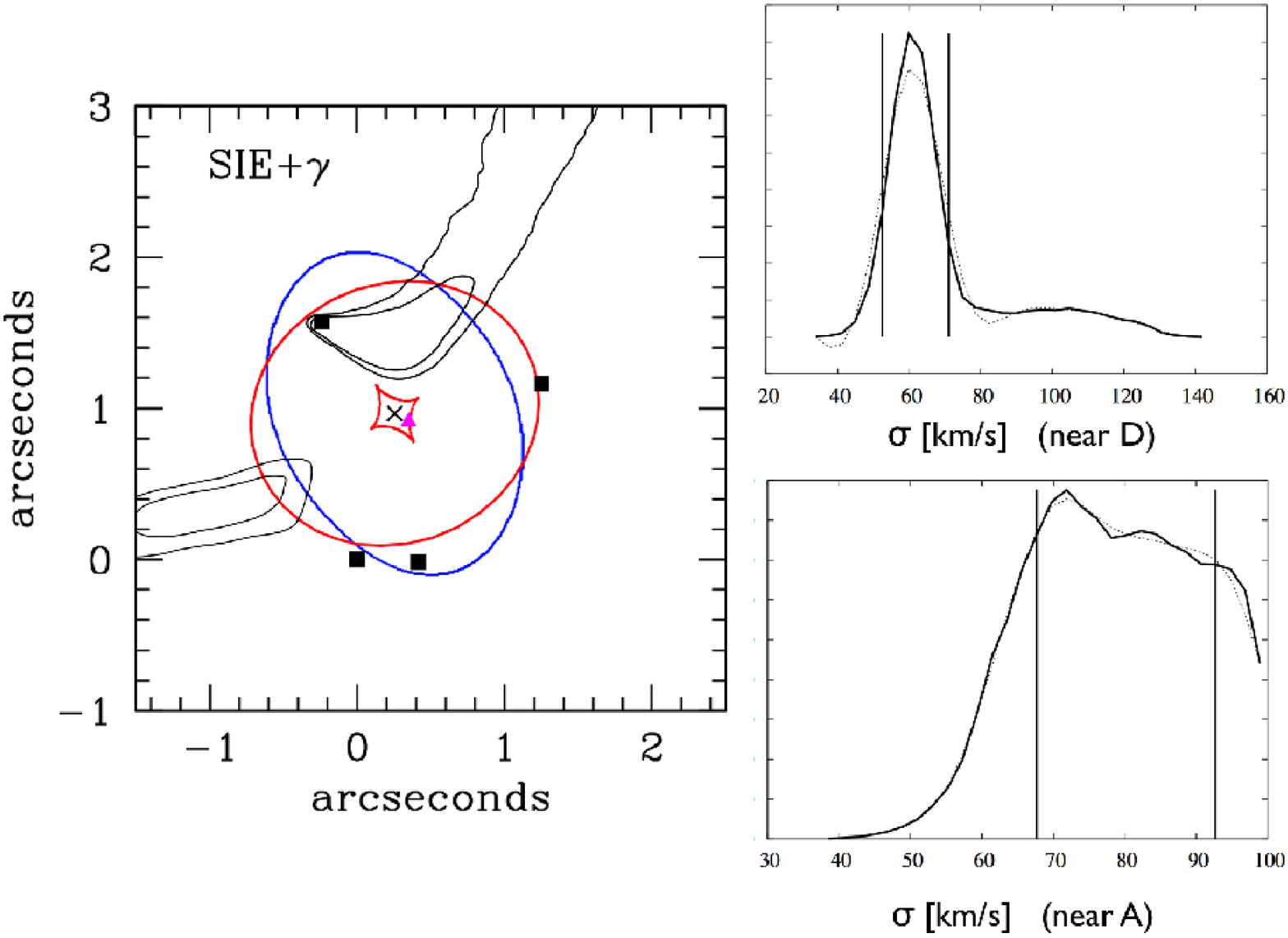}
\caption{ SIE$+\gamma$ $\mathit{(left)}$ model of SDSS~J1330+1810 with the location of a possible single substructure superposed. Contours are drawn from MCMC at 1$\sigma$ and 2$\sigma$. All other parameters are marginalised over. The two smaller plots represent velocity dispersion probabilities for the perturber. Vertical bars contain 68\% of the probability. \it{A colour version of this figure is available in the online version of the paper.}
\label{fig:1330substruct}}
\end{figure*}

For the SIE + $\gamma$ + substructure model, the position that a single perturber (satellite galaxy) would occupy in order to explain the observables was studied. There are two possible locations, in the vicinity of image A and D, respectively (Figure  \ref{fig:1330substruct}). The question is whether such substructure would be detected in the deep AO image. For comparison, if the satellite is compact, so as to be fit by a point-source, and has the same brightness as image D, it would have a velocity dispersion of $\sim95$ km/s. The most likely velocity dispersion of 60 km/s, in case the perturber is close to image D, translates to a virial mass of $6.5\times10^8 \mathrm{ M}_\odot$, and $\sim21$ mag ($\sim$2 mag fainter than image D). As the PSF was built from the images themselves, and the satellite galaxy could be located very close to image D, it is difficult to infer a detection magnitude. On the other hand, there might be multiple smaller perturbers causing the observed anomaly. \citet{fadely12} have considered models with a few as well as a population of substructure for HE 0435+1223, a quad with flux anomaly, and found that both are favoured by the data.

\begin{figure}
\includegraphics[width=75mm]{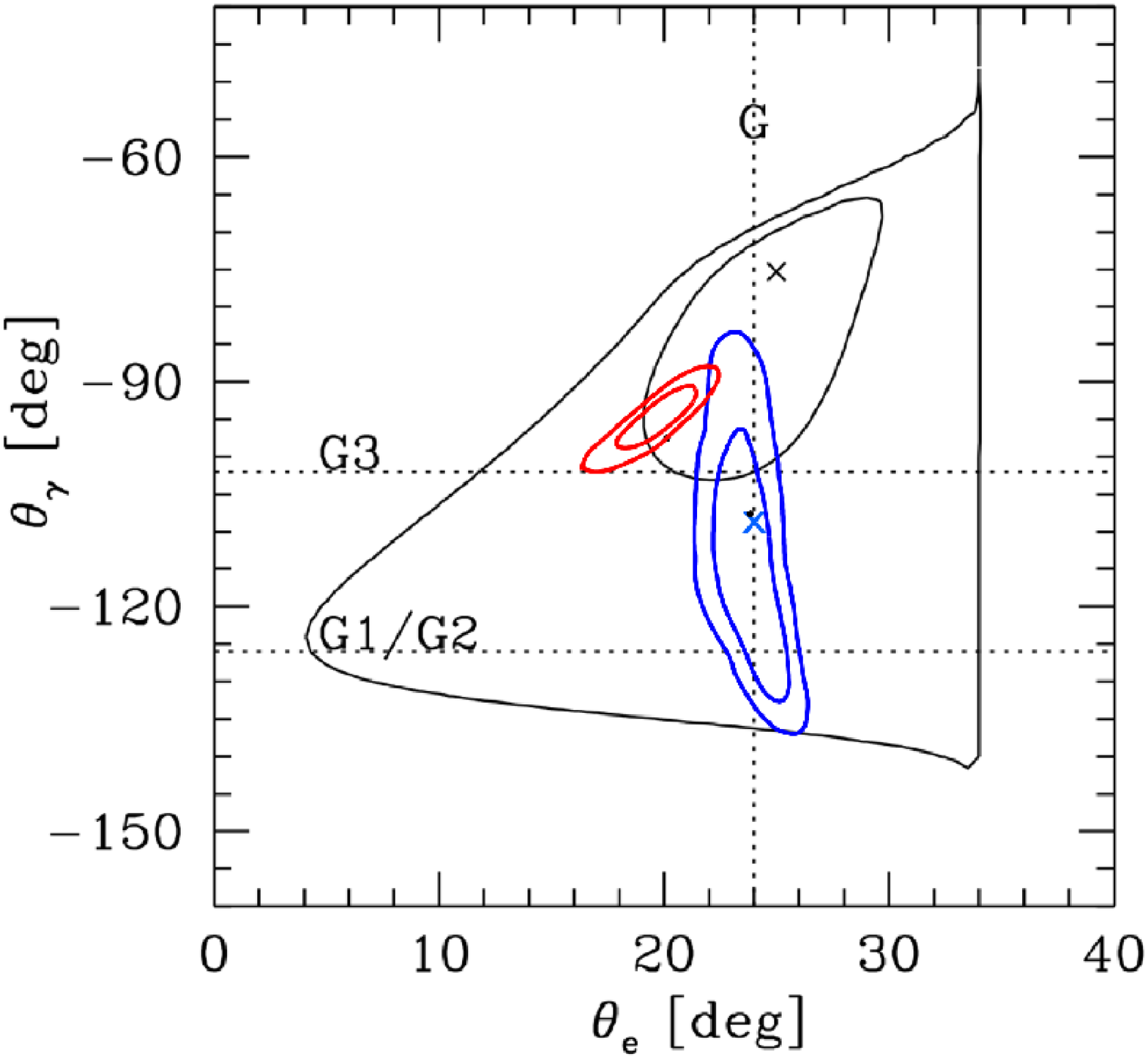}
\caption{ SDSS~J1330+1810 $1\sigma$ and $2\sigma$ contours of the position angles of the ellipticity and external shear. All other model parameters are marginalised over. Flux ratios are included as constraints. The lens models used are SIE$+\gamma$ from the original \citet{oguri08-4} observations $\mathit{(black)}$, as well as SIE$+\gamma$ $\mathit{(red)}$ and power law ellipsoid $+\gamma+\delta$ $\mathit{(blue)}$ respectively, using the IRCS+AO188 observations. Contours are estimated using $\Delta\chi^2=2.3$ and 6.2 for d.o.f. = 2 in case of the original data, and MCMC for the IRCS+AO188 observations. The crosses indicates the best-fit values. The vertical dotted line indicates the observed light orientation. Horizontal dotted lines are expected values when the external perturbations are caused by the centre of the group (G1/G2) or the nearby bright galaxy \citep[G3; following the notation in][]{oguri08-4}. \it{A colour version of this figure is available in the online version of the paper.}
\label{fig:1330PA}}
\end{figure}

As a last remark, Figure \ref{fig:1330PA} plots the orientation of the external shear for the SIE + $\gamma$ and the Power law ellipsoid + $\gamma + \delta$ model, in order to check which of G1 and G3 most affects the system. While the SIE + $\gamma$ clearly favours a shear direction towards G3, the Power law ellipsoid + $\gamma + \delta$ model does not distinguish between the two.

\begin{table}
 \centering
 \begin{minipage}{155mm}
  \caption{$\chi^2$ table for mass models of SDSS~J1330+1810}
  \begin{tabular}{@{}lcccc@{}}
  \hline
Model & 
w/ flux &
w/o flux  &
w/ flux  &
w/o flux\\ 
 & 
w/o field &
w/o field  &
w/ field  &
w/ field \\ 
\hline 
Sersic$+\gamma+\delta$ & 20/4 & 15/4 & 20/4 & 15/4 \\
Sersic$+\gamma$ & 43/6 & 29/3 &  & 29/3 \\
NFW+Sersic$+\gamma$ & 11/2 &  & 11/2 &  \\
Sersic$+\gamma+$substruct. & 43/3 & 0/0 &  &  \\
SIE$+\gamma+$substruct. & 0.9/1 & 0/-1 & 1/1 & 0/-1 \\
SIE$+\gamma$ & 46/4 & 17/1 & 20/4 & 14/1 \\
SIE$+\gamma$ (gal. pos. free) & 2/2 & 0/-1 & 0.6/2 &  \\
SIE$+\gamma+\delta$. & 15/2 & 0/0 & 29/2 &  \\
Pow. law ellip. $+\gamma$ & 11/3 & 0/0 & 7/3 & 0/0 \\
Pow. law ellip. $+\gamma+\delta$ & 1/1 &  & 0.4/1 &  \\
\hline
\end{tabular}
\\ 
{\footnotesize The values are reported as $\chi^2/\mathrm{d.o.f.}$}
\label{tab:1330chi}
\end{minipage}
\end{table}

To conclude, \citet{oguri08-4}, based on low-resolution data, notes that a standard SIE + $\gamma$ model can well reproduce both the astrometry and flux ratios in the system. Based on the new data, it was concluded that this is no longer the case. There is both astrometric and flux anomaly, but these can be solved with a smooth mass model, using a steep Power law ellipsoid + $\gamma + \delta$ model, where $\delta$ likely results from mass in the lens galaxy disk.  A satellite galaxy perturber cannot be ruled out, but it is not required by the data in this new model. Taking into account (albeit incompletely) the nearby lens environment does not modify these conclusions, but it does attenuate the discrepancies in the SIE + $\gamma$ model, particularly pertaining to the flux ratios. For comparison, \citet{sluse12} find 3 of the 14 quads with high-precision astrometry to have astrometric anomaly, when modelled with SIE + $\gamma$ and including the contribution of the nearest perturber. They do not consider more complicated models for the lensing galaxy. Spectroscopy of the field (which consists of at least two groups of galaxies at different redshifts), such as performed for other lenses by \citet{wong11}, is necessary to better account for the rich environment.


\subsection{SDSS~J1334+3315}\label{chapter:append1334}
   

The discovery of this double-image small-separation lensed quasar was published in \citet{rusu11}, based on spectroscopy, low-resolution imaging in the $zIJHK_s-$bands and LGSAO observations in the $JHKs-$bands. This was the first object observed in the present AO imaging campaign. It is one of only two sub-arcsecond systems in SQLS.

We noticed that the error bars on the fitted photometry in the non-AO $JHKs$ data also reported in \citet{rusu11} are more uncertain than the published error bars, which included just statistical errors reported by Galfit. This broadens the photometric redshift estimate of the lensing galaxy to $0.5 - 1.7$ (68\% confidence interval), for $\chi^2/\mathrm{d.o.f.}\sim1$, assuming the spectral template of an early-type galaxy. This is nonetheless lower than the source quasar redshift, in agreement with the gravitational lensing nature of the system. Although the $Iz-$ bands are particularly unreliable given the galaxy faintness, including them in the fit lowers the redshift to $0.53 - 0.96$ (68\% confidence interval), best-fit value 0.67 ($\chi^2/\mathrm{d.o.f.}\sim0.6$). 

\citet{rusu11} give an analysis of the AO data with a Gaussian + Moffat PSF built from image A, and quote only statistical errors. Here, the analysis was redone using Hostlens and an analytical PSF fit simultaneously on all components (A, B and G). The updated astrometry, photometry and morphology are summarised in Tables~\ref{tab:lensastrometry} and \ref{tab:lensmorphology}. 

\begin{figure*}
\includegraphics[width=140mm]{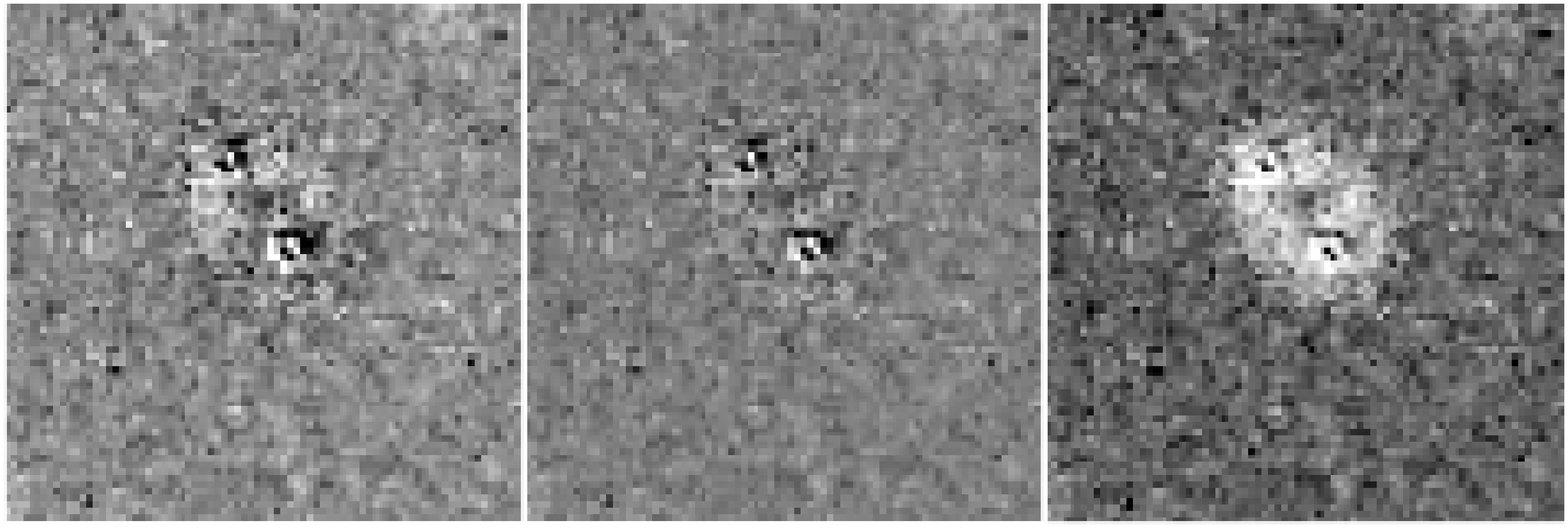}
\caption{Hostlens modelling of the SDSS~J1334+3315 AO data. North is up and East is to the left. \emph{Top left}: residuals after subtracting all components in $K'$ band ($4''\times4''$), without considering a host galaxy. \emph{top center}: residuals after modelling with an analytical PSF and  considering the host galaxy. \emph{Top right}: residuals after modelling with an analytical PSF, but not subtracting the host galaxy.
\label{fig:1334host}}
\end{figure*}

This is the only system in the current sample where the lensing galaxy is closer to the brighter image, which may make the PSF estimation more difficult. Also, the $K'-$band shows residuals around image B that may or may not be due to the quasar host galaxy (Figure \ref{fig:1334host}). Whether or not the (possible) host galaxy is incorporated in the models introduces significant changes in the estimated morphology and photometry ($\sim0.6$ mag) of the lensing galaxy.

In addition, the photometry/morphology of the lensing galaxy is not considered trustworthy in the AO data, due to the following caveat. There is a significant number of frames, particularly in the $HK'$-bands, that show a dark sky pattern. These were kept in the reduced data in order to increase the S/N, since overall a small number of frames were obtained, but may affect the photometric and morphological results. As a result, the SIE$+\gamma$ model was not used in Table \ref{tab:massmodel}. In particular, there are vast differences in the photometry of the lensing galaxy compared to the non-AO data, a fact which was also noted by \citet{rusu11}. The conclusions obtained in that paper regarding this system do not change.


\subsection{SDSS~J1353+1138}\label{section:1353}

SDSS~J1353+1138 was reported in \citet{inada06-1} as a small-separation $\sim1.4''$ double at $z_s=1.629$, with lens photometric redshift estimate of $z_l\sim0.25$. \citet{inada06-1} identify an additional component C in the vicinity of image A. They rule out several explanations for this component, concluding most likely that it is a foreground star, in agreement with its bluer colours.

\begin{figure*}
\includegraphics[width=145mm]{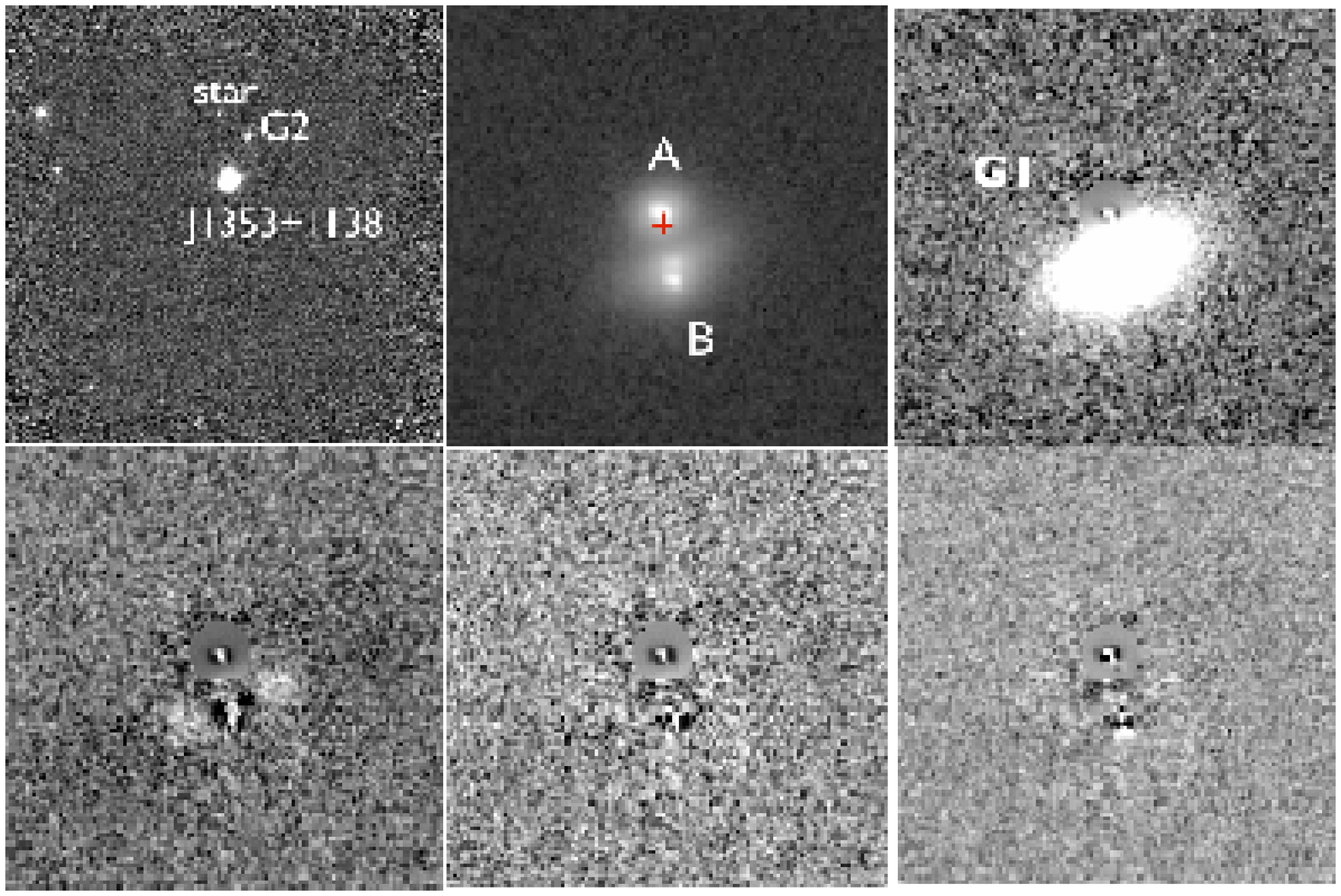}
\caption{Galfit/Hostlens modelling for SDSS~J1353+1138. North is up and East is to the left.  \emph{Top left}: complete FOV ($62''\times62''$), \emph{top centre}: original frame ($9''\times9''$); the possible component C is marked with a cross. \emph{Top right}: residuals after modelling with 2 Sersic profiles and a hybrid PSF; the lens galaxy is not subtracted. \emph{Middle left}: residuals after modelling with a 1 Sersic profile. \emph{Middle centre}: residuals after modelling with 2 Sersic profiles and subtracting all components. \emph{Middle right}: residuals after accounting for component C. \it{A colour version of this figure is available in the online version of the paper.}
\label{fig:1353resid}}
\end{figure*}

Since for this system image A is distinctly separated from B and G, the hybrid PSF method was employed. In the initial analytical modelling of all components, two case were considered, according to whether or not component C from the discovery paper is detected. In the case that component C is considered at the position inferred in \citet{inada06-1}, the $\chi^2$ value is lower than in the case that the component is not considered. However, comparing the results in both cases with subsequently generated hybrid PSFs reveals that in fact there are noticeably more residuals close to image B (Figure \ref{fig:1353resid}, bottom right), in the case that the hybrid PSF was built by first subtracting component C. Therefore, component C is considered undetected in the AO data, consistent with its known blue colours, in support of the foreground star hypothesis. 

For this object, as was the case with SDSS~J1313+5151, analytical PSF modelling produces a larger pixel standard deviation in the area containing B and G than the hybrid PSF does, therefore showing that the actual PSF is highly non-analytical. The scatter between the results when using a hybrid PSF core of 10 pixel radius (i.e. the region that was cut around image A, while the rest was replaced with an analytical wing), and another much larger radius, was included in the error budget. In the noise simulations, the system was created using a cut of image A as PSF, and remodelled using the hybrid PSF. As in the case of SDSS~J1254+2235, two concentric Sersic profiles were found to be necessary for a suitable modelling of the lensing galaxy. The galaxy has a large ellipticity, and the two Sersic components have orientations consistent with each other. 

The shear orientation in the SIS$+\gamma$ model differs from the orientation of G1 by only 10 deg, suggesting that the lens ellipticity is the main source of the quadrupole. However in the discovery paper, the two orientations differ by $\sim30$ deg, which led \citet{inada06-1} to the conclusion that they are inconsistent. For the SIE$+\gamma$ model, the ellipticity of G1a was used, this being the galaxy bulge component, with the smallest ellipticity. The effect of a nearby galaxy G2, assuming it is located at the same redshift, is small, changing the orientation of the shear by just 4 deg.


\subsection{SDSS~J1400+3134}\label{section:1400}

This system is reported in \citet{inada09} as a small-separation ($\sim1.7''$) double at $z_s=3.317$, with lens redshift estimate $\sim0.8$. 

\begin{figure*}
\includegraphics[width=165mm]{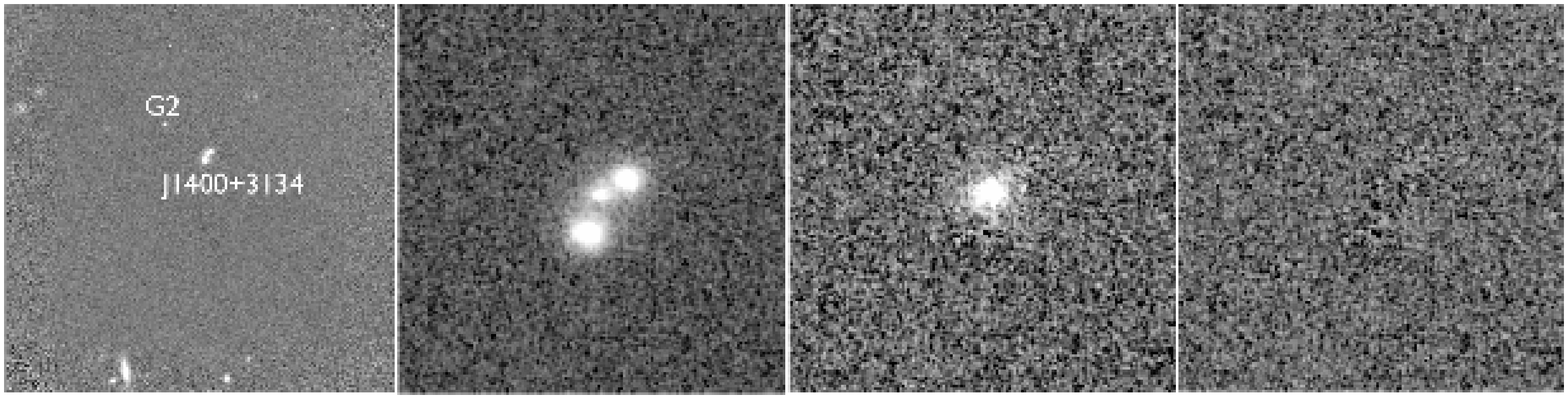}
\caption{Galfit/Hostlens modelling for SDSS~J1400+3134. North is up and East is to the left. \emph{Left}: $62''\times62''$ FOV.  \emph{centre left}: original frame ($10''\times10''$), \emph{centre right}: residuals after modelling with Sersic index 4 and an analytical PSF; lens galaxy is not subtracted. \emph{Right}: residuals after subtracting all components.}
\label{fig:1400resid}
\end{figure*}

 Although images A, B, and lensing galaxy G1 are sufficiently separated, the S/N is small, and as a result there are no significant residuals when the PSF is modelled analytically (Figure \ref{fig:1400resid}). If the Sersic index of G1 is a free parameter, the preferred value is very large, $\sim10$. Therefore, for this object, the Sersic index has been fixed at the fiducial value of 4, typical for an elliptical galaxy. This produces virtually no visual increase in the residuals at the location of the galaxy. Simulations were performed with the star in the SDSS~J0743+2457 FOV and its subsequent analytical fit. For the nearby galaxy G2, its position and magnitude were determined using the IRAF PHOT task.

The orientation of the shear in the SIS$+\gamma$ model is different at only $\sim14$ deg from the orientation of G1. While this suggests that the intrinsic ellipticity of G1 is the main source of the quadrupole, it is likely that the nearby galaxy G2 also affects the model. Indeed, by adding G2 as an SIS perturber to the SIS$+\gamma$ model, the shear position angle is in perfect agreement with the orientation of G1.

\subsection{SDSS~J1405+0959}\label{section:1405}

This system was discovered independently as a two-image lensed quasar candidate by \citet{jackson09} and \citet{inada14}. The AO observations, subsequent analysis and results are presented in a separate paper, \citet{rusu14}. There it is shown that the system is likely to be a triple, produced by two galaxy lenses. 


\subsection{SDSS~J1406+6126}

This system is reported in \citet{inada07}, as a double with $z_s=2.13$, $z_l=0.27$ and small image separation 1.08\arcsec. The morphology of the lensing galaxy is not presented in that paper.
 
\begin{figure*}
\includegraphics[width=165mm]{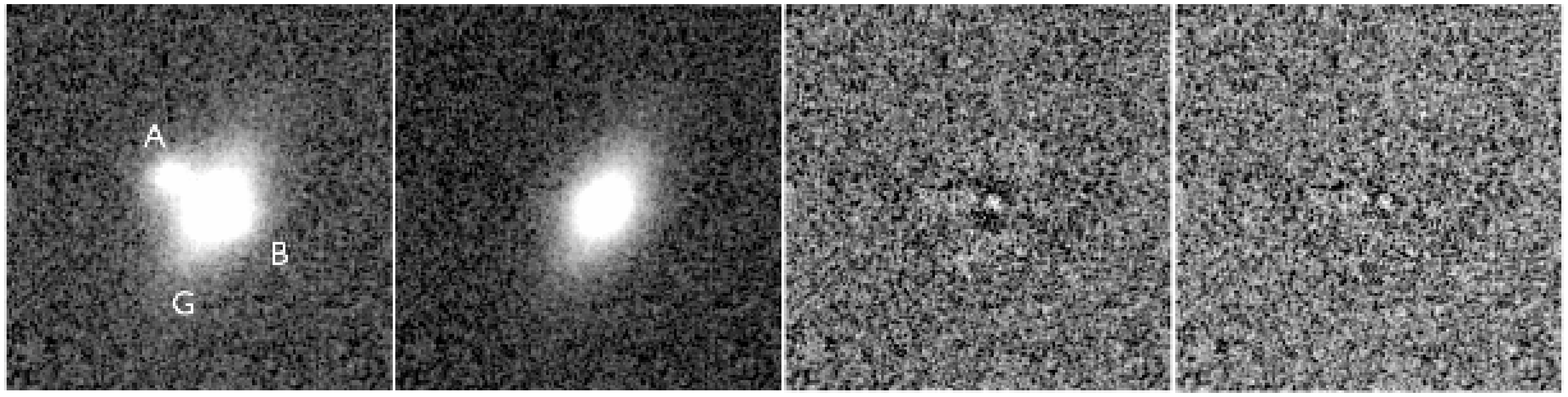}
\caption{Galfit/Hostlens modelling for SDSS~J1406+6126. North is up and East is to the left. \emph{Left}: original frame cut ($10''\times10''$), showing the location of G2 to the South. \emph{Centre left}: Residuals after subtracting images A and B, modelled with an analytical PSF. \emph{Centre right}: residuals after subtracting all components and modelling the lens with a Sersic index of 4. \emph{Right}: residuals after subtracting all components and modelling the lens with a free Sersic index. 
\label{fig:1406resid}}
\end{figure*}

The AO correction in the present campaign was very poor, with PSF $\sim 0.75\arcsec$. The PSF at image A is well approximated by as single Moffat profile. The system was simulated using this analytical PSF. The lens requires 2 concentric Sersic profiles ($n=1$ and 4, respectively; Figure \ref{fig:1406resid}). The flux ratio is in agreement with the discovery paper (V and I bands). The orientation of shear in the SIS+shear model is in agreement with the observed light orientation.

\subsection{SDSS~J1455+1447}

\begin{figure*}
\includegraphics[width=145mm]{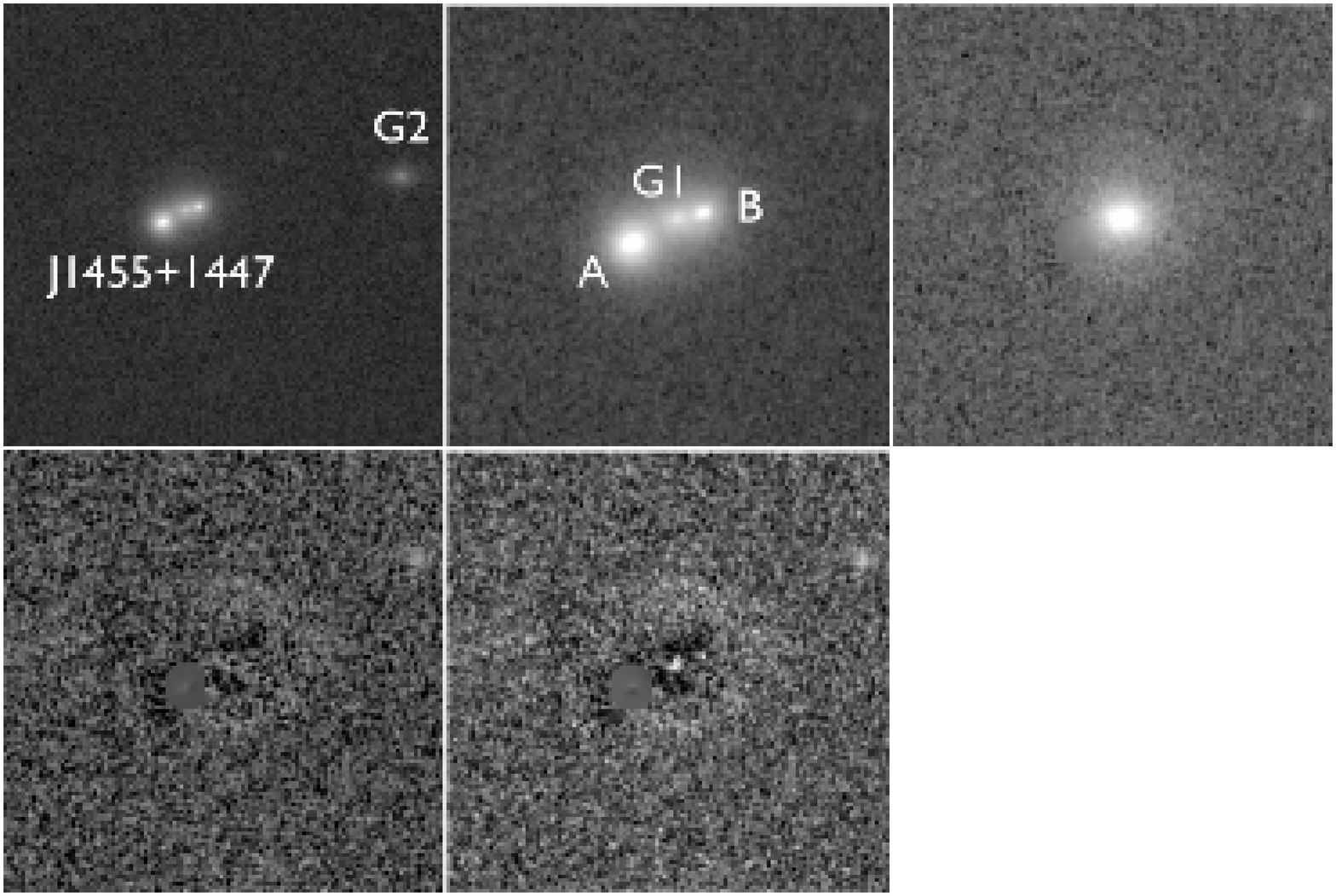}
\caption{Galfit/Hostlens modelling SDSS~J1455+1447. North is up and East is to the left. \emph{Top left}: original frame cut ($20''\times20''$), showing the location of G2 to the West. \emph{Top centre}: original frame cut ($10''\times10''$). \emph{Top right}: residual after modelling with a hybrid PSF, without subtracting the lens galaxy. \emph{Bottom left}: residual after subtracting the lens galaxy as well, modelled with a free Sersic index. \emph{Bottom centre}: residual after subtracting the lens galaxy as well, modelled with Sersic index 4. 
\label{fig:1455resid}}
\end{figure*}

The system was reported in \citet{kayo10} as a double at $z_s=1.424$, with image separation 1.73\arcsec. The AO data is suitable to construct a hybrid PSF from image A. Simulations were performed with the separate PSF star from SDSS~J0832+0404. The inferred relative astrometry is severely incompatible with the discovery paper, with all components being clearly resolved in the AO data. This also translates to large discrepancies in the inferred mass models. The orientation of the shear in the SIS$+\gamma$ model is close to both the orientation of the luminous profile of the lensing galaxy (at $\sim7$ deg) and the direction to the nearby component G2 (at $\sim17$ deg; Figure \ref{fig:1455resid}), and an even better agreement (at $\sim1$ deg) is obtained with the light orientation if G2 is explicitly accounted for in the mass model.

\subsection{SDSS~J1515+1511}\label{section:1515}
\sectionmark{SDSS~J1515+1511}

The discovery of this object is reported in \citet{inada14}. It is a small-separation ($\sim2.0''$) double at $z_s=2.054$, with spectroscopically determined lens redshift $z_l=0.742$.

This is the only object for which a suitable star to use as PSF exists in the FOV. The LGS was centred between the system and the nearby bright star, to insure as much as possible the same AO correction. To discard pixels that might be unreliable due to Poisson noise, the 2$\times$2 pixels at the peaks of A and B were masked, resulting in a quasar host galaxy that is three times as extended but half as luminous compared to the case in which the objects are not masked. 100 simulations with noise were also performed and modelled in the same way. The parameters of the host galaxy are shown in Table \ref{tab:host}.

A question that needs to be addressed is to what extent the star is a suitable PSF, especially considering the claim that the host galaxy is discovered. First, the analytical parameters of the star and the fitted system PSF in Table \ref{tab:analpsf} are nearly identical. Second, the extended residuals seen in case that the host galaxy is not considered cannot be caused by anisokineticism, since the tip-tilt star is located closer to the system, and not the star; therefore any effect due to anisokineticism would be expected to broaden the star, and not the system. Third, the host model produces a good fit which results in improved residuals, and the morphological parameters of the host are reasonable.

A nearby bright galaxy G2 is located in the frame, and its position and magnitude was measured with the IRAF PHOT task. The same was done for a faint galaxy located closer to the system ($20.18\pm0.07$ mag). 

The lens appears to be an edge-on disk-like galaxy, with very large ellipticity. The shear position angle in the SIS$+\gamma$ model is consistent with the galaxy orientation at about $\sim10$ deg, therefore the shear is likely caused by the galaxy intrinsic ellipticity. 

We checked that the expected influence of G2 and an even closer galaxy to the West are minimal, as including them scaled by the Faber-Jackson relation changes the shear orientation by only $\sim$1 and $\sim$2 deg, respectively. The SDSS photometric redshift of G2 is consistent with the redshift of G1, while the nearby perturber appears to be a disk-like galaxy.

\begin{figure*}
\includegraphics[width=145mm]{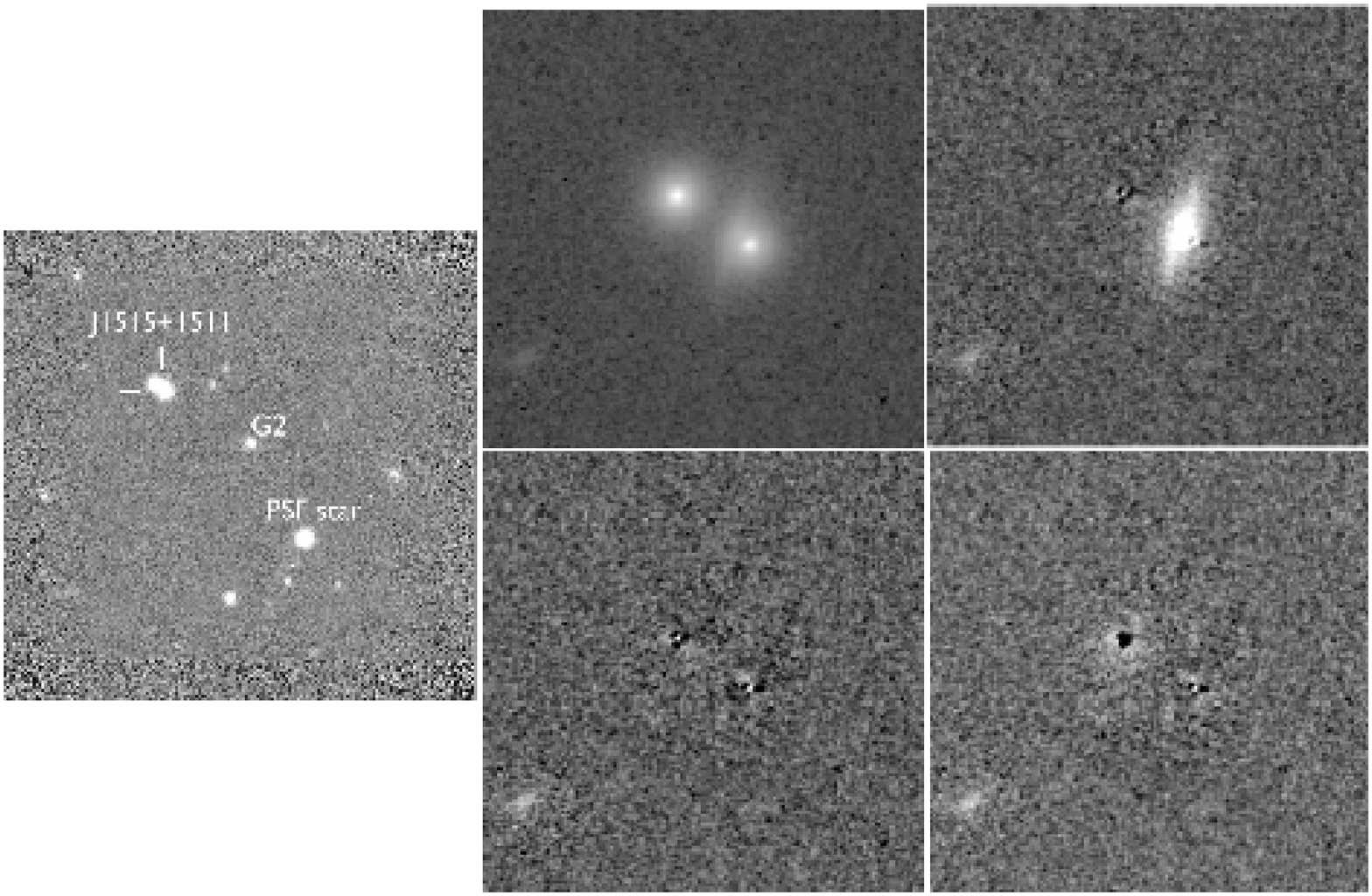}
\caption{Hostlens modelling for SDSS~J1515+1511. North is up and East is to the left. \emph{Left}: $66\arcsec \times 66\arcsec$ FOV. \emph{Top centre}: original frame ($10\arcsec \times 10\arcsec$), \emph{top right}: residuals after modelling with a free Sersic index and the nearby PSF star. \emph{Bottom centre}: residuals after modelling with a host galaxy. \emph{Bottom right}: residuals after modelling without a host galaxy.}
\label{fig:1515resid}
\end{figure*}


\subsection{SDSS~J1620+1203}\label{section:1620}

The discovery of SDSS~J1620+1203 is reported in \citet{kayo10}. It is an intermediate-separation ($\sim2.8''$) double at $z_s=1.158$, with spectroscopic lens redshift $z_l=0.398$.

\begin{figure*}
\includegraphics[width=165mm]{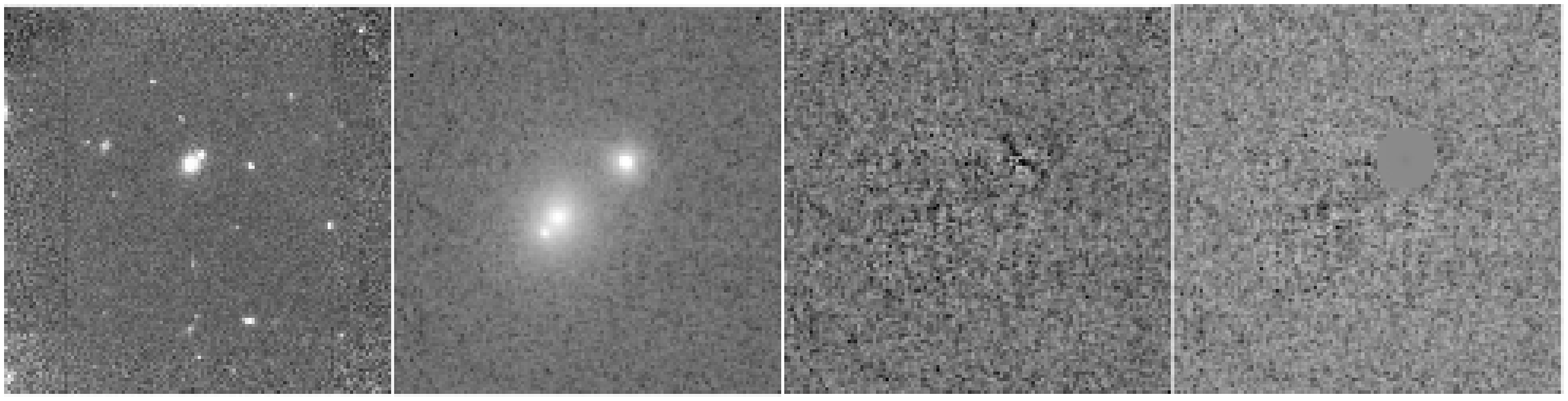}
\caption{Galfit/hostlens modelling for SDSS~J1620+1203. North is up and East is to the left. \emph{Left}: SDSS~J1620+1203 FOV ($64''\times64''$). \emph{Centre left}: original frame ($10''\times10''$), \emph{centre right}: residuals after modelling with a free Sersic index and an analytical PSF. \emph{Right}: residuals after modelling with a Sersic index 4 and a hybrid PSF.
\label{fig:1620resid}}
\end{figure*}

As the 2011 data consisted of only two frames, and therefore is significantly noisy, in addition to containing dark detector pattern, it was discarded entirely, except to check that the A/B flux ratio is consistent with the newer observations.

Since the system was observed both times in non-photometric conditions, we proceeded as follows. The system is detected in 2MASS with $K=15.203+0.136$; the image separation is $2.8''$, but the 2MASS pixel size is 2$''$, and the seeing was $2.7''$, so the system is not resolved in 2MASS. The photometric the zero-point was determined by equating the integrated magnitude of the objects, measured with Hostlens and Galfit, with the 2MASS value. Of course, this does not take into account any changes in the integrated magnitude of the quasar images since the 2MASS observations on 24 March 1999. Such effects are unlikely to be large, since the galaxy is $\sim 2$ mag brighter than the brightest quasar image. In addition, the lens is an elliptical galaxy with known redshift. From the Faber-Jackson relation, with the velocity dispersion inferred from the image separation, and calibrated with the \citet{rusin03} formula, the predicted magnitude of the lens is $K=15.05$. The lens magnitude estimated with the previous method is $K=15.50\ \pm \sim 0.04$. Adding to this the uncertainty in the 2MASS observation as zero-point, and the scatter of $\sim 0.5$ mag in the Rusin formula, these estimates are in agreement.


The morphological modelling of this system was perform
ed by creating a hybrid PSF with the core radius of 15 and 30 pixels. Simulations were made using the PSF star of SDSS~J1405+0959.

The observed light orientation is different from the shear direction in the SIS$+\gamma$ model. There are several galaxies around the system in the IRCS+AO188 FOV, and in addition there are several galaxies in the SDSS in the vicinity of the system (Figure \ref{fig:environ}), towards south and south-west, roughly the direction of the estimated shear.



\section{Summary of binary quasars and systems with unreliable data}\label{section:appendobjects}


\subsection{SDSS~J0926+3100}

\begin{figure*}
\includegraphics[width=175mm]{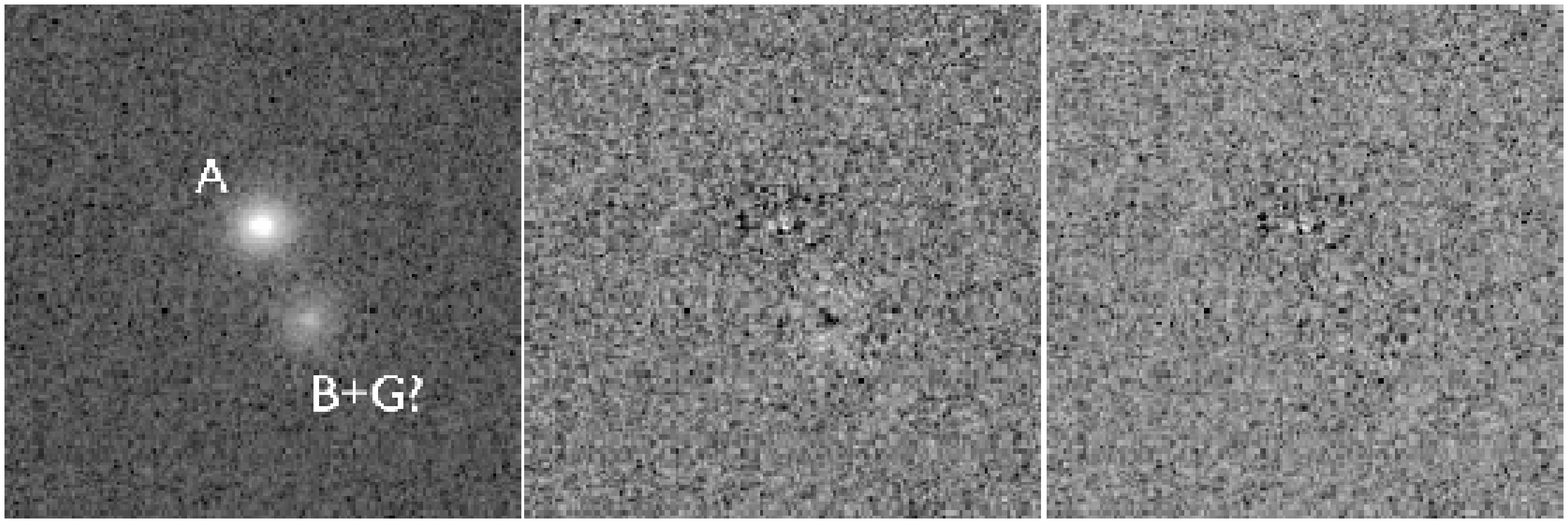}
\caption{Galfit/Hostlens modelling for the gravitationally lensed quasar candidate SDSS~J0926+3100. North is up and East is to the left. \emph{Left}: original frame cut ($10''\times10''$). \emph{Centre}: residuals after modelling with an analytical PSF, without fitting a lensing galaxy. \emph{Right}: residual after modelling with an analytical PSF, with a lensing galaxy.
\label{fig:0926resid}}
\end{figure*}

SDSS~J092634.560+305945.90 has been selected by SQLS as a gravitationally lensed quasar candidate from SDSS DR9\footnotemark \footnotetext{\url{https://www.sdss3.org/dr9/}}. The system does not appear to have a published spectrum in SDSS, and no follow-up spectrum was obtained. AO imaging of this system is presented in Figure \ref{fig:0926resid}. The system appears to consists of two well-resolved components separated by $\sim 2.00\arcsec$. Faint extended emission remains around the faint component after subtraction of an analytical PSF built on the bright object. Whereas the extended emission can be accounted for with an additional Sersic profile, it cannot be interpreted as a lensing galaxy, as its preferred location is just $\sim 1$ pixel offset from the point source, in the direction opposite to the bright component. Therefore we interpret this system to consist of a galaxy (the faint component to south-west) and either a quasar or a star (the bright component).

\subsection{SDSS~J1002+4449}\label{section:1002}

This system has been reported in \citet{inada08} as a two-image lensed quasar candidate. Several follow-up observations have been inconclusive. Spectroscopy in the SDSS is available for only one of the two quasars, which is located at $z=1.998$. The separation between the two components is $\sim0.7''$, which would make it the smallest-separation system in SQLS.

\begin{figure*}
\includegraphics[width=175mm]{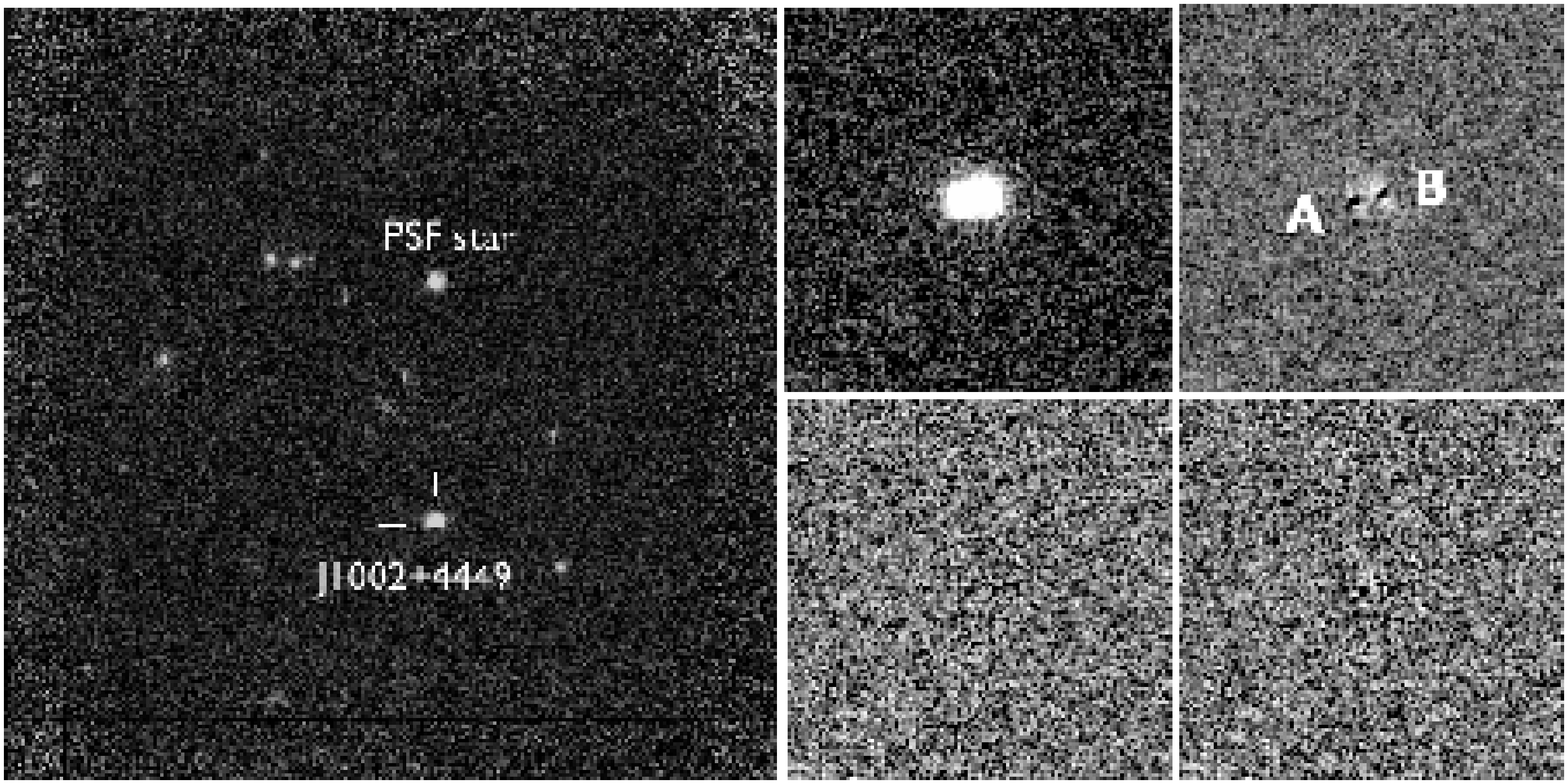}
\caption{Galfit/Hostlens modelling for SDSS~J1002+4449. North is up and East is to the left. \emph{Left}: ($64''\times64''$) FOV. \emph{Top centre}: original frame ($10''\times10''$), \emph{top right}: residuals after modelling the two quasars with the PSF star. \emph{Bottom centre}: residuals after modelling the system with a galaxy of Sersic index 4 and two analytical PSFs. \emph{Bottom right}: residuals after modelling with an analytical PSF, without a lens galaxy. 
\label{fig:1002resid}}
\end{figure*}

The observations were conducted with the FOV (and therefore the LGS) centered in between the system and the bright star to the North. However after eliminating bad-quality frames the configuration (PSF star, LGS, system) is no longer axially symmetric with respect to the LGS. As a result, the bright star has proven to be a bad PSF match (Figure \ref{fig:1002resid}, top right). Analytical PSF fitting with Hostlens produces much improved results (Figure \ref{fig:1002resid}, bottom). 

For this unconfirmed lensed candidate, the goal was to check if there is a lensing galaxy in between the quasars. Therefore the system was modelled both as two point-like objects, and as two point-like objects with a Sersic profile in between.
When modelling the morphology of the system with a lens galaxy, $\chi^2/\mathrm{d.o.f.}=102359/90579 = 1.1300$ was obtained, and  $\chi^2/\mathrm{d.o.f.}=102612/90585 = 1.1328$ with just two quasars, for a difference of $-0.0028$. The system was simulated 100 times as two point-like objects, using the bright star as a PSF, and then remodelled with analytical PSFs, both with and without a galaxy. This produced an average difference of $-0.0027\pm0.005\ (1\sigma$ of the distribution from the 100 values) between the models with and without a lensing galaxy. This strongly suggests that the galaxy detection is spurious. 

In addition, if the system is modelled with an SIS$+\gamma$ model, on the assumption that the galaxy detection is real and its derived relative astrometry is accurate, a shear of $\sim0.27$ is derived. This is the largest value obtained in the present sample for a system with a single lensing galaxy. There is no environmental clues in the FOV to suggest an explanation for this large shear. 

We conclude that the detection of a lensing galaxy cannot be claimed, making this object most likely a binary quasar. In principle, if this is a lens system, an estimate of the upper limit of the galaxy luminosity may be compared to its expected luminosity, derived from the observed quasar separation through the Faber-Jackson law, for a reasonable redshift range. However, the low S/N, the unavailability of an accurate PSF and the small separation does not permit a reliable estimate, and the best-fit estimate for the galaxy magnitude, $\sim17.15$ mag, is brighter than some of the other lenses in this paper, which means that the lens hypothesis cannot be entirely disregarded.

\subsection{SDSS~J1054+2733}

\begin{figure*}
\includegraphics[width=165mm]{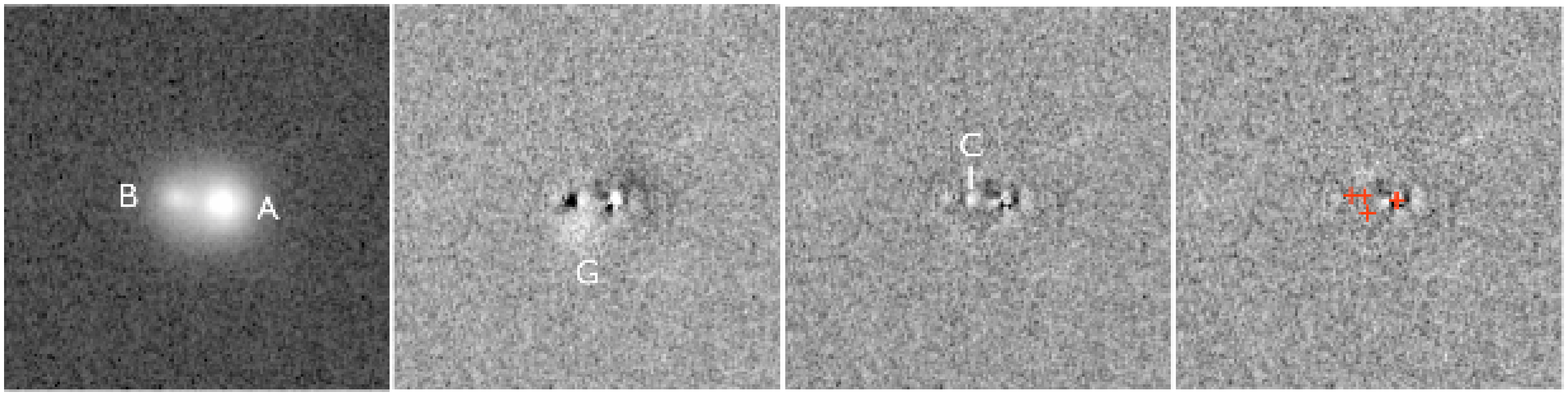}
\caption{Galfit/Hostlens modelling for SDSS~J1054+2733. North is up and East is to the left. \emph{Left}: original frame cut ($10''\times10''$). \emph{Centre right}: residuals after modelling with an analytical PSF, without fitting the lens galaxy. \emph{Centre right}: residual after including the lens galaxy in the fit. \emph{Right}: residual after modelling A, B, C, G simultaneously. Crosses mark the relative positions of all four objects. \it{A colour version of this figure is available in the online version of the paper.}
\label{fig:1054resid}}
\end{figure*}

The system was reported in \citet{kayo10} as a two-image lensed quasar at $z=1.452$, with an image separation $\sim1.27\arcsec$. The new AO data is shown in Figure \ref{fig:1054resid}. There are prominent residuals left at the centre of images A after the subtraction of an analytical PSF, as well as a flux excess in the vicinity of image B, in between A and B. Since a suitable independent PSF is unavailable, the hypothesis that the flux excess is due to non-analytical PSF residuals at image B cannot be discarded. If on the other hand this additional component is real (component C in Figure \ref{fig:1054resid}), it has not been detected at shorter wavelengths in \citet{kayo10}. The new component cannot still remains if the lensing galaxy is modelled with two sersic profiles, and it requires a point source to model satisfactorily (Figure \ref{fig:1054resid}, bottom right). On the other hand, the lensing galaxy appears to be offset towards south. 

The astrometry of this system, if accounting for component C, is given in Table \ref{tab:1054}. Due to the alignment of A, B, C, and the offset of the lensing galaxy, we were not able to fit a reasonable lensing model that can account for component C either as an extra quasar image, a central third image (with a non-singular isothermal sphere), or as part of a four-image system. If C was a chance superposition of a star, it would be expected to be brighter at visible wavelengths, although its non-detection in the discovery paper may be attributed to the poor resolution and small image separation. We conclude that the AO data of this system is not well characterised. 

\begin{table}
 \centering
 \begin{minipage}{80mm}
  \caption{Inferred Relative Astrometry and Absolute Photometry for SDSS~J1054+2733}
  \begin{tabular}{@{}cccc@{}}
  \hline 
  Object & 
  $\Delta X$ [$''$] & 
  $\Delta Y$ [$''$] & 
  $K'$ [mag] \\ 
  \hline 
A & 0.000 & 0.000 & 15.54 \\
B & $-1.226$ & 0.144 & 17.01 \\
C & $-0.869$ & 0.086 & 18.25 \\
G & $-0.717$ & $-0.389$ & 17.65\\
\hline
\end{tabular}
\\ 
\label{tab:1054}
\end{minipage}
\end{table}

\subsection{SDSS~J1128+2402}

\begin{figure*}
\includegraphics[width=175mm]{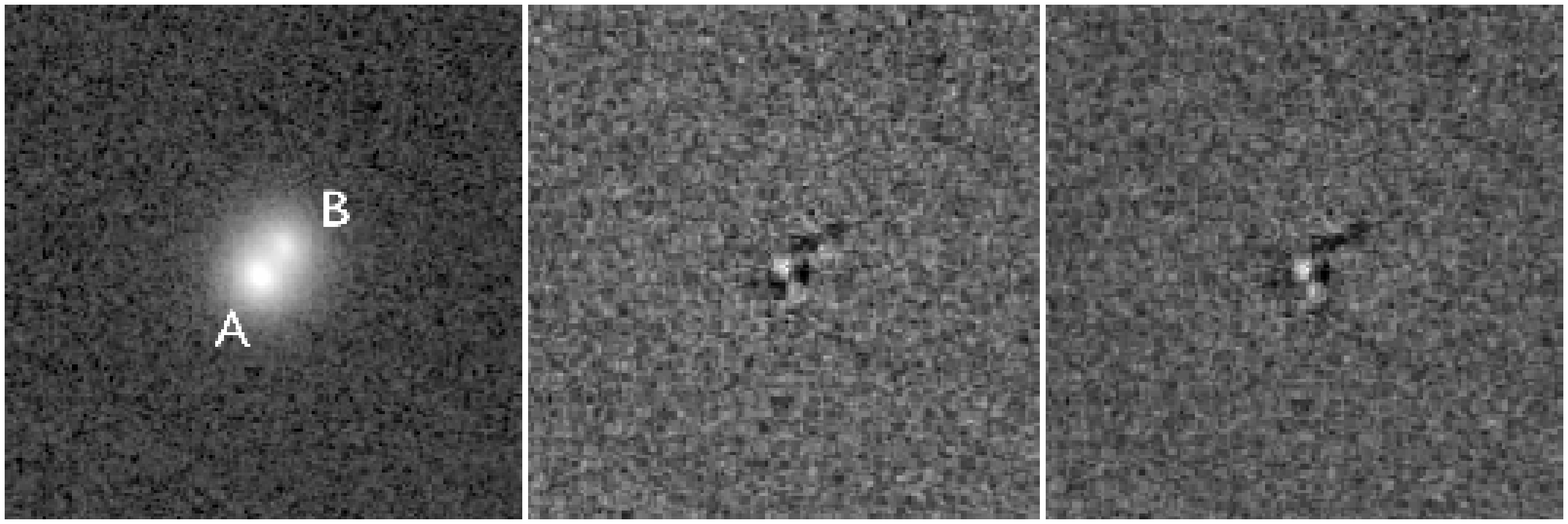}
\caption{Galfit/Hostlens modelling for SDSS~J1128+2402. North is up and East is to the left. \emph{Left}: original frame cut ($10''\times10''$). \emph{Centre}: residuals after modelling without a lens galaxy. \emph{Right}: residuals after modelling with a lens galaxy.  
\label{fig:1128resid}}
\end{figure*}

This system is reported in \citet{inada14} as a two-image lensed quasar at $z=1.608$, with an image separation $\sim0.84\arcsec$. In the new AO data, the inferred image separation is $0.75\arcsec$, making this system the smallest-separation lensed quasar in the SQLS.           
However, the lensing galaxy is undetected in the new observations (Figure \ref{fig:1128resid}), with the residuals being identical whether a lens galaxy is modelled or not. Without visual confirmation of improvement in the residuals, the extracted physical parameters of the lens are unreliable, as it is likely that noise is being fitted. Considering the very small separation of this system and the fact that there are non-analytical residuals which cannot be well fit, we conclude that the faint lensing galaxy, visible in \citet{inada14}, cannot be reliably resolved from the bright quasar images with the current technique.

\begin{figure*}
\includegraphics[width=130mm]{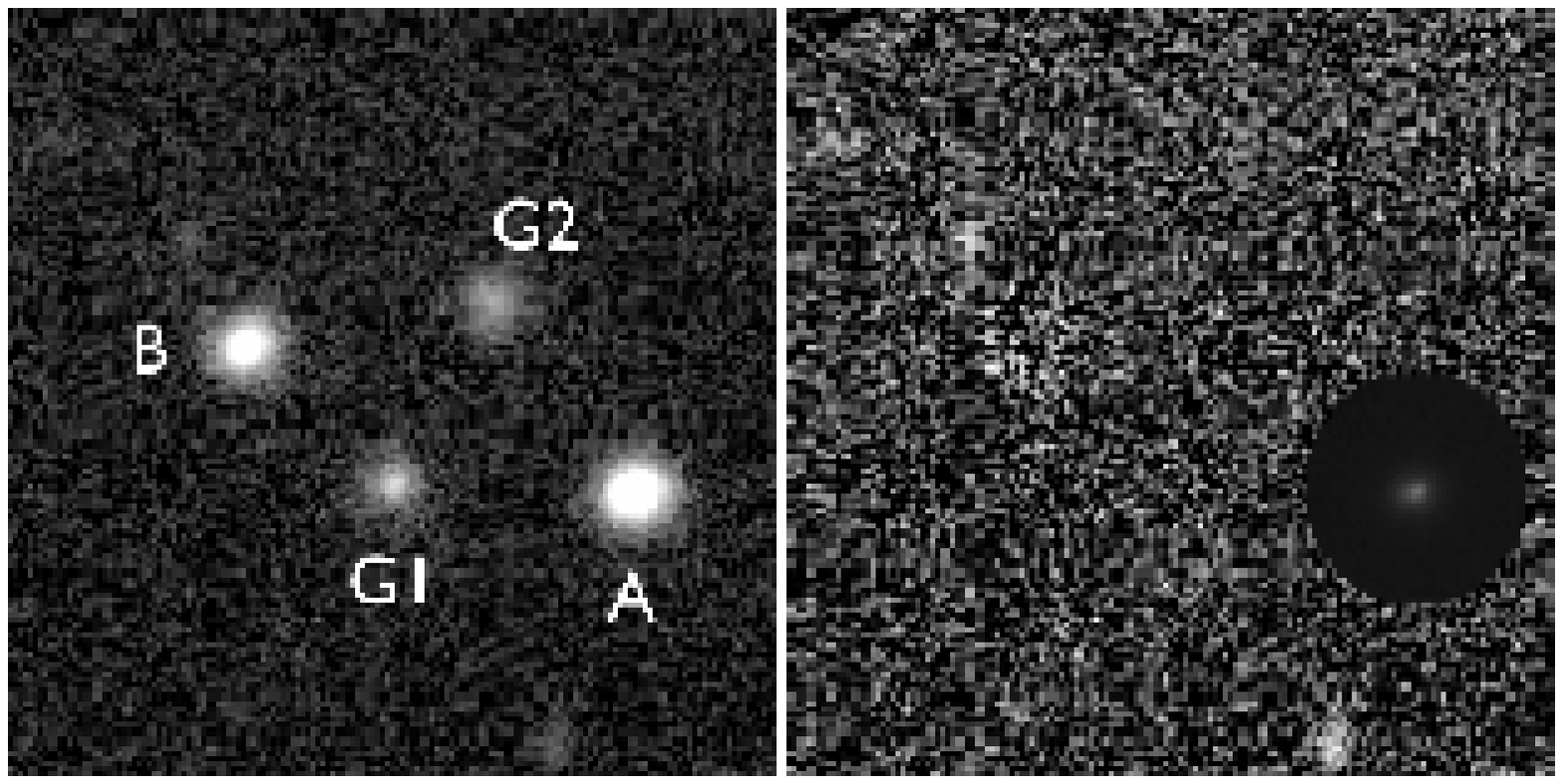}
\caption{$16\arcsec\times16\arcsec$ cut around SDSS~J1320+1644. \emph{Left}: original frame. Object names correspond to those in \citet{rusu13}. \emph{Right}: residuals after modelling A, B, G1 and G2 using a hybrid PSF built on image A. 
\label{fig:1320resid}}
\end{figure*}


\subsection{SDSS~J1320+1644}\label{section:1320}

The discovery of SDSS~J1320+1644 and subsequent data analysis based on non-AO data is reported in \citet{rusu13}. This is a large-separation quasar pair, likely to be images of a lensed quasar source at $z_s=1.502$, by a lensing galaxy group/cluster at $z_l=0.899$. However the lensing nature of this object is not secure. \citet{rusu13} noted that the most efficient way to test the lensing hypothesis is by performing AO observations sensitive enough to detect the quasar host galaxy, and search for evidence of lensed arcs. 

The acquired AO data was affected by cloud coverage, and as such did not reach the intended depth, being in fact shallower than the original non-AO data. Even after $2\times2$ pixel binning to increase S/N, there are no visible signs of a host galaxy detection (Figure \ref{fig:1320resid}). Both an analytical PSF and a PSF built on image A were used, leading to clean residuals for image B. 

The only new insight brought by the new AO data is on the quasar flux ratio. This is A/B $\sim1.8$, and is significantly larger than the values of 1.4 and 1.0, considered in \citet{rusu13}. While this has little impact on their inferred lens models, since the main model constraints were astrometric, it does impact their discussion on the intrinsic flux ratio. \citet{rusu13} argued that, since typical changes due to intrinsic variability and microlensing are small in the near-infrared, the flux ratio $\sim 1.4$ observed in the spectra at long wavelengths and in the $JHK_s-$bands is representative of the true flux ratio. An alternative hypothesis was that the flux ratio $\sim 1.0$ found at longer infrared wavelengths is more accurate, and the changes due to microlensing at shorter wavelengths occur on time scales longer than the $\sim1$ year spanned by the observations available at that time. The new flux ratio supports this second hypothesis in the sense that larger flux variations do indeed occur on time scales of $\sim3$ years, assuming, of course, that the gravitational lens hypothesis is correct for this system. 

\bsp

\label{lastpage}
\end{document}